\documentclass[]{jfm}

\usepackage{graphicx}
\usepackage{epstopdf, epsfig}
\usepackage{newtxtext}
\usepackage{float}
\usepackage{adjustbox}
\usepackage{newtxmath}
\usepackage{natbib}
\usepackage{hyperref}
\hypersetup{
    colorlinks = true,
    urlcolor   = blue,
    citecolor  = blue,
}

\usepackage{amsmath,amsfonts,mathtools}
\usepackage{subfigure}
\usepackage{bm}

\usepackage{array}
\newcolumntype{P}[1]{>{\centering\arraybackslash}p{#1}}
\usepackage{makecell}
\usepackage[table]{xcolor}

\shorttitle{ Orientation dynamics of a spheroid in the simple shear flow of a weakly elastic fluid}
\shortauthor{P. K. Singeetham, D. Madival, P. Garg and G. Subramanian}

\title{Orientation dynamics of a spheroid in the simple shear flow of a weakly elastic fluid}

\author{Pavan Kumar Singeetham\aff{1},
	Deepak Madival\aff{1}, 
        Piyush Garg\aff{2} 
	\and  Ganesh Subramanian\aff{1}
 \corresp{\email{sganesh@jncasr.ac.in}} }
 
\affiliation{\aff{1} Engineering Mechanics Unit, Jawaharlal Nehru Centre for Advanced Scientific Research, Bangalore-64, India.
\aff{2} Complex Fluids and Flows Unit, Okinawa Institute of Science and Technology Graduate University, Okinawa 904-0495, Japan.
}

\begin{document}

\maketitle

\begin{abstract}
We investigate the orientation dynamics of a neutrally buoyant spheroid, of an arbitrary aspect ratio\,($\kappa$), freely rotating in a weakly viscoelastic fluid undergoing simple shear flow. Weak elasticity is characterized by a small but finite Deborah number\,($De$), and the suspending fluid rheology is therefore modeled as a second-order fluid, with the constitutive equation involving a material parameter $\epsilon$ related to the ratio of the first and second normal stress differences; polymer solutions correspond to $\epsilon\in[-0.7,-0.5]$. Employing a reciprocal theorem formulation, along with expressions for the relevant disturbance fields in terms of vector spheroidal harmonics, we obtain the spheroid angular velocity to $O(De)$. In the Newtonian limit, a spheroid rotates along Jeffery orbits parametrized by an orbit constant $C$, although this closed-trajectory topology is structurally unstable, being susceptible to weak perturbations. For $De$ well below a threshold, $De_c(\kappa)$, weak viscoelasticity transforms the closed-trajectory topology into a tightly spiralling one. A multiple-scales analysis is used to interpret the resulting orientation dynamics in terms of an $O(De)$ orbital drift. The drift in orbit constant over a Jeffery period $\Delta C$, when plotted as a function of $C$, identifies four different orientation dynamics regimes on the $\kappa-\epsilon$ plane. For $\epsilon$ in the polymeric range, prolate spheroids always drift towards the spinning mode. Oblate spheroids drift towards the tumbling mode for $\kappa > \kappa_c(\epsilon)$, but towards an intermediate kayaking mode for $\kappa < \kappa_c(\epsilon)$; $\kappa_c$ increases from $0.2$ to $0.454$ as $\epsilon$ varies from $-0.5$ to $-0.7$. The rotation of spheroids of extreme aspect ratios, either slender prolate spheroids\,($\kappa \gg 1$) or thin oblate ones\,($\kappa \ll 1$), about the vorticity axis, is arrested for $De \geq De_c(\kappa)$. The resulting rotation-arrested states exist within the flow-gradient plane in both cases, being nearly aligned with the flow direction for the prolate case, and oriented in the vicinity of the gradient direction for the oblate one. While rotation-arrested states are unstable to off-plane perturbations in the former instance, secondary bifurcations beyond the primary rotation-arrest threshold lead eventually to stable arrested states for thin oblate spheroids.
\end{abstract}
\begin{keywords}
Spheroids, Viscoelasticity, Orientation dynamics, Second-order fluid, Rotation arrest
\end{keywords}
\section{\bf Introduction}\label{intro}	
Anisotropic particle suspensions are commonly encountered in both natural and industrial applications. Paper manufacture involves processing of aqueous cellulose fiber suspensions, and the quality of the finished product is partly determined by the orientation distribution of the high-aspect-ratio fibers. The degree of improvement in the mechanical characteristics of polymer composites, consisting of embedded fiber-like inclusions in a polymer matrix, is known to be substantially influenced by the distribution of fiber orientations; the latter being determined by flow conditions in the molten state. Gravity-based concentration of naturally occurring flake-shaped mineral particles, the settling and aggregation of large crystallites in magma chambers \citep{koyaguchi1990,schwindinger1999}, and the centrifugal determination of haematocrit are all directly affected by both the sedimentation of anisotropic particles\,(more generally, migration under the action of a body force) \citep{caro2012}, and rotation by an ambient shearing flow. The orientation distribution of anisotropic ice crystals in Cirrus clouds is controlled by the conflict between sedimentation-induced orientation drift and turbulence-induced dispersion \citep{cho1981,klett1995,garrett2003,hogan2012}, which influences the cloud scattering and absorption properties, in turn affecting the earth's radiation budget \citep{ramanathan1989}. 

While the above examples highlight the general importance of anisotropic particle suspensions, the present study specifically concerns the dynamics of viscoelastic suspensions of such particles, in the infinitely dilute limit, when inter-particle interactions may be neglected, and the suspension microstructure and rheology are entirely determined by the orientation dynamics of a single particle. Relevant scenarios include red blood cell transport in blood vessels \citep{ye2016,beris2021}, swimming of slender microorganisms in non-Newtonian fluid environments \citep{arratia2022life, spagnolie2023swimming}, and viscoelasticity-based cell separation in microfluidic applications \citep{lin2015,davino2017,li2022}, In the last example, separation is achieved via the differential rates of migration induced by elastic lift forces; like the rheology, shear-induced migration is sensitively dependent on the single-particle orientation distribution. Thus, the present study examines the orientation dynamics of a neutrally buoyant spheroid, of an arbitrary aspect ratio $\kappa$, in an ambient simple shear flow of a viscoelastic fluid, with the aim of finding how weak viscoelasticity influences the long-time orientation distribution. We consider the case of the Deborah number\,($De$) being small but finite, $De = \dot{\gamma}\tau$ being the dimensionless measure of elasticity, with $\tau$ being the (longest)\,relaxation time of the underlying microstructure\,(for instance, polymer molecules) and $\dot{\gamma}$ being a characteristic ambient shear rate. In light of $De$ being small, viscoelasticity is modeled using the constitutive equation of a second-order fluid. Within this analytical framework, the nature of departure from Newtonian rheology also depends on an additional parameter $\epsilon$, that is related to the ratio of the normal stress differences; $N_{2}/N_{1}=-\left( 1+1/2\epsilon\right)$, $N_1$ and $N_2$ here being the primary and secondary normal stress differences. While $\epsilon$ can vary over a reasonably wide range for complex fluids \citep{vivek2015}, $-0.7 \le \epsilon \le -0.5$ for the important case of polymer solutions \citep{ganesh2006,ganesh2007}.
 
\begin{figure}
		\centering
		\subfigure[]{\includegraphics[scale=0.17]{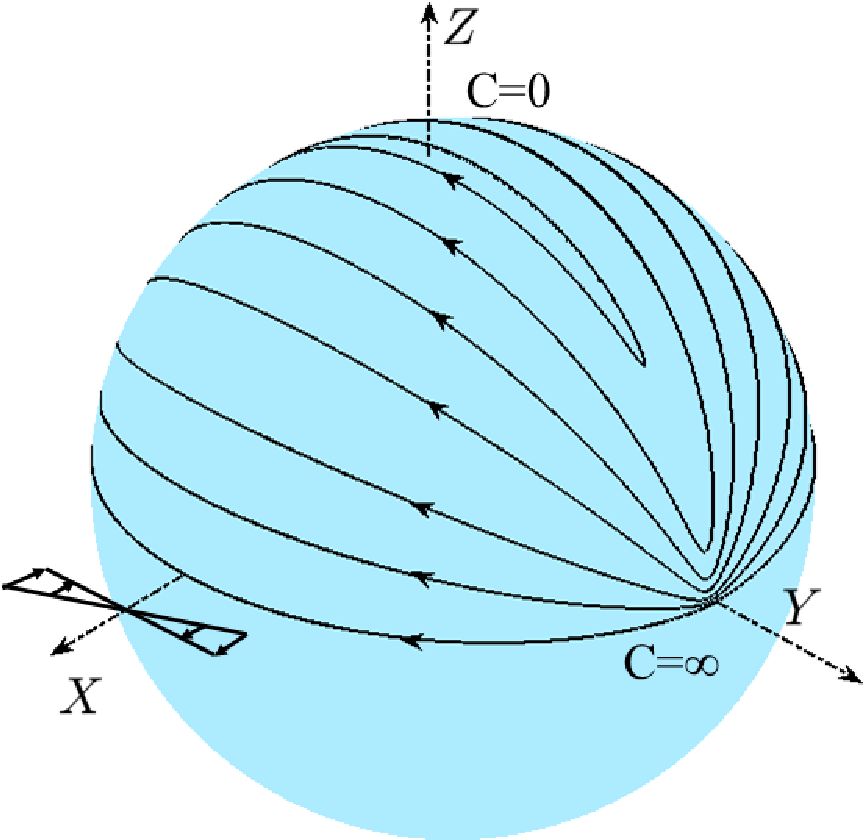} \label{fig_jeffy_Prolate_a}}
        	\subfigure[]{\includegraphics[scale=0.17]{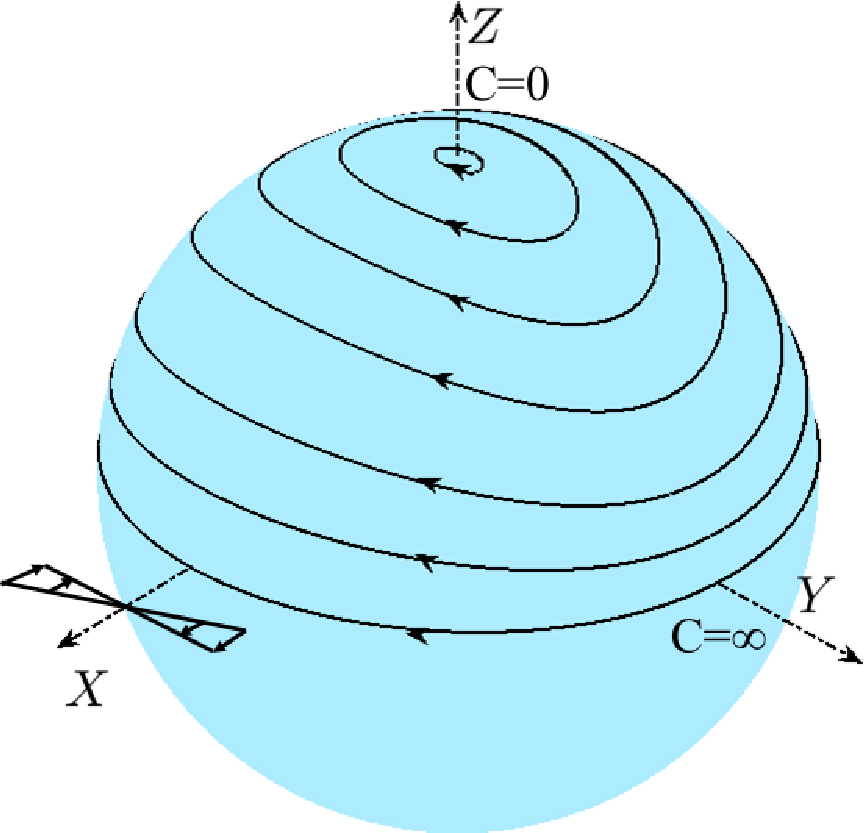} \label{fig_jeffy_Prolate_b}}
		\subfigure[]{\includegraphics[scale=0.17]{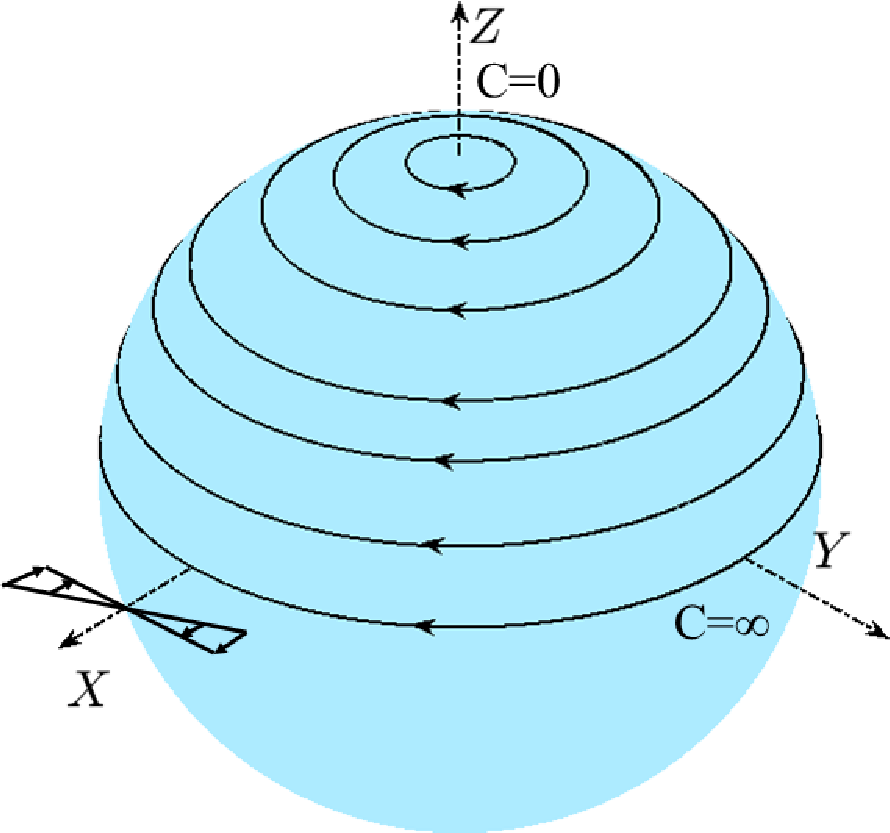}\label{fig_jeffy_Prolate_c}}
        	\subfigure[]{\includegraphics[scale=0.17]{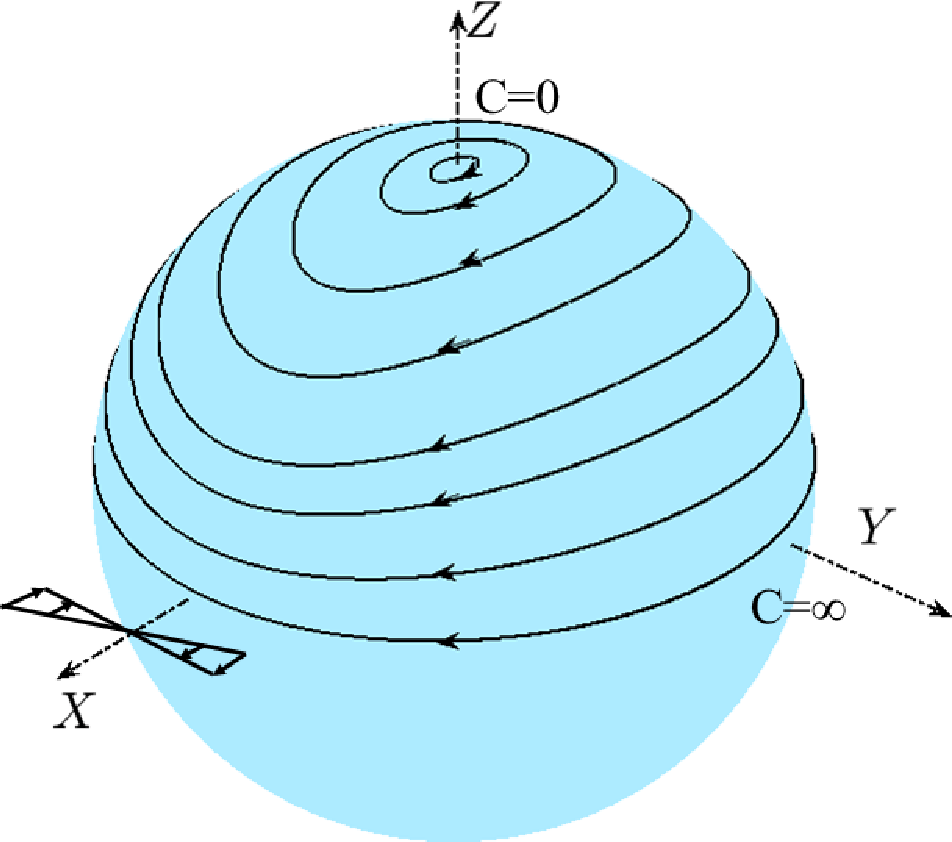}\label{fig_jeffy_Prolate_d}}
            \subfigure[]{\includegraphics[scale=0.17]{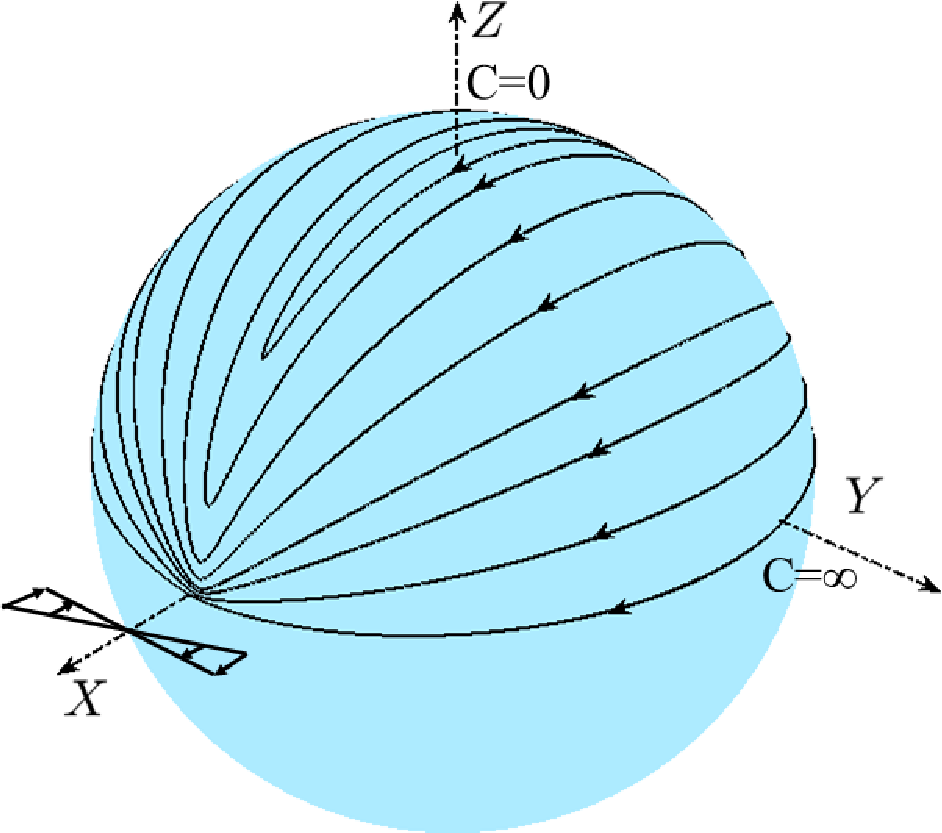}\label{fig_jeffy_Prolate_e}}
		\caption{Jeffery orbits on the unit hemisphere for spheroids of different aspect ratios; $\kappa=$ (a) $0.045$ (b) $0.6$ (c) $1$ (d) $1.67$ and (e) $22.37$. $C=0$ and $C = \infty$ correspond to the spinning\,(log-rolling) and tumbling modes, respectively; the interval $0 < C < \infty$ correspond to three-dimensional kayaking modes.}\label{fig_jeffy_Prolate}
\end{figure}

The orientation vector of an inertialess spheroid, in simple shear flow of a Newtonian fluid, rotates along any of a one-parameter family of spherical ellipses, that are now known as Jeffery orbits \citep{jeffery1922}. Jeffery orbits for a prolate spheroid have their major axes aligned with the flow direction, with the ones for an oblate spheroid being obtained via a $90^{\circ}$ rotation about the ambient vorticity direction. Figure \ref{fig_jeffy_Prolate} depicts the Jeffery orbits for spheroids over a wide interval of aspect ratios, ranging from a thin oblate spheroid\,($\kappa \ll 1$), through a sphere\,($\kappa = 1$), to a slender prolate spheroid\,($\kappa \gg 1$). While the orbits for the sphere are circles in the plane transverse to vorticity, those for extreme aspect ratios are pronouncedly eccentric, asymptoting to a meridional character in the limits $\kappa \rightarrow 0$ and $\infty$, and accompanied by a divergence of the rotation period. The inertialess rotation of both spheroids and more general fore-aft symmetric bodies of revolution, along such closed orbits, has been confirmed in multiple experimental investigations \citep{anczurowski1968,Pittman1975}. Spiralling trajectories are precluded in figures \ref{fig_jeffy_Prolate_a}-\ref{fig_jeffy_Prolate_e} owing to the reversibility of the Stokes equations. Research efforts that analyze the emergence of such trajectories and the associated breakdown of the closed Jeffery-orbit topology, under the influence of weak irreversible effects, have a long history. In what follows, we restrict ourselves to reviewing those that have examined viscoelasticity of the suspending fluid as the underlying cause for irreversible dynamical behavior; the analogous role of inertia has recently been reviewed in \citet{Ganesh_Navaneeth_2022}.

Although the focus here is on rotation of non-spherical particles, it is worth briefly recounting work on the effect of suspending fluid elasticity on a neutrally buoyant sphere. Experiments have shown an elasticity-induced slowdown of sphere rotation 
in a wide variety of viscoelastic fluids\,(these include polybutene-based Boger fluids and worm-like micellar solutions), with the magnitude of the decrease in the angular velocity increasing with $De$ \citep{snijkers2011}. This slowdown is, in fact, analogous to that induced by inertia\,\citep{lin1970,morris2004}, with the associated streamline patterns bearing a resemblance to each other - both finite-$Re$ and finite-$De$ patterns are characterized by the emergence of regions of recirculating streamlines on either side of the rotating sphere\,\citep{dvino2008}. It is important to note that the second-order fluid approximation, corresponding to the $O(De)$ term in the retarded motion expansion for weakly elastic fluids\,\citep{Bird_book}, does not yield an angular velocity correction \citep{brunn1976b,brunn1976a} - a correction at this order would have to be quadratic in the ambient velocity gradient, on account of the regular nature of the small-$De$ expansion. However, the only permissible quadratic combination that gives a pseudo-vector is $\bm{\omega} \cdot \bm{E}$, $\bm{\omega}$ being the ambient vorticity vector, this being identically zero for all ambient planar linear flows\,(simple shear being a special case). Thus, the first correction to the angular velocity occurs at $O(De^2)$, requiring a third-order fluid approximation, and has been derived by \cite{housiadas2011}; in contrast, the inertial slowdown mentioned above, first occurs at $O(Re^{\frac{3}{2}})$, on account of the singular nature of the small-$Re$ limit \citep{lin1970,marath2017}. The rotational slowdown persists for finite $De$, having been confirmed across multiple computational efforts \citep{hwang2004, dvino2008, snijkers2009,dvino2015}. The elasticity-induced slowing down of sphere rotation is important since, as will be seen below, this slowdown is magnified by a departure from sphericity, and manifests as a complete arrest of a tumbling spheroid above a $\kappa$-dependent threshold $De$.

\subsection{\bf Experiments} \label{introsec:expt}
Following the preliminary observations of \cite{Karnis1966}, \cite{gauthier1971} examined in detail the rotation of non-spherical particles in the approximate simple shearing flow of aqueous polyacrylamide\,(PAA) solutions of varying concentration, realized in the narrow gap between counter-rotating circular cylinders. The authors found rod-like and disk-like particles to spiral towards the log-rolling and tumbling modes, the orbits of minimum dissipation in the respective cases. In a followup effort using more elastic polymer solutions at higher shear rates, but with particles of moderate aspect ratios, \cite{bartram1975} found, for the first time, a stable intermediate limit cycle\,(kayaking mode) for a disk-like particle. The authors also found progressively longer periods of rotation with increasing viscoelasticity. Further, unlike a sphere, rotation eventually ceased at a finite $De$, leading to near-flow-aligned and near-gradient-aligned rotation-arrested states for rod and disk-shaped particles, respectively. Arrested states for the latter particles had already been seen by \cite{gauthier1971}; the arrested states for rod-like particles in \cite{bartram1975} were unstable to off-plane perturbations, which resulted in an initial drift away from the flow-gradient plane, followed by an eventual spiralling approach towards ambient vorticity. \cite{johnson1990}, and later \cite{gunes2008}, used rheo-optical means to examine the combined influence of Brownian motion\,(as characterized by the rotary Peclet number $Pe_r$) and viscoelasticity, on the orientation distributions in dilute non-interacting suspensions of sub-micron prolate spheroids, over a range of aspect ratios. Synthetic haematite particles were used in both efforts, with the former effort also using larger hardened red blood cells of a prolate geometry; the suspending liquids used included both shear-thinning polymer solutions and elastic Boger fluids, with experiments performed in Couette and parallel-plate geometries. Based on variation of the scattering dichroism as a function of the shear rate, and direct observations, \cite{gunes2008} inferred an initial transition to a vorticity-aligned distribution, and then from a vorticity-aligned distribution to a flow-aligned one, for large $Pe_r$, with increasing $De$; Boger fluids exhibited only the first transition.

\cite{iso1996a} and \cite{iso1996b} examined index-matched semi-dilute suspensions of non-Brownian slender fibers\,($\kappa \gtrsim 19$) in solutions of PAA dissolved in a polyethylene-glycol-glycerine mixture, and subject to a simple shear flow, again realized in a cylindrical Couette apparatus; the elasticity being varied by changing PAA concentration. The aim was to assess competing effects of a viscoelasticity-induced drift and inter-fiber hydrodynamic interactions, the latter being dominant over excluded volume interactions in the semi-dilute regime. The authors conducted both single-fiber experiments, and those involving hydrodynamically interacting suspensions, with observation times being much longer than in earlier efforts. For the weakly elastic cases\,($100$ppm PAA) examined by \cite{iso1996a}, single-fiber experiments revealed a spiralling approach towards eventual near-vorticity-alignment, similar to the experiments of Mason and coworkers described in the previous paragraph. In the suspension experiments, the increasing importance of hydrodynamic interactions led to a shift in peak probability from the vorticity to the flow direction, with the orientation distributions at the highest volume fraction being similar to those for a Newtonian fluid with a flow-aligned maximum. For moderate elasticities\,($200-500$ppm PAA), single-fiber experiments yielded stable stationary orientations in the flow-vorticity plane, with the inclination to the flow direction decreasing with increasing elasticity; the initial approach towards these stationary orientations had a spiralling character, this being the opposite of the \citet{bartram1975} observations\,(alignment followed by spiralling) above. The orientation distributions in the suspension experiments were also qualitatively different from the weakly elastic cases - increasing volume fraction led to a shift in peak probability from an orientation close to that obtained in the single-fiber experiments above, to a more nearly vorticity-aligned one, or to a flow-aligned one, depending on the elasticity. For strongly elastic fluids\,($1000$ and $2000$ppm PAA; $De \gtrsim O(1)$) examined by \cite{iso1996b}, the stationary orientations were either flow-aligned, or in the flow-vorticity plane making a modest angle\,($\approx 20^\circ$) to the flow direction; the approach towards these orientations being rapid and direct. The dominance of the viscoelastic drift also meant that the orientation distributions remained largely invariant to the changing importance of hydrodynamic interactions, with peak probabilities occurring at orientations observed in the single-fiber experiments. 

\subsection{\bf Theory} \label{introsec:theory}
Unlike spheres, there can and will be an elasticity-induced modification of anisotropic particle rotation within the second-order fluid framework. The availability of the spheroid orientation vector\,($\bm{p}$) allows one to construct contributions other than $\bm{\omega} \cdot \bm{E}$, at quadratic order in the ambient velocity gradient, that alter spheroid rotation for all orientations other than the vorticity-aligned one. The first theoretical effort in this regard was that of \cite{leal1975} who examined the orientation dynamics of a neutrally buoyant slender fiber in simple shear flow of a second-order fluid. Using a reciprocal theorem formulation in conjunction with viscous slender body theory, he showed that weak viscoelasticity leads to an orbital drift that causes the fiber to spiral in towards an eventual vorticity-aligned configuration. The $O(De)$ drift was found to be proportional to $(1+2\epsilon)$, and therefore scaled with $N_2$. This implied a vanishing drift for Boger fluids that closely obey the Oldroyd-B constitutive relation, and correspond to $\epsilon = -0.5$ in the weakly elastic limit. Further, above a threshold $De$, the viscoelastic torque became comparable to the viscous Jeffery torque, leading to rotation arrest within the flow-gradient plane - the fiber adopted a stationary nearly flow-aligned configuration. This is an unstable equilibrium, however, and any perturbation from the flow-gradient plane leads to an initially monotonic, and eventually spiralling, approach towards vorticity-alignment. This prediction qualitatively agrees with experiments of \cite{gauthier1971} and \cite{bartram1975} in the weakly elastic limit. Our analysis, in the limit of large $\kappa$, confirms the essential correctness of the above conclusions, including specifically the $\epsilon$-dependence, while correcting the magnitude of the numerical pre-factor\,(by a factor of $4$). 

Using symmetry arguments, \cite{brunn1977} generalized Leal's effort to transversely isotropic particles; almost all such particles rotate along Jeffery orbits in simple shear flow, as originally demonstrated by \cite{bretherton1962}, with the orbit eccentricity now governed by an equivalent aspect ratio rather than the geometric one. One conclusion, already evident from symmetry, is that the tumbling and spinning modes always correspond to possible steady states, regardless of particle shape\,(the longitudinal cross-section). The existence and stability of a limit cycle\,(kayaking mode), intermediate between these modes, depends on the detailed particle geometry, however. For the particular case of a rigid tri-dumbbell, the author showed the absence of such a limit cycle. Identifying tri-dumbbells of aspect ratios greater and less than unity, respectively, with the rod and disk-shaped particles used by \cite{gauthier1971}, led to a broad agreement between theory and experiment; the limiting case of dumbbells was similar to that of the fiber above, although the $O(De)$ drift towards vorticity was proportional to $(1+4\epsilon)$, and therefore, non-zero even for Boger fluids. Our own calculations detailed below show that, for the narrow $\epsilon$-range specific to polymer solutions, a stable intermediate limit cycle does arise for oblate spheroids below a threshold $\kappa$ that depends sensitively on $\epsilon$. Further, in \S\ref{Results:epsilon_kappa}, considering a larger interval $\epsilon\,\in\,[-5,5]$, pertinent to a broader range of complex fluids \citep{vivek2015}, we map out the existence and stability of this limit cycle on the $\kappa-\epsilon$ plane. 

\cite{cohen1987} and \cite{chung1987} examined fiber orientation distributions in simple shear flow, arising under the combined influence of the viscoelastic drift obtained by \cite{leal1975} and rotational Brownian motion. For sufficiently small $De$, one may exploit the separation of time scales between Jeffery rotation, and the $O(De)$ drift, to obtain an orbit-averaged drift-diffusion equation, with the distribution across Jeffery orbits being a function of $DePe_r$, Based on such an analysis, the authors observed a transition from a flow-aligned distribution to a vorticity-aligned one, that involves a sequence of bimodal distributions with peaks at both flow and vorticity-aligned orientations. 

\cite{harlen1993} theoretically investigated the orientation dynamics of a slender fiber in a dilute polymer solution, in the complementary large-$De$ limit, with the polymer concentration\,($c$) assumed to be small, so the disturbance velocity field is still given by its Stokesian approximation. Owing to its logarithmically small\,(in $\kappa$) magnitude at distances of order the fiber length, and the large relaxation time, the polymeric stress was approximated as being convected along the streamlines of the ambient shear, while developing due to the $O(1/\ln \kappa)$ disturbance velocity gradient. For fibers nearly aligned with the flow-vorticity plane, the viscoelastic torque was found to lead to an irreversible drift across Jeffery orbits although, unlike the small-$De$ analysis of \cite{leal1975} above, the drift was not proportional to $N_2$.  Recently, \citet{arjun2023} have again analyzed rotation of a nearly aligned fiber, in simple shear, for small polymer concentrations. The authors included $O(1/\ln \kappa)$ terms, missed out by the original \citet{harlen1993} analysis, while also including the contributions of higher-order singularities, that become comparable to the leading order line distribution of Stokeslets, during the aligned phase. In addition, the authors allowed for $De$ to be arbitrary, provided only that $c\,De$ was small, and thereby, identified a rich range of orientational dynamics in the $\kappa-c\,De$ plane. Of particular importance was the emergence of stable rotation-arrested states in the flow-gradient plane beyond a threshold $c\,De$. We comment further on the limiting form of these results, for small $De$, in \S\ref{conclu}. 

\subsection{\bf Computations} \label{introsec:comp}
The restriction to weak elasticity in any theoretical endeavor\,(due to either an asymptotically small relaxation time or polymer concentration), has motivated a number of computational efforts that have examined single-particle orientation dynamics in a viscoelastic shearing flow. \cite{phan2002} employed a version of a boundary element method to examine the rotation of a neutrally buoyant prolate spheroid, with $\kappa =2$, in simple shear flow of an Oldroyd-B fluid. For $De = 0.7$, the authors found both an increase in the rotation period and a slow drift towards ambient vorticity, consistent with the small-$De$ experiments in \S\ref{introsec:expt}; although, computational expense meant that the duration of the simulations was short. Using simulations based on a finite element arbitrary Lagrangian–Eulerian domain method, \cite{lv2011} examined the motion of a rectangular particle, in a wall-bounded plane Couette flow, under the combined effects of inertia and elasticity. Although the focus was on shear-induced migration, the authors nevertheless found that elasticity led to a slowdown of particle rotation relative to that in a Newtonian fluid. In the absence of an exact solution for the general axisymmetric particle, \cite{borzacchiello2016} proposed an ad-hoc simplification for the rate of change of $\bm{p}$, by including an effective velocity gradient with an $O(De)$ correction, instead of the ambient one, in the Jeffery equation. An additional constant denoting the amplitude of $O(De)$ correction was tuned for agreement with the actual dynamics; although, an exact match is precluded due to differing angular dependencies of the $O(De)$ terms in the exact and simplified equations. 

\cite{davino2014} simulated the rotation of prolate spheroids with $\kappa\,\in\,[2,16]$ in a viscoelastic simple shear flow, again using a finite element method with constraints enforced via Lagrange multipliers. The same set of authors, in a more recent followup effort\,\citep{davino2019}, have examined the elasticity-induced migration of spheroids in pressure-driven flow through a micro-channel. In the 2014 paper above, using the Giesekus constitutive relation, the authors find the spinning mode to be globally stable for sufficiently small $De$, consistent with the \cite{gauthier1971} findings. For $De$ above a threshold, a pair of stable equilibria\,(by symmetry) within the flow-vorticity plane appear to bifurcate from the spinning mode, that simultaneously turns unstable; their emergence\,(with superposed small-amplitude oscillations) being broadly consistent with the findings of \cite{iso1996a,iso1996b} for moderately and strongly elastic fluids. With increasing $De$, these intermediate equilibria move towards the pair of stationary unstable orientations in the flow-gradient plane. Interestingly, there exists a second threshold $De$, above which the latter flow-aligned equilibria turn stable, implying the emergence of additional (unstable)\,fixed points whose invariant manifolds separate basins of attraction corresponding to the flow-aligned and intermediate equilibria. In effect, with increasing $De$, prolate spheroids were found to transition from stable spinning, to either flow-aligned or near-flow-aligned states, via intermediate aligned equilibria in the flow-vorticity plane; the thresholds corresponding to the bifurcations, that mediate these transitions, being functions of $\kappa$. 

\cite{wang2019} have again examined prolate spheroid\,($\kappa = 4$) rotation in a viscoelastic simple shear flow, but under conditions where the separation between the moving walls is not much greater than the spheroid size. The spheroid orientation dynamics, when restricted to the mid-plane, was similar to that described above. Although, confinement transforms the log-rolling motion at the smallest $De$ into a kayaking one, and also appears to slow down the movement of the intermediate aligned equilibrium towards the flow-gradient plane; thus, the spheroid exhibited an intermediate aligned orientation, in the flow-vorticity plane, even at the highest $De\,(=4)$ examined. Spheroids not starting at the mid-plane were found to drift towards ambient vorticity, while simultaneously migrating towards the nearer wall.  The very recent effort of \cite{li2023} investigates the orientation dynamics of a neutrally buoyant prolate spheroid, under the combined influences of inertia and viscoelasticity, using an immersed boundary method; the fluid rheology again being modeled using the Giesekus relation, with confinement effects being less severe than in \citet{wang2019}. A comparison of the theoretically predicted orientation dynamics with the said simulations, as a function of the elasticity number $El = De/Re$, will be undertaken in a followup paper, and herein we only note that the results in the elasticity-dominant limit are consistent with those of \cite{davino2014} above. While the latter limit corresponds, in principle, to $El \rightarrow \infty$, it was realized for $El \gtrsim 0.3$ in the said simulations.

It is worth ending this survey by acknowledging a large number of studies that have focused on shear-induced (transverse)\,migration of spheroidal particles in nonlinear shearing flows; the importance of this problem, in the context of microfluidic applications\,(particle-sorting/separation in micro-channels), was noted at the beginning. On account of a separation of time scales pertaining to positional and orientational dynamics, a migration calculation in wide channels requires the local-linear-flow orientation dynamics as an input \citep{prateek2023}. Herein, we only note the efforts of Narsimhan and co-workers and the study of \cite{li2024}. The former have, in a series of efforts \citep{tai2020a,tai2020b,wang2020,tai2022}, theoretically examined migration of non-spherical particles in unbounded quadratic flows of a second-order fluid. To begin with, the focus was on the co-rotational limit\,($\epsilon = -0.5$) when the Stokesian flow field remains unaltered, with the viscoelastic correction only comprising an $O(De)$ Giesekus pressure. Later articles moved beyond this restriction, with the authors recently also having performed experiments on migration of both spherical and spheroidal particles in channels. On the whole, the flow geometry appears to strongly influence the orientation dynamics, although migration in all cases is towards the centerline. The study of \cite{li2024} is one of very few that examines oblate spheroids - the authors study the inertio-elastic migration of such a spheroid\,($\kappa = 0.5$) in a square duct, using an immersed boundary method; viscoelasticity being modeled using the Oldroyd-B relation, so as to avoid transverse secondary flows. Along expected lines, the authors find a transition from wall-ward migration, to migration towards the centerline, with increasing $El$; although, the rate of migration exhibits a non-monotonic dependence on $El$. The computations also indicate the appearance of a kayaking mode or an intermediate rotation-arrested state, at higher $El$, which is broadly consistent with our findings here.\\

The manuscript is organized as follows. In \S\ref{Problemform:recipro}, we formulate the generalized reciprocal theorem to obtain the angular velocity of a neutrally buoyant spheroid freely rotating in a viscoelastic fluid undergoing simple shear flow. The $O(De)$ contribution to the angular velocity is written in terms of surface and volume integrals which may be evaluated using the Stokesian approximations for the relevant fields. \S\ref{Problemform:velocity} details the expressions for the Stokesian disturbance fields, for both the actual and test problems involved in the formulation, while also introducing the Jeffery-orbit-based $(C,\tau)$ coordinate system. In \S\ref{Problemform:visco}, based on symmetry arguments, we write down the general form of the spheroid angular velocity in a viscoelastic ambient linear flow, to $O(De)$, which forms the basis of the discussion in subsequent sections. The aspect-ratio functions that appear in this equation are evaluated analytically in a spheroidal coordinate system, for an ambient simple shear flow; the resulting expressions, with associated limiting forms, are tabulated in Appendix \S\ref{appA}. \S\ref{Orientation_dynamics} describes the spheroid orientation dynamics for $De$ well below the rotation arrest threshold. \S\ref{Orientation_dynamics:multi} presents a multiple scales analysis, which exploits the separation between the Jeffery period and the time scale characterizing the viscoelastic drift, to derive a one-dimensional description of the orientation dynamics in terms of the evolution of the orbit constant $C$. This  orbital drift interpretation allows for the use of $De^{-1}\Delta C/(C^2+1)$ vs $C/(C+1)$ plots, in \S\ref{Orientation_dynamics:deltaC}, to identify four different orientation dynamics regimes\,(Regimes 1 to 4); here, $\Delta C(C;\kappa,\epsilon)$ is the change in orbit constant in a single Jeffery time period. \S\ref{Results:epsilon_kappa} presents plots of $De^{-1} \Delta C/(C^2+1)$ vs $C/(C+1)$ for various $\kappa$ and $\epsilon$, for both prolate and oblate spheroids, allowing for an organization of Regimes 1 to 4 on the $\kappa-\epsilon$ plane in figure \ref{fig_kappa_epsilon}. The orbital drift interpretation breaks down when the $O(De)$ angular velocity correction becomes comparable to the leading order Jeffery rotation. The resulting rotation arrest of slender prolate and thin oblate spheroids, in the flow-gradient plane, is examined in \S\ref{Rotation_arrest}. Rotation arrested states of prolate spheroids arise for $De$ exceeding an $O(1/\kappa)$ threshold, but are unstable to off-plane perturbations. For the oblate case, the saddle-node bifurcation associated with rotation arrest occurs for $De$ exceeding an $O(\kappa)$ threshold, but is followed by two more bifurcations which  lead to stable rotation-arrested within the flow-gradient plane. \S\ref{conclu} summarizes our findings, while also comparing them with some of the previous theoretical and computational efforts discussed in \S\ref{introsec:theory} and \S\ref{introsec:comp} above. Appendix \S\ref{appB} revisits the slender-body analysis of \citet{leal1975} using a simpler Fourier space approach, with the results for the large-$\kappa$ drift serving as a validation of the exact results obtained in the main manuscript and \S\ref{appA}.

\section{\bf Problem formulation : viscoelastic drift}\label{Problemform}
\subsection{\bf The generalized reciprocal theorem}\label{Problemform:recipro}
We use the generalized reciprocal theorem to obtain an expression for the angular velocity of a neutrally buoyant spheroid in a viscoelastic simple shear flow. 
The theorem relates two sets of velocity and stress fields - $(\bm{\sigma}^{(1)}, \bm{u}^{(1)} )$ pertains to the problem of interest above, a freely rotating spheroid of an arbitrary aspect ratio in simple shear flow of a second-order fluid, while $(\bm{\sigma}^{(2)}, \bm{u}^{(2)} )$ corresponds to the test problem, taken to be a spheroid rotating about an axis transverse to its symmetry axis, in an otherwise quiescent Newtonian fluid, at zero Reynolds number. In terms of the disturbance fields\,(indicated by primes) of the actual problem, the theorem yields the identity:
\begin{equation}\label{eq1}
\int_{S_{p}} \left(\bm{\sigma}^{(2)} \cdot \bm{u}^{\prime (1)}-\bm{\sigma}^{\prime (1)} \cdot \bm{u}^{(2)} \right)\cdot \bm{n} ~dS=De \int_{V} \bm{\sigma}^{\prime (1)}_{NN} : \nabla \bm{u}^{(2)} ~dV,
\end{equation}
where $\bm{n}$ is the unit outer normal to the spheroid surface\,($S_p$). The disturbance fields have been assumed to decay away sufficiently rapidly at large distances, so as to ensure the absence of any contributions from the surface integral at infinity. The latter is typically true in the small-$De$ limit, owing to the ordered fluid constitutive equation being obtained from a truncation of the retarded motion expansion\,(see below), and the latter being essentially a regular perturbation expansion about the Newtonian limit \citep{Bird_book}. 

The non-dimensional equations of motion and the continuity equation, for the problem of interest, are given by:
\begin{equation}\label{eq2}
\nabla \cdot \bm{\sigma}^{(1)}=0, \quad \nabla \cdot \bm{u}^{(1)}=0,
\end{equation}
with 
\begin{equation}\label{eq3}
 \left.
	\begin{array}{ll}
\bm{u}^{(1)}=\bm{\Omega}^{(1)} \wedge \bm{x} \quad \text{for}  \quad \bm{x} \in S_{P},\\
\bm{u}^{(1)} \rightarrow \bm{\Gamma} \cdot \bm{x} \quad \text{for}  \quad \bm{x} \rightarrow \infty.
\end{array}
\right\}
\end{equation}
The stress field in (\ref{eq2}) is defined by 
\begin{equation}\label{eq4}
   \bm{\sigma}^{(1)}=\bm{\sigma}_{N}^{(1)}+De~ \bm{\sigma}_{NN}^{(1)}, 
\end{equation}
where $\bm{\sigma}_{N}^{(1)}=-p^{(1)}\bm{I}+2 \bm{e}^{(1)}$ and $\bm{\sigma}_{NN}^{(1)}$ are the Newtonian and non-Newtonian stress contributions, respectively. For $De \ll 1$, the retarded motion expansion can be employed to incorporate weak elastic effects in a perturbative manner. The first correction to Newtonian rheology corresponds then to a second-order fluid, and at this order, $\bm{\sigma}_{NN}^{(1)}$ in (\ref{eq4}) may be written as the sum of corotational ($\bm{\sigma}_{NNC}^{(1)}$) and quadratic ($\bm{\sigma}_{NNQ}^{(1)}$) contributions. Thus, $\bm{\sigma}^{(1)}_{NN}= \bm{\sigma}^{(1)}_{NNC} +\bm{\sigma}^{(1)}_{NNQ}$, with
\begin{equation}\label{eq5}
\bm{\sigma}^{(1)}_{NNC}=2 \epsilon \left( \frac{\partial \bm{e}^{(1)}}{\partial t}+\bm{u}^{(1)} \cdot \nabla\bm{e}^{(1)} +\bm{w}^{(1)} \cdot \bm{e}^{(1)}+(\bm{w}^{(1)} \cdot \bm{e}^{(1)})^{\dagger} \right),
\end{equation}
\begin{equation}\label{eq6}
\bm{\sigma}^{(1)}_{NNQ}=4 (1+\epsilon) \left( \bm{e}^{(1)} \cdot \bm{e}^{(1)} \right),
\end{equation}
where $\bm{e}^{(1)}=(\nabla \bm{u}^{(1)}+\nabla \bm{u}^{(1)^{\dagger}})/2$ and $\bm{w}^{(1)}=(\nabla \bm{u}^{(1)}-\nabla \bm{u}^{(1)^{\dagger}})/2$ are the rate of strain and vorticity tensors, respectively, with $\epsilon$ being the second-order fluid parameter that, as mentioned in \S\ref{intro}, is related to the ratio of the normal stress differences \citep{ganesh2006,ganesh2007}. The disturbance velocity field in (\ref{eq1}) is defined by $\bm{u}^{\prime (1)}= \bm{u}^{(1)}-\bm{\Gamma} \cdot \bm{x}$, where $\bm{\Gamma} \cdot \bm{x}$ is the ambient simple shear flow, with $\bm{\Gamma}$ being the transpose of the ambient velocity gradient tensor. On $S_{p}$,  $\bm{u}^{\prime (1)}= \bm{\Omega}^{(1)} \wedge \bm{x}-\bm{\Gamma} \cdot \bm{x}$, with $\bm{\Omega}^{(1)}$ being the spheroid angular velocity determined from a torque-free constraint. 

In the above, $\bm{\Gamma}=\bm{1}^{\prime}_{x} \bm{1}^{\prime}_{y}$ in a space-fixed coordinate system with axes oriented along the flow ($X^{\prime}$), gradient ($Y^{\prime}$) and vorticity ($Z^{\prime}$) directions of the ambient simple shear; $\bm{E}=(\bm{\Gamma} +\bm{\Gamma}^{\dagger})/2$ and $\bm{W}=(\bm{\Gamma}^{\dagger}-\bm{\Gamma})/2$ denote the (non-dimensional)\,ambient rate of strain and vorticity tensors, respectively. One may now write the governing equations for the actual problem as:  
\begin{equation}\label{eq7}
\begin{aligned}
\nabla \cdot \bm{\sigma}^{ \prime (1)}&=-\nabla p^{\prime (1)} + \nabla^{2} \bm{u}^{ \prime (1)} + De ~\nabla \cdot \bm{\sigma}_{NN}^{ \prime (1)}=0,
\end{aligned}
\end{equation}
\begin{equation}\label{eq8}
\nabla \cdot \bm{u}^{ \prime (1)}=0,
\end{equation}
in terms of the disturbance fields. The disturbance stress field in (\ref{eq1}) and (\ref{eq7}) is defined as $\bm{\sigma}^{\prime (1)}= \bm{\sigma}^{ (1)}-\bm{\sigma}^{ \infty}$, where $\bm{\sigma}^{\infty}=\bm{\sigma}_{N}^{ \infty} + De \bm{\sigma}_{NN}^{ \infty}$;  $\bm{\sigma}_{N}^{ \infty}=-p^{\infty} \bm{I}+2 \bm{E}$ and $\bm{\sigma}_{NN}^{ \infty}= \bm{\sigma}_{NNC}^{ \infty} +\bm{\sigma}_{NNQ}^{ \infty}$, with $\bm{\sigma}_{NNC}^{ \infty}= 2 \epsilon \left( \bm{W} \cdot \bm{E}+(\bm{W} \cdot \bm{E})^{\dagger} \right)$ and $\bm{\sigma}_{NNQ}^{ \infty}= 4 (1+\epsilon) \left( \bm{E} \cdot \bm{E} \right)$, are the Newtonian and non-Newtonian ambient stress components. The corotational and quadratic components of $\bm{\sigma}_{NN}^{ \prime (1)}$ are therefore given by
\begin{eqnarray}\nonumber
\bm{\sigma}^{\prime (1)}_{NNC}&=&\bm{\sigma}^{(1)}_{NNC}-\bm{\sigma}^{\infty}_{NNC}\\\nonumber
&=&  2 \epsilon   \Big(\frac{\partial \bm{e}^{\prime(1)}}{\partial t}+\bm{u}^{(1)} \cdot \nabla\bm{e}^{\prime(1)}+\bm{w}^{\prime(1)} \cdot \bm{e}^{\prime(1)}+(\bm{w}^{\prime(1)} \cdot \bm{e}^{\prime(1)})^{\dagger}+\bm{W} \cdot \bm{e}^{\prime(1)}+(\bm{W} \cdot \bm{e}^{\prime(1)})^{\dagger} \\ \label{eq9}
&&+\bm{w}^{\prime(1)} \cdot \bm{E}+(\bm{w}^{\prime(1)} \cdot \bm{E})^{\dagger}  \Big), \\ \nonumber
\bm{\sigma}_{NNQ}^{\prime (1)}&=& \bm{\sigma}^{(1)}_{NNQ}-\bm{\sigma}^{\infty}_{NNQ}\\ \label{eq10}
&=&4 (1+\epsilon) \left(\bm{e}^{\prime(1)} \bm{\cdot} \bm{e}^{\prime(1)}+\bm{e}^{\prime(1)} \cdot \bm{E}+\bm{E} \bm{\cdot e}^{\prime(1)}\right). 
\end{eqnarray}
Note that $\epsilon=-1$, with $\bm{\sigma}^{\prime (1)}_{NNQ}=0$, corresponds to the corotational limit, when viscoelasticity leaves the Stokesian velocity field unchanged\,\citep{Bird_book}.
\begin{figure}
	\centering
	\includegraphics[scale=0.35]{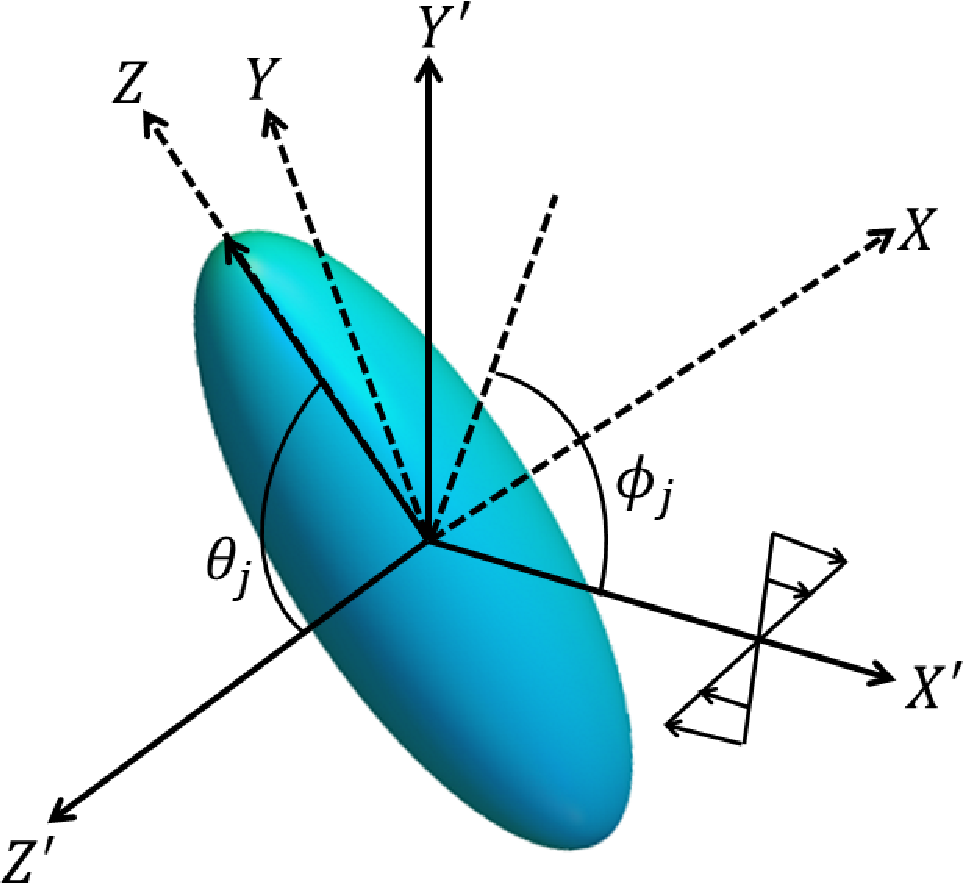}
	\caption{The body-fixed\,($XYZ$) and space-fixed\,($X'Y'Z'$) coordinate systems for a spheroid in a simple shear flow; the $Y$-axis being constrained to lie in the $X'Y'$-plane at all times. The polar angle between the spheroid symmetry axis and ambient vorticity\,($\theta_j$), and the dihedral angle between the flow-vorticity and orientation-vorticity planes\,($\phi_{j}$) define the spheroid orientation in the space-fixed reference frame.}\label{fig_geometry} 
\end{figure}

The Stokesian test problem is defined by 
\begin{equation}\label{eq11}
\nabla \cdot \bm{\sigma}^{(2)}=0,
\quad
\nabla \cdot \bm{u}^{(2)}=0,
\end{equation}
with 
\begin{equation}\label{eq12}
\left.
\begin{array}{ll}
\bm{u}^{(2)}=\bm{\Omega}^{(2)} \wedge \bm{x} \quad \text{for}  \quad \bm{x} \in S_{P},\\ 
\bm{u}^{(2)} \rightarrow 0 \quad \text{for}  \quad \bm{x} \rightarrow \infty,
\end{array}
\right\}
\end{equation}	
where $\bm{\Omega}^{(2)}$ is the spheroid angular velocity constrained to be orthogonal to its symmetry axis\,($\bm p$). On using the boundary conditions for the actual and test problems, (\ref{eq1}) takes the form:
\begin{equation}\label{eq13}
\begin{aligned}
\bm{\Omega}^{(1)} \cdot \mathcal{\bm{L}}^{(2)} = \bm{\Gamma}  \bm{:} \int_{Sp}  \bm{x} \left(\bm{\sigma}^{(2)} \cdot  \bm{n}  \right)  ~dS +De ~ \int_{V} (\bm{\sigma}_{NNC}^{\prime (1)}+\bm{\sigma}_{NNQ}^{\prime (1)}) : \nabla \bm{u}^{(2)} dV. 
\end{aligned}
\end{equation}
Since there exists a Giesekus pressure field $p_{G}^{\prime}$ in the corotational limit such that
$\nabla \cdot \bm{\sigma}^{\prime (1)}_{NNC} -\nabla p_{G}^{\prime (1)}=0$, with
\begin{equation}\label{eq14}
p_{G}^{\prime (1)}=\epsilon \left( \frac{\partial p^{\prime (1)}}{\partial t} +\bm{u}^{(1)} \bm{\cdot} \nabla p^{\prime (1)} + \bm{e}^{\prime (1)} : \bm{e}^{\prime (1)}+ 2 (\bm{e}^{\prime (1)} : \bm{E})\right),
\end{equation}
the corotational term in (\ref{eq13}) may be rewritten in terms of a surface integral as:
\begin{equation} \label{eq15}
\int_{V} \bm{\sigma}_{NNC}^{\prime (1)} :  \nabla \bm{u}^{(2)} dV=~\bm{\Omega}^{(2)} \cdot \int_{S}\bm{x} \times  \left\{ \left(\bm{\sigma}_{NNC}^{\prime(1)} - p^{\prime (1)}_{G} \bm{I}  \right) \cdot \bm{n} \right\}  dS .
\end{equation}
Further, noting that the test velocity field (\ref{eq11})-(\ref{eq12}) is linear in $\bm{\Omega}^{(2)}$, one may write $\bm{u}^{(2)} = \bm{U}^{(2)} \cdot \bm{\Omega}^{(2)}$, $\mathcal{\bm{L}}^{(2)} = \bm{L}^{(2)} \cdot \bm{\Omega}^{(2)}$ and $\bm{\sigma}^{(2)} = \bm{\Sigma}^{(2)} \cdot \bm{\Omega}^{(2)}$, where $\bm{U}^{(2)}$ and $\bm{L}^{(2)}$ are second-order tensors, and $ \bm{\Sigma}^{(2)}$ a third-order tensor, dependent on position and the spheroid aspect ratio. All three tensors are known in closed form - see \citet{vivek2016}. Using (\ref{eq14}) and (\ref{eq15}), and the fact that $\bm{\Omega}^{(2)}$ is arbitrary but for being transverse to $\bm{p}$, (\ref{eq13}) takes the final form:
\begin{equation}\label{eq16}
\begin{aligned}
\bm{\Omega}^{(1)} \cdot \bm{L}^{(2)} &= \bm{\Gamma}  \bm{:} \int_{Sp}  \bm{x} \left(\bm{\Sigma}^{(2)} \cdot  \bm{n}  \right)  ~dS\\
&+De~\left(\int_{S}\bm{x} \times  \left\{ \left(\bm{\sigma}_{NNC}^{\prime (1)} - p^{\prime (1)}_{G} \bm{I}  \right) \cdot \bm{n} \right\}  dS + \int_{V} \bm{\sigma}_{NNQ}^{\prime (1)} : \nabla \bm{U}^{(2)} dV \right). 
\end{aligned}
\end{equation}
The first term in (\ref{eq16}) is the Stokesian torque that leads to the Jeffery orbits depicted in figure \ref{fig_jeffy_Prolate}, while the second $O(De)$ term is the viscoelastic torque contribution. To $O(De)$, the fields involved in the associated surface and volume integrals may be replaced by their Stokesian analogs\,(the resulting integrals are readily shown to be convergent), leading to additive contributions from the corotational and quadratic components. Although not analyzed here, for the same reason, inertial and viscoelastic contributions to the spheroid angular velocity are also additive in the small $Re$, $De$ limit, so the inertio-elastic trajectory topology is only a function of $El$.

Following \citet{vivek2016}, in \S\ref{Problemform:velocity} below, we write down the disturbance velocity and pressure fields, that determine the various fields involved in (\ref{eq16}), in terms of superpositions of the appropriate vector spheroidal harmonics. The latter are defined in a body-fixed coordinate system\,($XYZ$) with the $Z$-axis instantaneously aligned with the spheroid symmetry axis, and rotating with it such that the $Y$-axis lies within the flow-gradient\,($X'Y'$) plane at all times; see figure \ref{fig_geometry}. As a result, the angular velocity components of the coordinate system, orthogonal to $\bm{p}$, coincide with those of the spheroid, both being given by $\bm{\Omega}_{jeff}\cdot ({\bm I} - {\bm p}{\bm p}) \equiv (\dot{\phi}_{j} \sin\theta_{j},\dot{\theta}_{j},0)$; $\dot{\theta}_j$ and $\dot{\phi}_j$ are defined below by (\ref{eq37}). But, the spin about the $Z$-axis\,($\dot{\phi}_{j} \cos\theta_{j}$) differs from the actual spheroid spin\,($\frac{1}{2}\bm \omega \cdot \bm p = -\frac{1}{2}\cos\theta_j$, $\bm \omega =-\bm{1}_{z'}$ being the ambient vorticity vector). Since the unsteady terms in (\ref{eq9}) and (\ref{eq14}) need to be evaluated in the aforementioned coordinate system, we rewrite the material-derivative combination in the said equations as:
\begin{equation}\label{eq17}
\begin{aligned}
    \frac{\partial \bm{e}^{\prime (1)}}{\partial t} &+\bm{u}^{(1)} \cdot \nabla \bm{e}^{\prime (1)} 
    = \left(\frac{\partial \bm{e}^{\prime (1)}}{\partial t}+ ( \bm{\Omega}_{b} \times \bm{x} ) \cdot \nabla \bm{e}^{\prime (1)} + \bm{\Omega}_{bt} \cdot  \bm{e}^{\prime (1)} + (\bm{\Omega}_{bt} \cdot  \bm{e}^{\prime (1)})^{\dagger} \right) \\
    &+\bm{u}^{(1)} \cdot \nabla \bm{e}^{\prime (1)} - ( \bm{\Omega}_{b} \times \bm{x} ) \cdot \nabla \bm{e}^{\prime (1)}+ \bm{e}^{\prime (1)}\cdot \bm{\Omega}_{bt}+(\bm{e}^{\prime (1)} \cdot \bm{\Omega}_{bt})^{\dagger}\\
    &= \left(\frac{\partial \bm{e}^{\prime (1)}}{\partial t}\right)_{r} 
    +\left(\bm{u}^{(1)}  - ( \bm{\Omega}_{b} \times \bm{x} ) \right) \cdot \nabla \bm{e}^{\prime (1)}+ \bm{e}^{\prime (1)}\cdot \bm{\Omega}_{bt}+(\bm{e}^{\prime (1)} \cdot \bm{\Omega}_{bt})^{\dagger},
    \end{aligned}
\end{equation}
 \begin{equation}\label{eq18}
     \begin{aligned}
  \frac{\partial p^{\prime (1)}}{\partial t}+\bm{u}^{(1)} \cdot \nabla p^{\prime (1)}      &= \left(\frac{\partial p^{\prime (1)}}{\partial t} + ( \bm{\Omega}_{b} \times \bm{x} ) \cdot \nabla p^{\prime (1)} \right)+\bm{u}^{(1)} \cdot \nabla p^{\prime (1)} - ( \bm{\Omega}_{b} \times \bm{x} ) \cdot \nabla p^{\prime (1)}\\
  &= \left(\frac{\partial p^{\prime (1)}}{\partial t}  \right)_{r}+\left( \bm{u}^{(1)} - ( \bm{\Omega}_{b} \times \bm{x} )  \right) \cdot \nabla p^{\prime (1)},
     \end{aligned}
 \end{equation}
 where $\bm{\Omega}_{b}=(-\dot{\phi}_{j} \sin\theta_{j},\dot{\theta}_{j},\dot{\phi}_{j} \cos\theta_{j} )$ in (\ref{eq17})-(\ref{eq18}) is the coordinate system angular velocity mentioned above. The terms with suffix $r$ in (\ref{eq17}) and (\ref{eq18}) denote the rates of change in the body-fixed frame. While the expression for the transformation of the scalar pressure field, to the said frame, is obvious, that for the tensor field\,($\bm{e}^{\prime (1)}$) involves a Jaumann-like contribution, $\bm{e}^{\prime (1)}\cdot \bm{\Omega}_{bt}+(\bm{e}^{\prime (1)} \cdot \bm{\Omega}_{bt})^{\dagger}$, where $\bm \Omega_{bt}=\frac{d \bm{R(t)}}{dt} \cdot {\bm R}^{\dagger}(t)$ is the anti-symmetric tensor corresponding to the pseudovector $\bm{\Omega}_b$, with $\bm{R}(t)= \left[\begin{smallmatrix}
\cos(\theta_{j}) \cos\phi_{j} & \cos\theta_{j} \sin\phi_{j} & -\sin\theta_{j} \\
-\sin\phi_{j}& \cos\phi_{j} & 0\\
	\sin\theta_{j} \cos\phi_{j}& 	\sin\theta_{j} \sin\phi_{j} & \cos\theta_{j}\\
\end{smallmatrix} \right]$ being the rotation tensor that connects the body-fixed and space-fixed frames. Using (\ref{eq17}) and (\ref{eq18}), one may write down the following final expressions for evaluation of the viscoelastic stress components and the Giesekus pressure field in (\ref{eq16}):
\begin{eqnarray}\nonumber
    \bm{\sigma}^{\prime (1)}_{NNC}&=&  2 \epsilon \Big( \left(\frac{\partial \bm{e}^{\prime (1)}}{\partial t}\right)_{r} +\left(\bm{u}^{(1)}  - ( \bm{\Omega}_{b} \times \bm{x} ) \right) \cdot \nabla \bm{e}^{\prime (1)}+ \bm{e}^{\prime (1)}\cdot \bm{\Omega}_{bt}+(\bm{e}^{\prime (1)} \cdot \bm{\Omega}_{bt})^{\dagger}+\bm{w}^{\prime (1)} \cdot \bm{e}^{\prime (1)}\\ \label{eq19}
&&+(\bm{w}^{\prime (1)} \cdot \bm{e}^{\prime (1)})^{\dagger} +\bm{W} \cdot \bm{e}^{\prime (1)}+(\bm{W} \cdot \bm{e}^{\prime (1)})^{\dagger}+\bm{w}^{\prime (1)} \cdot \bm{E}+(\bm{w}^{\prime (1)} \cdot \bm{E})^{\dagger}  \Big), \\  \label{eq20}
\bm{\sigma}_{NNQ}^{\prime (1)}&=&4 (1+\epsilon) \left(\bm{e}^{\prime (1)} \bm{\cdot} \bm{e}^{\prime (1)}+\bm{e}^{\prime (1)} \cdot \bm{E}+\bm{E} \bm{\cdot e}^{\prime (1)}\right), \\  \label{eq21}
p_{G}^{\prime (1)}&=&\epsilon \left( \left( \frac{\partial p^{\prime (1)}}{\partial t}  \right)_{r}+\left( \bm{u}^{(1)} - ( \bm{\Omega}_{b} \times \bm{x} )  \right) \cdot \nabla p^{\prime (1)} + \bm{e}^{\prime (1)} : \bm{e}^{\prime (1)}+ 2 (\bm{e}^{\prime (1)} : \bm{E}) \right).
\end{eqnarray}
The time derivatives that appear in (\ref{eq19}-\ref{eq21}) are evaluated as $\left(\frac{\partial}{\partial \,t}\right)_{r}=\dot{\theta}_{j} \frac{\partial}{\partial \theta_{j}}+\dot{\phi}_{j} \frac{\partial}{\partial \phi_{j}}$ where, in keeping with the use of a Stokesian approximation for the viscoelastic integrand, $\dot{\theta}_j$ and $\dot{\phi}_j$ are evaluated using (\ref{eq37}), which corresponds to the spheroid rotating along Jeffery orbits.

\subsection{\bf The Stokesian disturbance fields in spheroidal coordinates: Jeffery orbits at $De = 0$}\label{Problemform:velocity}

As already indicated, to $O(De)$, the integrals in (\ref{eq16}) may be evaluated using the Stokesian approximations for the disturbance fields in the actual problem. One therefore needs the Stokes disturbance fields associated with a torque-free spheroid in a simple shear flow\,($\bm{u}^{(1)}_{s}, p^{(1)}_s$), and that induced by a rotating spheroid in an otherwise quiescent fluid ($\bm{u}^{(2)}, p^{(2)}$). The disturbance in the former case arises only due to the straining component of the ambient shear. Following \cite{vivek2016}, one may decompose the ambient rate-of-strain tensor into five canonical components, in the body-fixed frame, as $\bm{E}=  \cos^{2} \theta_{j} \sin \phi_{j} \cos\phi_{j} \bm{1}_{x}\bm{1}_{x} + \frac{1}{2}\cos \theta_{j} (\cos2 \phi_{j} )(\bm{1}_{x}\bm{1}_{y}+\bm{1}_{y}\bm{1}_{x} )+  \sin \theta_{j} \cos \theta_{j} \sin \phi_{j} \cos \phi_{j} ( \bm{1}_{x}\bm{1}_{z} +  \bm{1}_{z}\bm{1}_{x} )- \cos \phi_{j}  \sin \phi_{j}  \bm{1}_{y}\bm{1}_{y}+\frac{1}{2}\sin \theta_{j} \cos2 \phi_{j} (\bm{1}_{y}\bm{1}_{z}+\bm{1}_{z}\bm{1}_{y} )+ \sin^{2} \theta_{j} \sin \phi_{j} \cos \phi_{j}  \bm{1}_{z}\bm{1}_{z}$. The disturbance velocity\,($\bm{u}_{1s}-\bm{u}_{5s}$) and pressure fields\,($p_{1s}-p_{5s}$), corresponding to each of the component flows, are given by the following expressions:
{\allowdisplaybreaks
\begin{align} \label{eq:distvel_actual}
    &\bm{u}_{1s}=\frac{-d \bar{\xi}_{0}}{Q^{1}_{1}(\xi_{0})-\xi_{0} Q^{1}_{2}(\xi_{0})} (\sin^{2} \theta_{j} \sin \phi_{j} \cos\phi_{j} ) \bm{S}_{2,0}^{(3)},\\
    &\bm{u}_{2s}=\frac{-d \bar{\xi}_{0}}{3Q^{1}_{1}(\xi_{0})-\xi_{0} Q^{1}_{2}(\xi_{0})} (\sin \phi_{j} \cos \phi_{j}(1+\cos^{2} \theta_{j}) ) (\bm{S}_{2,2}^{(3)}+\bm{S}_{2,-2}^{(3)} ),\\
     &\bm{u}_{3s}=\frac{i d \bar{\xi}_{0}}{3Q^{1}_{1}(\xi_{0})-\xi_{0} Q^{1}_{2}(\xi_{0})} (\cos \theta_{j} \cos 2 \phi_{j} ) (\bm{S}_{2,2}^{(3)}-\bm{S}_{2,-2}^{(3)} ),\\
     &\bm{u}_{4s}=\frac{2 d \xi_{0} \bar{\xi}_{0}}{Q^{1}_{2}(\xi_{0})(2 \xi_{0}^2-1)} (\sin \theta_{j} \cos \theta_{j} \sin \phi_{j} \cos \phi_{j} ) (\bm{S}_{2,1}^{(3)}-\bm{S}_{2,-1}^{(3)} ),\\
      &\bm{u}_{5s}=\frac{-i  d \xi_{0} \bar{\xi}_{0}}{Q^{1}_{2}(\xi_{0})(2 \xi_{0}^2-1)} (\sin \theta_{j} ( \cos^{2} \phi_{j}-\sin^{2} \phi_{j}) ) (\bm{S}_{2,1}^{(3)}+\bm{S}_{2,-1}^{(3)} ),\\
    &p_{1s}=\frac{-2 d \bar{\xi}_{0}}{Q^{1}_{1}(\xi_{0})-\xi_{0} Q^{1}_{2}(\xi_{0})} (\sin^{2} \theta_{j} \sin \phi_{j} \cos\phi_{j}) D_{3}F_{1}^{0},\\
      &p_{2s}=\frac{-2d \bar{\xi}_{0}}{3Q^{1}_{1}(\xi_{0})-\xi_{0} Q^{1}_{2}(\xi_{0})} (\sin \phi_{j} \cos \phi_{j}(1+\cos^{2} \theta_{j})) (D_{3} F_{1}^{2}+D_{3} F_{1}^{-2}),\\
     &p_{3s}=\frac{ 2i d \bar{\xi}_{0}}{3Q^{1}_{1}(\xi_{0})-\xi_{0} Q^{1}_{2}(\xi_{0})} (\cos \theta_{j} \cos 2 \phi_{j}) (D_{3} F_{1}^{2}-D_{3} F_{1}^{-2}),\\   
      &p_{4s}=\frac{4 d \xi_{0} \bar{\xi}_{0}}{Q^{1}_{2}(\xi_{0})(2 \xi_{0}^2-1)} (\sin \theta_{j} \cos \theta_{j} \sin \phi_{j} \cos \phi_{j}) (D_{3} F_{1}^{1}-D_{3} F_{1}^{-1}),\\
       &p_{5s}=\frac{-  2 i d \xi_{0} \bar{\xi}_{0}}{Q^{1}_{2}(\xi_{0})(2 \xi_{0}^2-1)} (\sin \theta_{j} ( \cos^{2} \phi_{j}-\sin^{2} \phi_{j}) ) (D_{3} F_{1}^{1}+D_{3} F_{1}^{-1}).
\end{align}}
The test velocity and pressure fields, corresponding to transverse rotations about the $X$ and $Y$ axes, are given by:
\begin{equation}
    \bm{u}^{(2)}_{sx}=\frac{i d (2 \xi_{0}^{2}-1)}{2 \xi_{0} Q^{0}_{1}(\xi_{0})-\bar{\xi}_{0} Q^{1}_{1}(\xi_{0})}(\bm{S}^{(2)}_{1,1}-\bm{S}^{(2)}_{1,-1}) +\frac{i d (\xi_{0} Q^{1}_{1}(\xi_{0})+2\bar{\xi}_{0} Q^{0}_{1}(\xi_{0}) ) }{Q^{1}_{2}(\xi_{0})(2 \xi_{0} Q^{0}_{1}(\xi_{0})-\bar{\xi}_{0} Q^{1}_{1}(\xi_{0}))}(\bm{S}^{(3)}_{2,1}+\bm{S}^{(3)}_{2,-1}),
\end{equation}
\begin{equation}
        \bm{u}^{(2)}_{sy}=\frac{d (2 \xi_{0}^{2}-1)}{2 \xi_{0} Q^{0}_{1}(\xi_{0})-\bar{\xi}_{0} Q^{1}_{1}(\xi_{0})}(\bm{S}^{(2)}_{1,1}+\bm{S}^{(2)}_{1,-1}) +\frac{ d (\xi_{0} Q^{1}_{1}(\xi_{0})+2\bar{\xi}_{0} Q^{0}_{1}(\xi_{0}) ) }{Q^{1}_{2}(\xi_{0})(2 \xi_{0} Q^{0}_{1}(\xi_{0})-\bar{\xi}_{0} Q^{1}_{1}(\xi_{0}))}(\bm{S}^{(3)}_{2,1}-\bm{S}^{(3)}_{2,-1}),
\end{equation}
\begin{eqnarray} 
  p^{(2)}_{sx}&=&\frac{2 i d (\xi_{0} Q^{1}_{1}(\xi_{0})+2\bar{\xi}_{0} Q^{0}_{1}(\xi_{0}) ) }{Q^{1}_{2}(\xi_{0})(2 \xi_{0} Q^{0}_{1}(\xi_{0})-\bar{\xi}_{0} Q^{1}_{1}(\xi_{0}))}(D^{1}_{2}+D^{-1}_{2}), \label{eq:distpres_test1}\\
  p^{(2)}_{sy}&=&\frac{ 2 d (\xi_{0} Q^{1}_{1}(\xi_{0})+2\bar{\xi}_{0} Q^{0}_{1}(\xi_{0}) ) }{Q^{1}_{2}(\xi_{0})(2 \xi_{0} Q^{0}_{1}(\xi_{0})-\bar{\xi}_{0} Q^{1}_{1}(\xi_{0}))}(D^{1}_{2}-D^{-1}_{2}). \label{eq:distpres_test2}
\end{eqnarray}
The $\bm S^{(q)}_{t,s}$'s\,($q = 1-3$) in the above expressions are vector spheroidal harmonics, with $D_t^s$ denoting the pressure contribution associated with $\bm S^{(3)}_{t,s}$. The detailed expressions for these harmonics in spheroidal coordinates\,$(\xi,\eta,\phi)$, and their connections to the vector spherical harmonics in the classical Lamb expansion\,(Chapter $4$ in \cite{kim1991}) have been discussed in earlier papers\,\citep{vivek2015,vivek2016,marath2018}; $\xi_0$ in (\ref{eq:distvel_actual}-\ref{eq:distpres_test2}) is the coordinate label for the spheroid, with $d$ being half the inter-foci distance. It must be said that the above expressions are based on the pioneering results obtained by Kushch and coworkers\,\citep{kushch1995,kushch1997,kushch1998,kushch2000,kushch2004} in the context of linear elasticity. 

In the Newtonian limit, corresponding to $De=0$, (\ref{eq16}) reduces to 
\begin{equation}\label{eq36}
\bm{\Omega}_{jeff} \cdot \bm{L}^{(2)} = \bm{\Gamma}  \bm{:} \int_{Sp}  \bm{x} \left(\bm{\Sigma}^{(2)} \cdot  \bm{n}  \right)  ~dS=\bm{\Omega}^{(1)} \cdot \bm{L}^{(2)},
\end{equation}
where $\bm{\Omega}_{jeff}$ is the Jeffery angular velocity. On evaluating the surface integral on the RHS using the known expression for $\bm{\Sigma}^{(2)}$ in spheroidal coordinates\,\citep{VivekThesis2009,vivek2016}, one obtains the components of $\bm{\Omega}_{jeff}$, orthogonal to $\bm{p}(\equiv  \bm{1}_{z}) =\sin \theta_{j} \cos\phi_{j} \bm{1}_{x^{\prime}} +\sin \theta_{j} \sin\phi_{j} \bm{1}_{y^{\prime}} +\cos \theta_{j} \bm{1}_{z^{\prime}}$, as:
\begin{equation}\label{eq37}
\dot{\phi}_{j}=-\frac{1}{2}+\frac{(\kappa^2 -1)}{2 (\kappa^2+1)} \cos 2 \phi_{j}, \qquad 
\dot{\theta}_{j} =\frac{(\kappa^2 -1)}{4 (\kappa^2+1)} \sin 2 \phi_{j} \sin 2 \theta_{j}. 
\end{equation}

The above equations lead to the Jeffery orbits in figure \ref{fig_jeffy_Prolate}. These orbits are spherical ellipses; the orientation and shape of the ellipse as a function of $\kappa$, for prolate and oblate spheroids, was described in \S\ref{intro}. Note that, in his original paper, \citet{jeffery1922} used ellipsoidal harmonics to first obtain the angular velocity of a triaxial ellipsoid, and then considered the simpler expression that results for a spheroid.
 
 In the absence of inertia, viscoelasticity, Brownian motion or interparticle interactions, a spheroid is confined to the Jeffery orbit, determined by its initial orientation, for all time. This is readily seen in $(C,\tau )$ coordinates defined by:
\begin{equation}\label{eq38}
C=\tan \theta_{j} \frac{\sqrt{\kappa^2 \sin^2  \phi_{j}+\cos^2 \phi_{j}}}{\kappa}; \quad \tau=\tan^{-1}\left(\frac{1}{\kappa \tan \phi_{j} } \right).
\end{equation}
Here, $C\,\in\,[0,\infty)$ is the orbit constant, with $C=0$ and $C=\infty$  corresponding to the spinning\,(log-rolling) and tumbling modes, respectively. All finite non-zero $C$ correspond to three-dimensional precessional motions termed the kayaking modes; $\tau$ is the phase along a given orbit. (\ref{eq37}), written in $(C,\tau)$ coordinates, take the simple forms: 
\begin{equation}\label{eq39}
     \frac{d C}{d t}=0, \qquad
     \frac{d \tau}{d t}=\frac{\kappa}{\kappa^2+1},
\end{equation}
where, unlike $\phi_{j}$, $\tau$ is seen to change at a uniform albeit $\kappa$-dependent rate. The period of rotation, $T_{jeff}=  2\pi (\kappa^2+1)/\kappa$, is independent of $C$. It is worth mentioning that, for inelastic fluids with a shear-rate-dependent viscosity, although reversibility constraints still lead to a spheroid moving along closed orbits, $T_{jeff}$ is nevertheless a function of $C$. For shear-thinning fluids, for instance, one expects the greater disturbance induced by a tumbling spheroid to lead to a smaller effective viscosity, and thereby faster rotation, compared to the kayaking modes. That $T_{jeff}$ is a increasing function of $C$ for power-law fluids with $n < 1$\,($n$ being the power-law index) has been confirmed in recent calculations \citep{abtahi2019}.

In the context of the results presented below, we note that  $\kappa=\xi_{0}/\sqrt{\xi_{0}^2-1}$ for a prolate spheroid, $\xi_{0}$ being the coordinate label mentioned above in the context of the expressions for the disturbance fields. In what follows, we will use both $\kappa$ and $\xi_0$ to refer to the spheroid aspect ratio; $\xi_0 \rightarrow \infty\,(\kappa \rightarrow 1)$ and $\xi_0 \rightarrow 1\,(\kappa \rightarrow \infty)$ correspond to the spherical and slender-fiber limits. One may rewrite (\ref{eq37}) in terms of $\xi_0$, in which case the orbit equations for an oblate spheroid are obtained, from those for a prolate one, by using the transformation  $\xi_{0} \leftrightarrow i \bar{\xi_{0}}$, $d \leftrightarrow -i d$ in the dimensional form of the equations. This transformation is used below, for nonzero $De$, to obtain the oblate orbital drift from the prolate one; the aspect ratio on the oblate side is given by $\kappa = \sqrt{\xi_0^2-1}/\xi_0$

\subsection{\bf The rotation rate in presence of weak viscoelastic effects: $De\ll 1$}\label{Problemform:visco}

Linearity of the Stokes equations translating to a linear dependence of the spheroid rotation rate $\dot{\bm p}$ on the ambient velocity gradient, the unit length constraint\,($\bm{p}\cdot \dot{\bm p}=0$) and the requirement of an affine response in the limit of solid body rotation\,($\dot{\bm p} = \bm{p} \cdot {\bm W}$), immediately leads to the general form $\dot{\bm p} = \bm{p} \cdot \bm{W} + B(\kappa)\left\{ \bm{E} \cdot \bm{p} - (\bm{E}:\bm{p}\bm{p})\bm{p}\right\}$, for $De = 0$, in an ambient linear flow. The detailed calculations show that the Bretherton constant $B(\kappa) = \frac{\kappa^2-1}{\kappa^2+1}$; the components of the resulting equation in spherical polar coordinates, for simple shear flow, are (\ref{eq37}). The regular nature of the viscoelastic correction, and the quadratic dependence of the second-order fluid stress on the velocity gradient, imply that the $O(De)$ correction to the above rate of change must be a quadratic function of $\bm{E}$ and $\bm{W}$. The other two constraints above then lead to the following form for the rotation rate, to $O(De)$ in an ambient linear flow:
\begin{equation}\label{eq40}
\begin{aligned}
    &\bm{\dot{p}}=\bm{p} \cdot \bm{W} +  B(\kappa) \left\{ \bm{E} \cdot \bm{p} - (\bm{E}:\bm{p}\bm{p})\bm{p} \right\}
    + De \bigg\{\beta_{1}(\xi_0,\epsilon)(\bm{E:pp})(\bm{I-pp}) \cdot (\bm{E \cdot p})\\
    &+\beta_{2}(\xi_0,\epsilon)(\bm{E:pp})\bm{W\cdot p}+\beta_{3}(\xi_0,\epsilon)\bm{(I-pp)\cdot W \cdot (E\cdot p)}+\beta_{4}(\xi_0,\epsilon)(\bm{I-pp})\bm{\cdot E \cdot }(\bm{E \cdot p}) \bigg\}. 
\end{aligned}
\end{equation}
The purpose of a detailed calculation is to determine the four independent functions of $\xi_0$ and $\epsilon$; the dependence on $\epsilon$ is linear, mirroring the dependence in the original constitutive equation; see (\ref{eq5}) and (\ref{eq6}). The quadratic form in (\ref{eq40}) is also valid for weak inertial effects, with $Re$ now being the proportionality factor, and the $\beta_i's$ only being functions of $\xi_0$. The closed-form expressions for these functions are implicit in the results of \cite{vivek2016} and \cite{marath2018}, and have been tabulated in \S\ref{shape_funs_inertia}.

 Returning to the viscoelastic correction and the specific case of simple shear flow, the components of the $O(De)$ correction in (\ref{eq40}), in terms of $\theta_j$ and $\phi_j$, assume the following forms:
\begin{eqnarray}\nonumber
    (\dot{\phi}_{j})_{De}= \cos\phi_{j} \sin\phi_{j} \Big(&&G_{1}(\xi_{0},\epsilon)+ G_{2}(\xi_{0},\epsilon)\cos2 \theta_{j} + G_{3}(\xi_{0},\epsilon)\cos 2 \phi_{j}\\ \label{eq41}
 &&+G_{4}(\xi_{0},\epsilon) \cos2\theta_{j} \cos 2\phi_{j}	\Big),\\\nonumber
(\dot{\theta}_{j})_{De}= \cos\theta_{j} \sin\theta_{j} \Big(&& F_{1}(\xi_{0},\epsilon) +F_{2}(\xi_{0},\epsilon) \cos 2 \phi_{j} +F_{3}(\xi_{0},\epsilon) \cos 2 \theta_{j} + F_{4}(\xi_{0},\epsilon) \cos 4\phi_{j}\\ \label{eq42}
 &&+F_{5}(\xi_{0},\epsilon) \cos(2\theta_{j}-4\phi_{j})+ F_{6}(\xi_{0},\epsilon) \cos(2\theta_{j}+4\phi_{j}) \Big),	
\end{eqnarray}
where the $F_{i}$'s and $G_{i}$'s are related to the $\beta_{i}$'s above through the following relations: $F_{1}=( \beta_{1} +4 \beta_{4})/16$, $F_{2}=-\beta_{3}/4$, $F_{3}=-\beta_{1}/16$, $F_{3}=F_{4}$, $F_{5}=F_{6}=-F_{3}/2$, $G_{1}=(\beta_{2}+2\beta_{3})/4$, $G_{2}=-\beta_{2}/4$ and $G_{3}=-G_{4}=\beta_{1}/4$. These functions have been obtained in closed form by analytically evaluating the surface integral involving $p'_G$, and the volume integral involving $\bm{\sigma}^{\prime (1)}_{NNQ}$, in (\ref{eq16}). The detailed expressions are given in \S\ref{shape_fun_visco}, where the co-rotational and quadratic contributions to each of these functions have been identified. Interestingly, the final expressions imply $\beta_2 = \beta_3 = 0$, so that the original invariant form, (\ref{eq40}), is independent of $\bm{W}$, this being consistent with the general expression originally proposed by \citet{brunn1977}\,(and evaluated by him for a tri-dumbbell); this is not true for the inertial case, which ends up involving both $\bm{E}$ and $\bm{W}$ \citep{vivek2016,marath2018}. We also note that there is no contribution from the nonlinear terms of the corotational component in (\ref{eq19}) and (\ref{eq21}); in contrast, both linear and nonlinear terms of the quadratic component in (\ref{eq20}) contribute to (\ref{eq41}) and (\ref{eq42}).

\section{\bf Viscoelastic orientation dynamics below rotation arrest: $De \ll 1$}\label{Orientation_dynamics}
Based on its definition in (\ref{eq38}), one obtains the time rate of change of $C$ as:
\begin{equation}\label{eq3p1}
\begin{aligned}
\frac{d C}{d t}=\,De\,C\Bigg\{\frac{1}{\sin \theta_{j} \cos \theta_{j} } \left(
\frac{d \theta_{j}}{d t}\right)_{De} + \frac{ (\kappa^2-1) \cos\phi_{j} \sin \phi_{j} }{\kappa^2 \sin^{2} \phi_{j}+ \cos^2 \phi_{j} } \left(\frac{d \phi_{j}}{d t} \right)_{De} \Bigg\},
\end{aligned}
\end{equation}
where $\left( \frac{d \theta_{j}}{dt}\right)_{De}$ and $\left( \frac{d \phi_{j}}{dt}\right)_{De}$ are given by (\ref{eq41}) and (\ref{eq42}), respectively. The non-zero $dC/dt$ implies a viscoelasticity-induced orbital drift. In this section, we restrict ourselves to the scenario where this drift occurs at a rate much slower than Jeffery rotation, so that the originally closed unit-sphere trajectory topology is transformed into a tightly spiralling one. For $\kappa$ fixed, and for $De$ sufficiently small, the time scale of $O(De^{-1}\dot{\gamma}^{-1})$, that characterizes the orbital drift, is asymptotically large compared to the Jeffery period of $O(\dot{\gamma}^{-1})$ that characterizes the evolutions of $\theta_j$ and $\phi_j$ on the RHS of (\ref{eq3p1}). In \S\ref{Orientation_dynamics:multi}, we show that a multiple time scales analysis\,\citep{bender1999} may be used in this limit to quantify the effect of viscoelasticity in terms of the evolution of $C$ alone. In \S\ref{Orientation_dynamics:deltaC}, we show how this leads to the small-$De$ orientation dynamics on the entire unit sphere being characterized by plots of $\Delta C$ vs $C/(C+1)$, $\Delta C$ here being the change in orbit constant during a single Jeffery period.

For a fixed $De$ however small, the aforementioned separation of time scales breaks down when $\kappa$ is sufficiently small or large. For slender prolate and thin oblate spheroids, the Jeffery angular velocity becomes asymptotically small in the neighborhood of the flow-gradient and gradient-vorticity planes, which correspond to the near-aligned phases of rotation in the respective cases. As a result, the viscoelastic correction becomes important at leading order, eventually leading to rotation arrest above a threshold $De$ that is still small. In the absence of a time scale separation, small-$De$ trajectories do not have a tightly spiralling character everywhere on the unit sphere, allowing for non-trivial bifurcations especially on the oblate side. This scenario is analyzed in \S\ref{Rotation_arrest}.

\subsection{\bf Multiple time scales analysis of spheroid rotation for $De \ll 1$}\label{Orientation_dynamics:multi}

Provided that the drift and Jeffery-rotation time scales are well separated, one may proceed via a multiple scales analysis with $t_1 = \dot{\gamma}t$ and $t_{2} = De\dot{\gamma}t$ as the fast and slow time variables\,\citep{bender1999,ganesh2004}. Expanding the time derivative as $d/dt = d/d\,t_1+De\,d/dt_2$, one formally writes (\ref{eq3p1}) in the form:
\begin{align}
\frac{dC}{dt_1} + De \frac{dC}{dt_2} =&\, De \mathcal{H}(C,\tau;\kappa,\epsilon), \label{eq3p2} \\
=&\, De \mathcal{H}(C,\frac{\kappa}{\kappa^2+1}t_1;\kappa,\epsilon), \label{eq3p3}
\end{align}
where the expression for $\tau$ from (\ref{eq38}) has been used in going from the (\ref{eq3p2}) to (\ref{eq3p3}), keeping in mind that this represents a fast-time dependence. The function ${\mathcal H}$ is obtained by first expressing $\theta_j$ and $\phi_j$ in terms of $C$ and $\tau$\,(this corresponds to inverting the $(C,\tau)$-relations given by (\ref{eq38})), substituting these in (\ref{eq41}) and (\ref{eq42}), which may then be substituted in (\ref{eq3p1}).
(\ref{eq3p3}) suggests writing $C$ in terms of a two-time-scale expansion of the form\,\citep{marath2018,banerjee2021}, $C =C_{0}(t_1,t_2)+De\, C_1(t_1,t_2)+O(De^2)$, which gives:
\begin{equation}\label{eq3p4} 
\frac{d C_{0}}{d t_1} = 0,
\end{equation}
 at leading order, so $C_{0} \equiv C_0(t_2)$ represents the slow drift due to viscoelasticity. At $O(De)$, one obtains:
\begin{equation}\label{eq3p5} 
\frac{d C_{1}}{d t_{1}} + \frac{d C_{0}}{d t_{2}} = \mathcal{H} \left(C_0, \frac{\kappa t_{1}}{\kappa^2+1}; \kappa, \epsilon \right),
\end{equation}
where $C_1$ accounts for the small amplitude fast fluctuations about the aforementioned drift. To determine $C_{0}$, we integrate (\ref{eq3p5}), with $t_2$ fixed, over $t_1\,\in\,[0,T_{jeff}]$, corresponding to a single fast-time period\,(a single Jeffery rotation). In doing so, $C_1$ is constrained to be a periodic function of $t_{1}$, which is equivalent to eliminating secular growth on the fast time scale. This leads to the first term in (\ref{eq3p5}) integrating out to zero, yielding:
\begin{equation}\label{eq3p6} 
\begin{aligned} \frac{d C_{0}}{d t_{2}} &= \frac{1}{T_{jeff}} \int_{0}^{T_{jeff}} \mathcal{H} \left(C_0, \frac{\kappa t_{1}}{\kappa^2+1}; \kappa, \epsilon \right) d t_{1}, \\
&= \frac{1}{2 \pi} \int_{0}^{2\pi} \mathcal{H} (C_0, \tau; \kappa, \epsilon) d\tau.
\end{aligned} 
\end{equation}
Using (\ref{eq3p6}) in (\ref{eq3p5}) gives the following expression for $C_{1}$:
\begin{equation}\label{eq3p7} 
C_{1} = \frac{T_{jeff}}{2 \pi} \left( \int_{0}^{\tau} \mathcal{H} (C_0,\tau^{\prime}; \kappa, \epsilon) d \tau^{\prime} - \frac{\tau}{2\pi}\int_{0}^{2\pi} \mathcal{H} (C_0, \tau; \kappa, \epsilon) d\tau \right),
\end{equation}
where $C_0$ is assumed constant during the $\tau$-integration.
Together, (\ref{eq3p6}) and (\ref{eq3p7}) provide a reduced one-dimensional description of the effect of viscoelasticity. The autonomous ODE governing $C_0$ may be solved numerically to obtain the slow drift of the orbit constant with time. This slow variation is then used in (\ref{eq3p7}), after the $\tau$-integration, to obtain the fast orbit-constant fluctuations that accompany the drift.  

\begin{figure}
		\centering
		\subfigure[]{\includegraphics[scale=0.3]{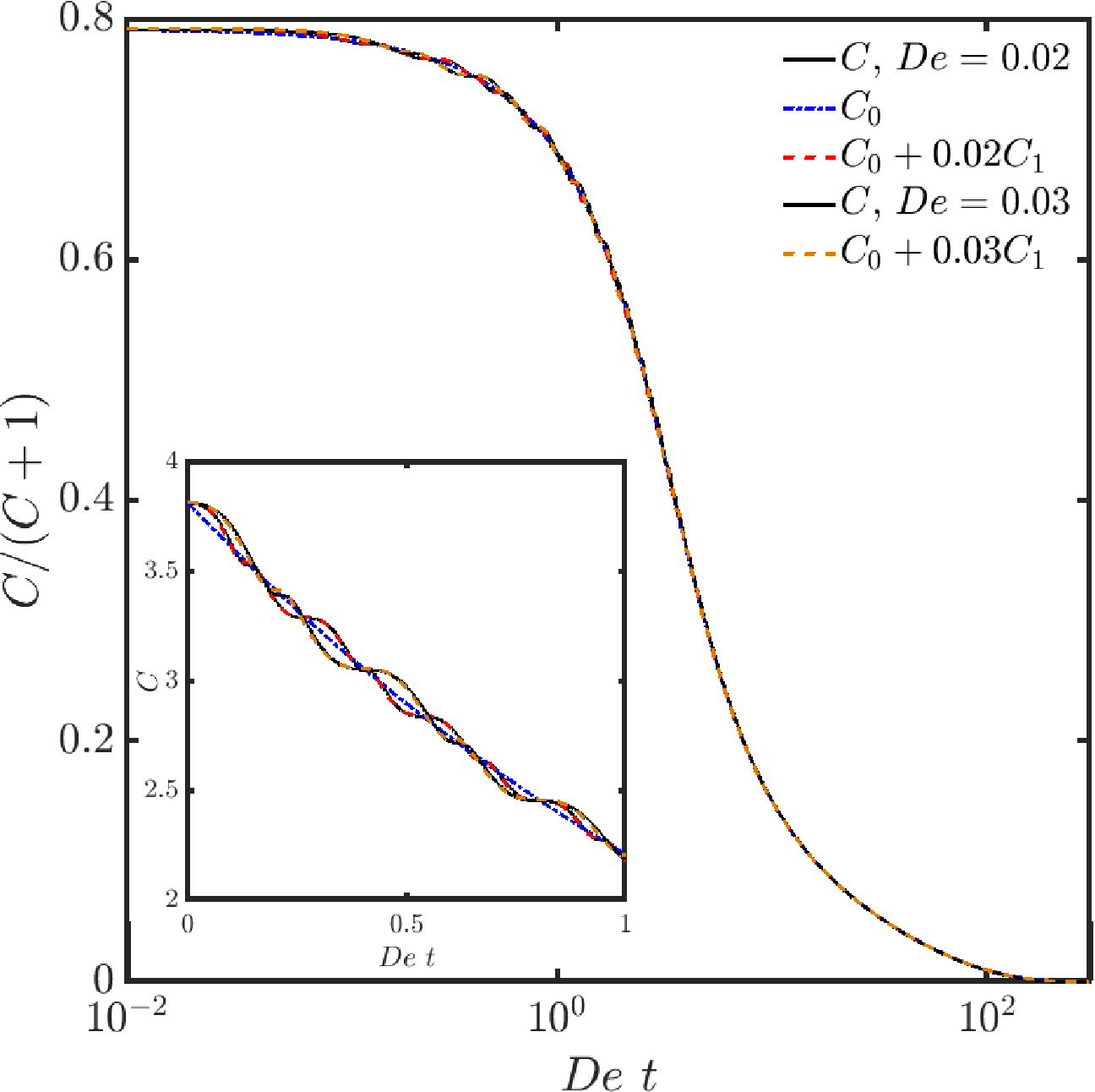}\label{fig_multi_a} }
		\subfigure[]{\includegraphics[scale=0.3]{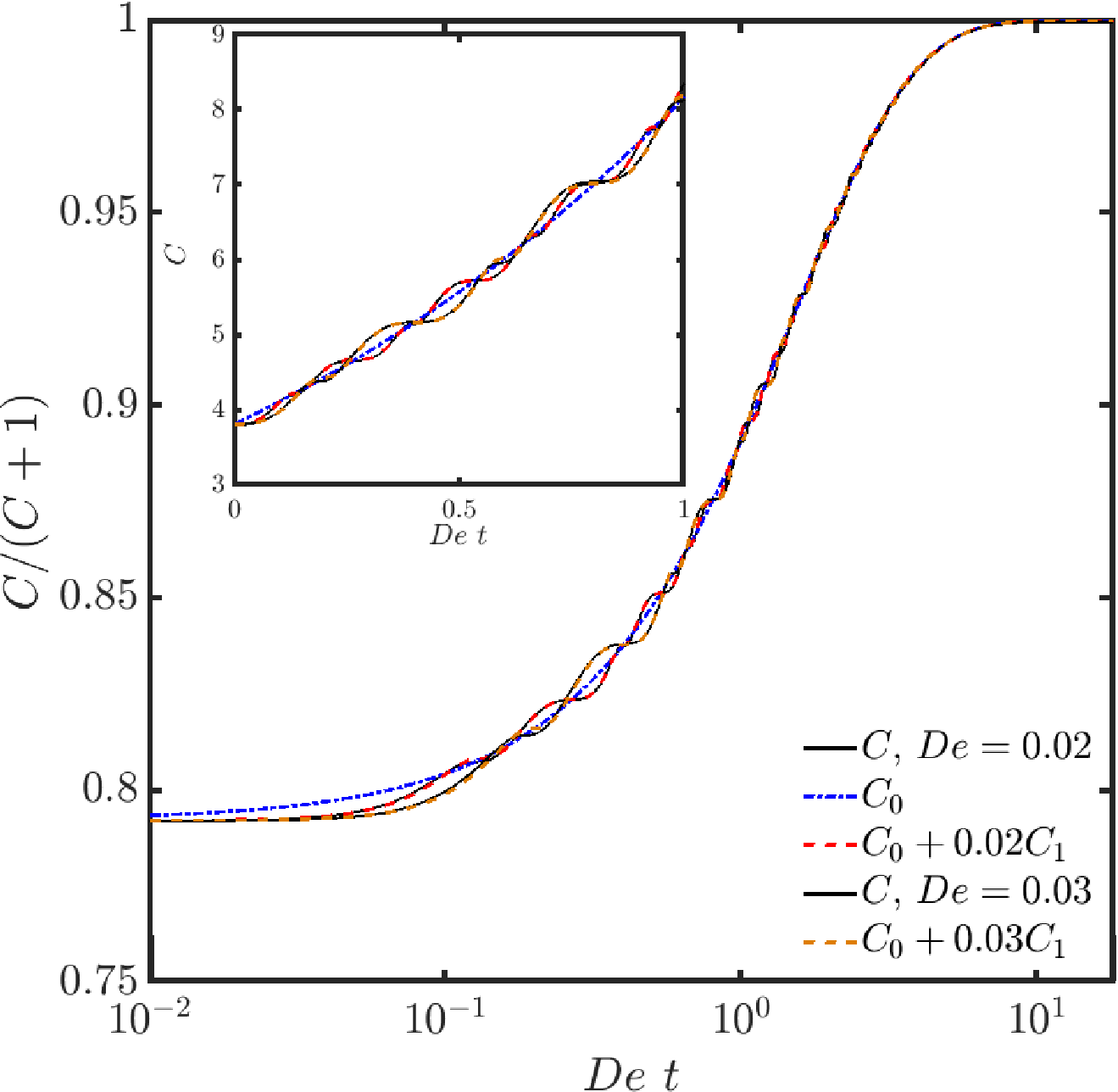}\label{fig_multi_b}}
		\caption{The orbital coordinate $C$ as a function of fast time scale $t$ starting from $(\theta_{j},\phi_{j})=(23 \pi/48,0)$ for spheroid of aspect ratio $\kappa=4$ for (a) $\epsilon=-2$ and (b) $\epsilon=2$. The calculations are based on (1) $C$ obtained from the numerical integration of the governing equations of $\theta_{j}$ and $\phi_{j}$ (solid line), (2) orbit drift alone obtained from solving (\ref{eq3p6}) (dash-dotted line), and (3) the drift together with the fluctuations (dashed line).  }\label{fig_multi}
\end{figure}

Figures \ref{fig_multi_a} and \ref{fig_multi_b} compare the orbit constant prediction\,($C_0 + De C_1)$, obtained from solving (\ref{eq3p6}) and (\ref{eq3p7}), to that obtained from directly solving the equations for $\theta_j$ and $\phi_j$, to $O(De)$, and then calculating $C$ using (\ref{eq38}). This is done for $De=0.02$ and $0.03$, for prolate spheroids with $\kappa =4$, and for $\epsilon = -2$ and $2$. In all cases, the spheroid starts from the initial orientation $(\theta_{j},\phi_j) \equiv (23\pi/48,0)$, which corresponds to $C(0) =3.814$ for $\kappa=4$. The main plots show $C$ decreasing to zero\,(Figure \ref{fig_multi_a}), and increasing to infinity\,(Figure \ref{fig_multi_b}), on a time scale of $O(De^{-1}\dot{\gamma}^{-1})$, for the two values of $\epsilon$ considered. This is consistent with the organization of the small-$De$ orientation dynamics regimes on the $\kappa-\epsilon$ plane, shown later in \S\ref{Results:epsilon_kappa}\,(see figure \ref{fig_kappa_epsilon}), where the spinning mode is the only stable mode for $\epsilon = -2$, in the interval $1 < \kappa \leq 9.324$, with the tumbling mode alone being stable for $\epsilon = 2$, in the interval $1 < \kappa \leq 6.087$. The inset views in figures \ref{fig_multi_a} and \ref{fig_multi_b} show that the fluctuations in $C$ are also captured accurately by the multiple scales analysis, for small $De$; as expected, the amplitude of these fluctuations scales as $De$.

\subsection{\bf Effect of viscoelasticity based on the orbital drift interpretation }\label{Orientation_dynamics:deltaC}

Although we have solved (\ref{eq3p6}) for the orbit constant evolution, characterizing the small-$De$ trajectory topology does not necessarily require solving the said ODE. The RHS of (\ref{eq3p6}) may be directly interpreted as the average rate of change of C\,(scaled by $De$) over a single Jeffery period, in the neighborhood of a Jeffery orbit with orbit constant $C$, and is therefore denoted as $\frac{\Delta C(C;\kappa,\epsilon)}{De\,T_{jeff}}$. Here, $\Delta C$ may also be regarded as the pitch of a tightly spiralling small-$De$ trajectory, measured in units of $C$. From the RHS of (\ref{eq3p6}), one obtains 
{\allowdisplaybreaks 
\begin{align}\nonumber
\Delta C =& De \int_{0}^{T_{jeff}} \mathcal{H} \left(C_0, \frac{\kappa t_{1}}{\kappa^2+1}; \kappa, \epsilon \right) d t_{1} \\ \nonumber
=&De~ C \frac{\kappa^2+1}{\kappa} \Bigg\{ \Big( I_{1} ~F_{1}(\xi_{0},\,\epsilon) + I_{2} ~F_{2}(\xi_{0},\,\epsilon)+ I_{3}~ F_{3}(\xi_{0},\,\epsilon) + I_{4} ~ F_{4}(\xi_{0},\,\epsilon)+ I_{5} ~F_{5}(\xi_{0},\,\epsilon)  \\\label{eq3p8}
&+ I_{6}~ F_{6}(\xi_{0},\,\epsilon) \Big) + \Big( J_{1} ~G_{1}(\xi_{0},\,\epsilon) + J_{2}~G_{2}(\xi_{0},\,\epsilon)+ J_{3}~ G_{3}(\xi_{0},\,\epsilon)
+ J_{4} ~ G_{4}(\xi_{0},\,\epsilon)  \Big)\Bigg\},
\end{align}}where the $I_1-I_6$ and $J_1-J_4$ are functions of $C$ and $\kappa$, that result from integrating the trigonometric functions in (\ref{eq41}) and (\ref{eq42}) over $\tau$. These expressions have been given in \cite{vivek2016}, and are tabulated in \S\ref{ItoJ_expr} for convenience.

\begin{figure}
		\centering
		\subfigure[]{\includegraphics[scale=0.15]{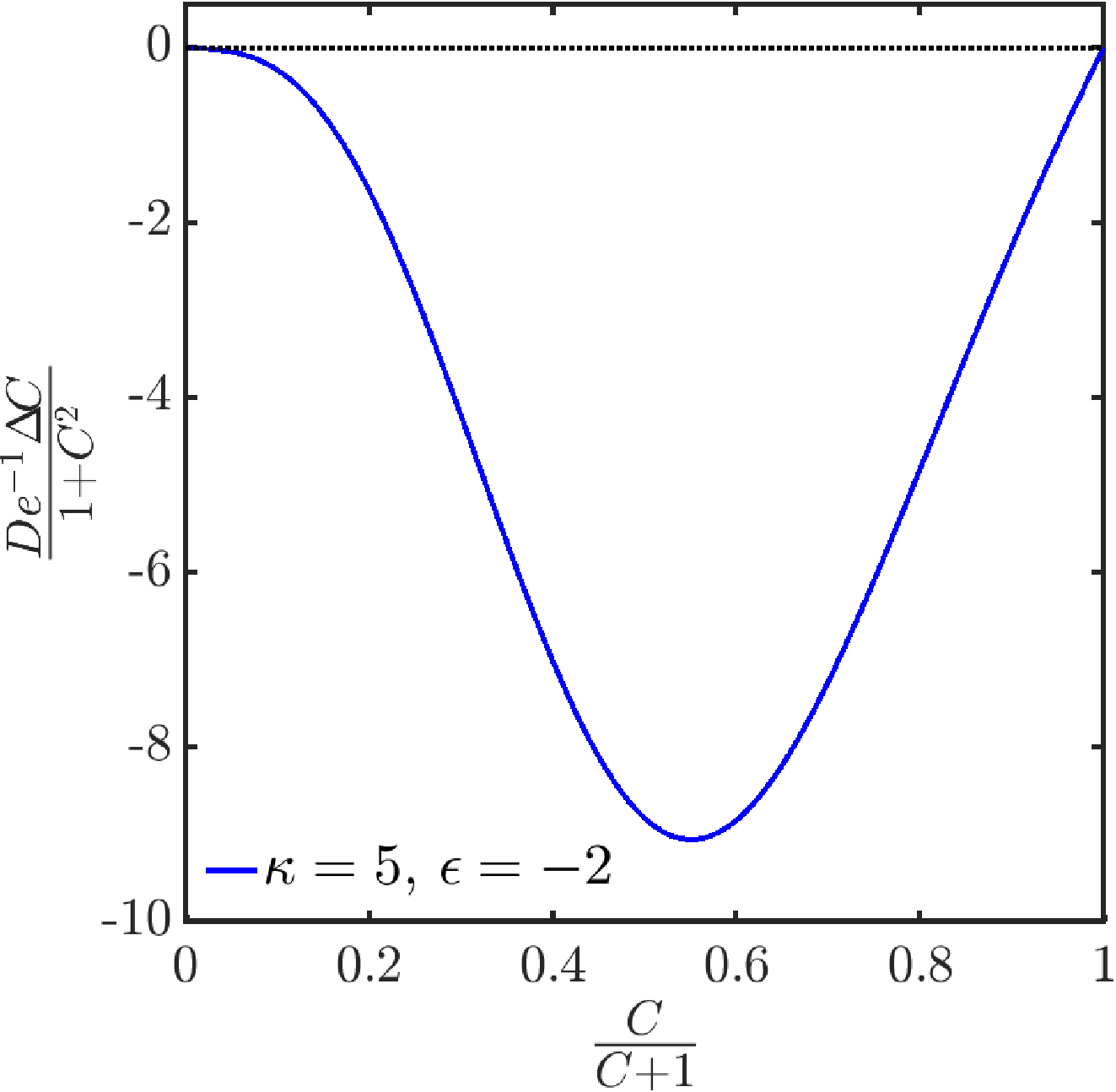}\label{fig3a}  }
		\subfigure[]{\includegraphics[scale=0.21]{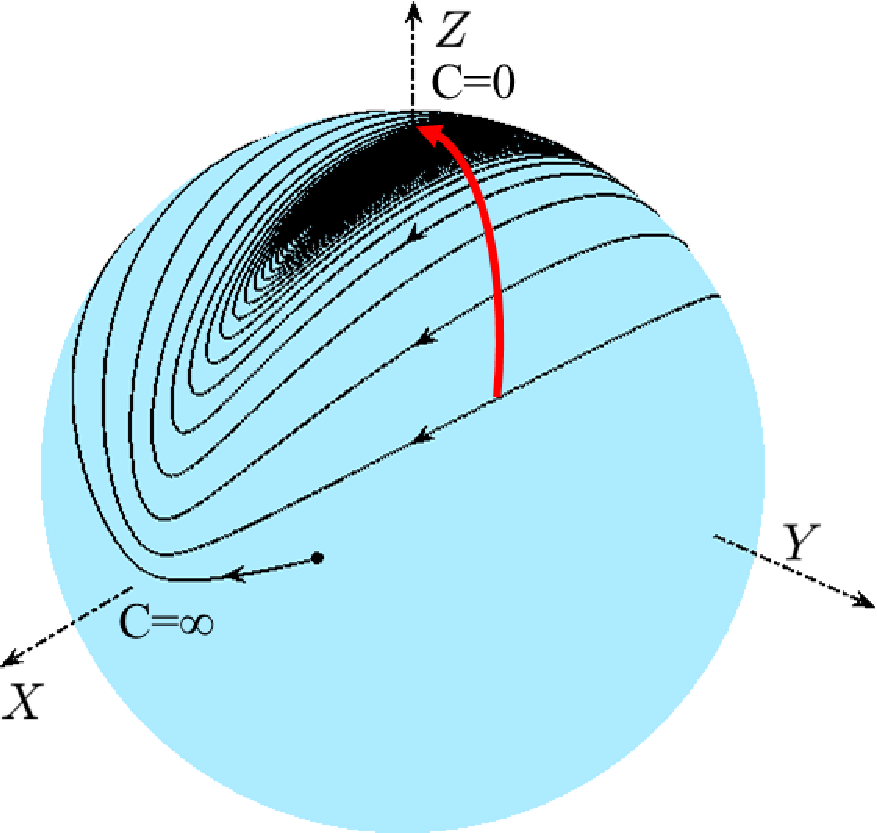}\label{fig3b}}
          \subfigure[]{\includegraphics[scale=0.15]{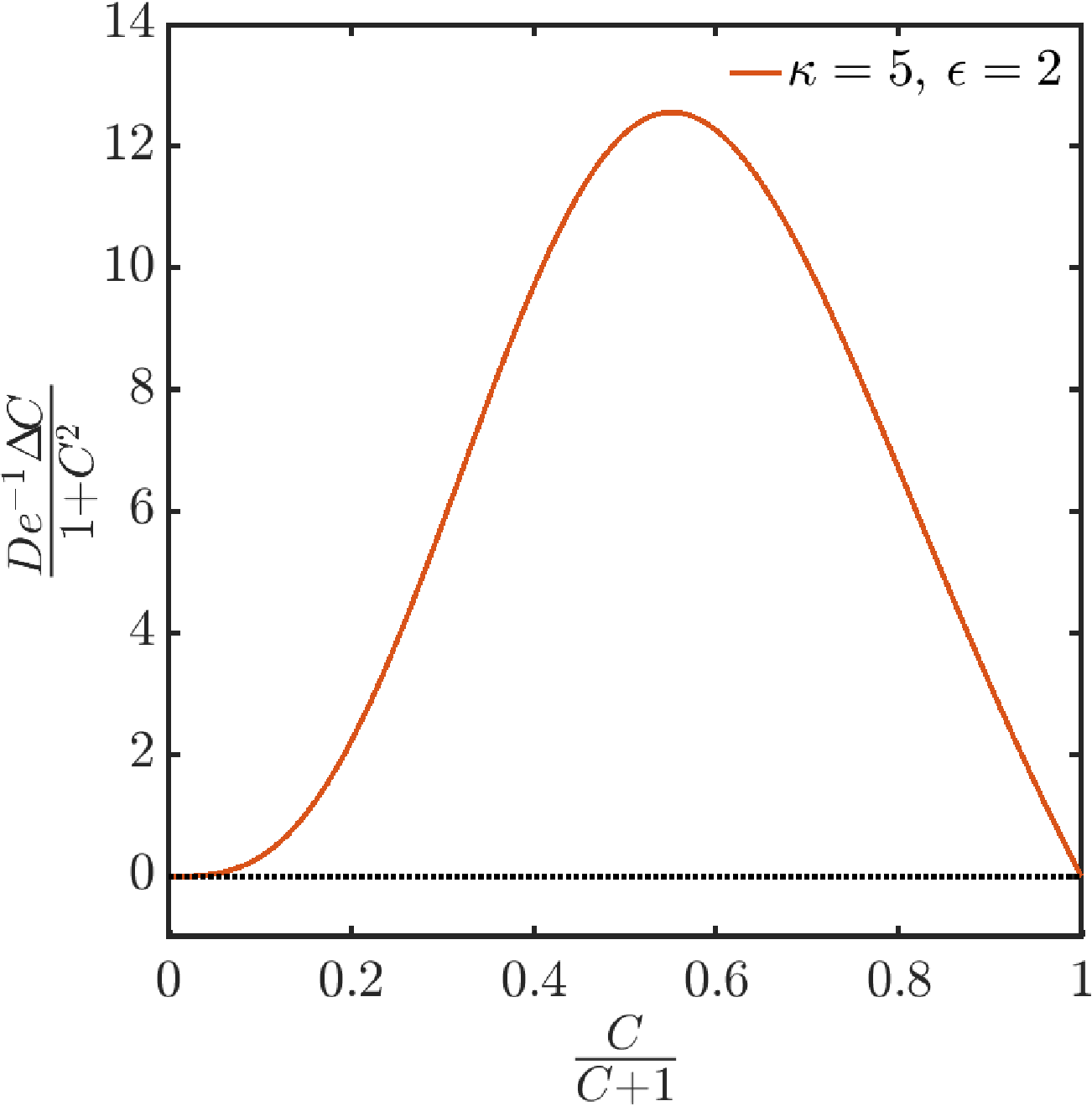}\label{fig3c} }
        \subfigure[]{\includegraphics[scale=0.21]{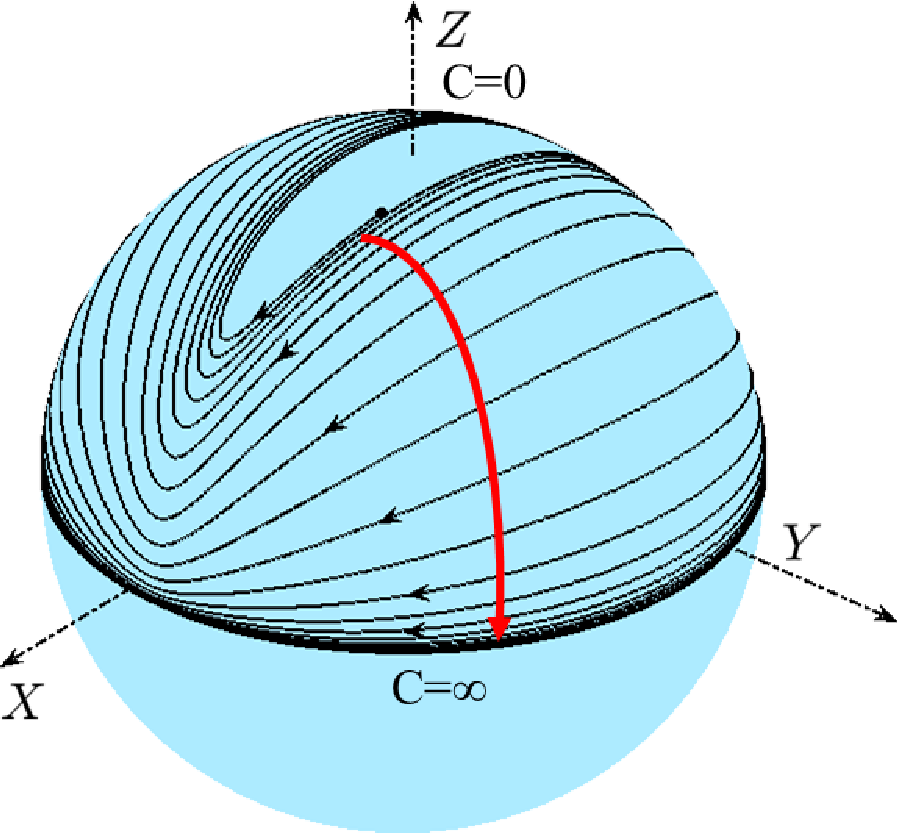}\label{fig3d} }
		\caption{The drift due to elasticity, $De^{-1} \Delta C /(C^2+1)$, as characterized by the normalized change in the orbit constant in a single Jeffery period, plotted as a function of $C/(C + 1)$, for a prolate spheroid; $C/(C + 1)=0$ and $C/(C+ 1)=1$ correspond to the spinning and tumbling modes, respectively. In (a) and (c)  for $\Delta C<0$ and $\Delta C>0$ $\forall\,C$, and the corresponding trajectories are in (b) and (d) with $De=0.01$. The black arrow indicates the direction of drift. } \label{fig3}
\end{figure}

\begin{figure}
		\centering
		\subfigure[]{\includegraphics[scale=0.15]{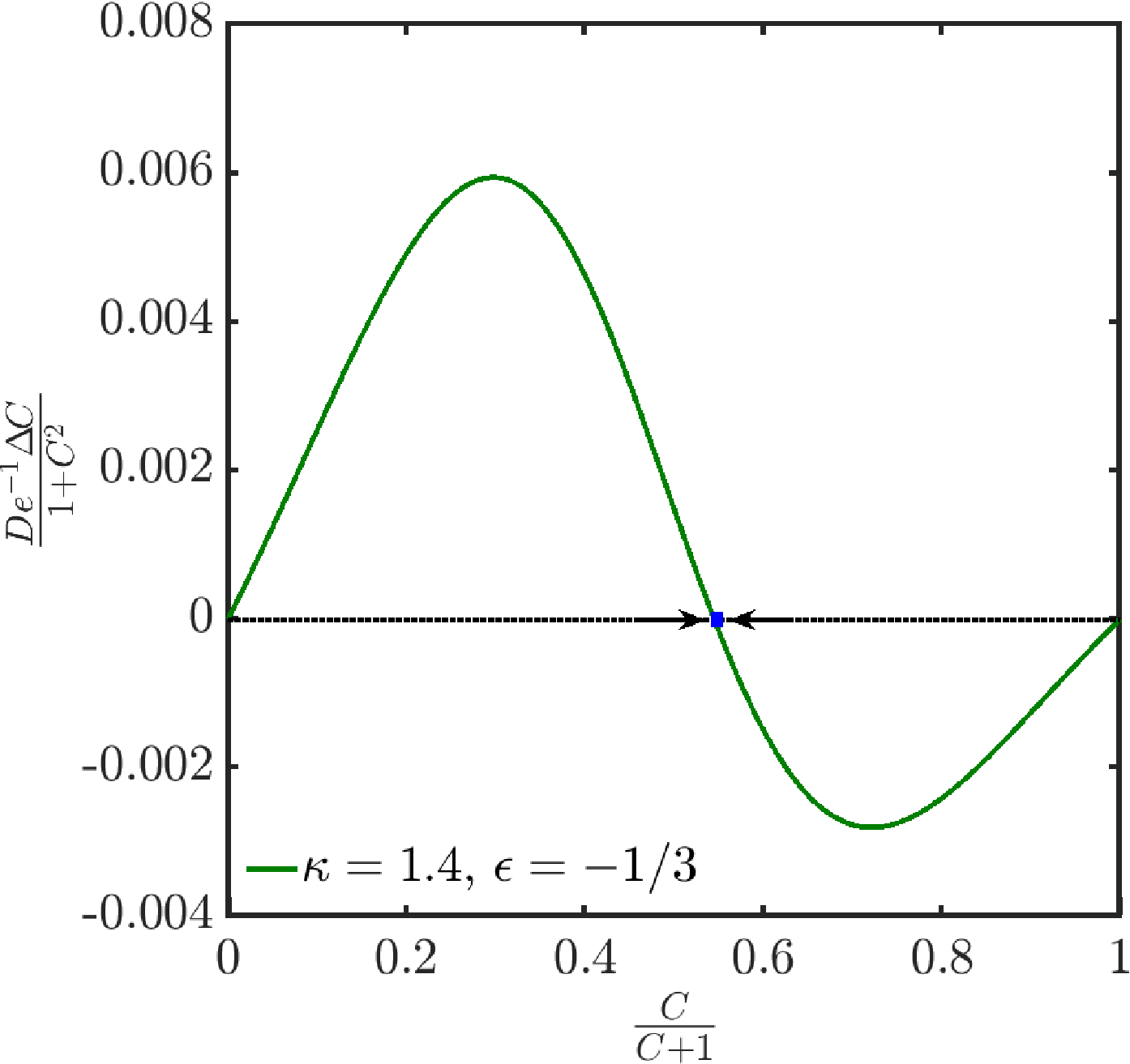}\label{fig4a}  }
		\subfigure[]{\includegraphics[scale=0.202]{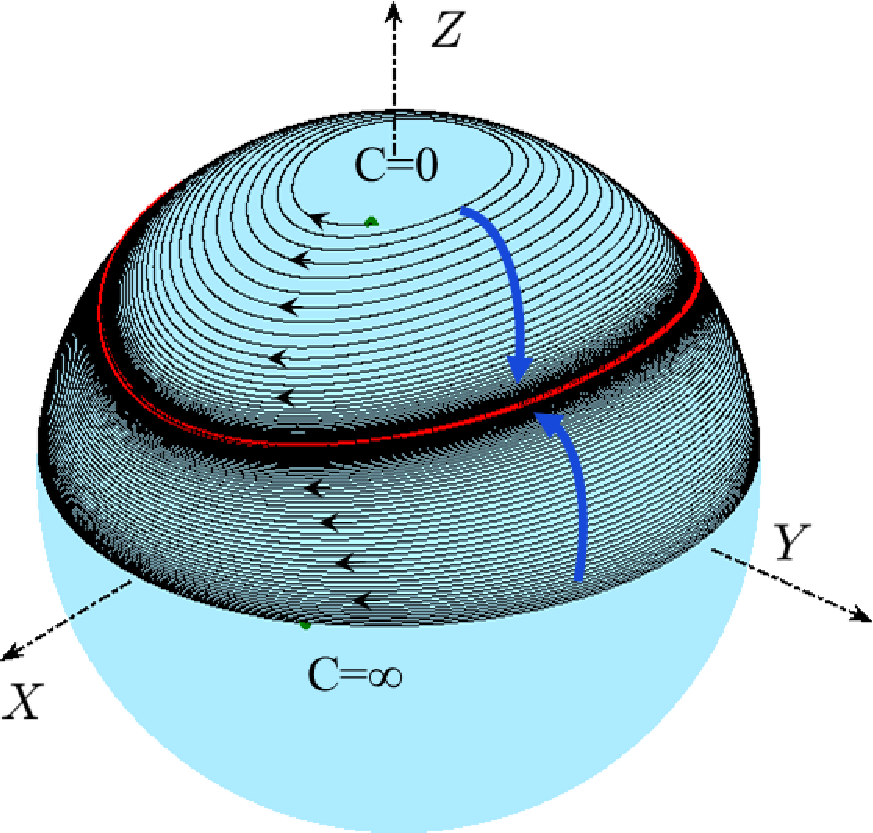}\label{fig4b}}
          \subfigure[]{\includegraphics[scale=0.15]{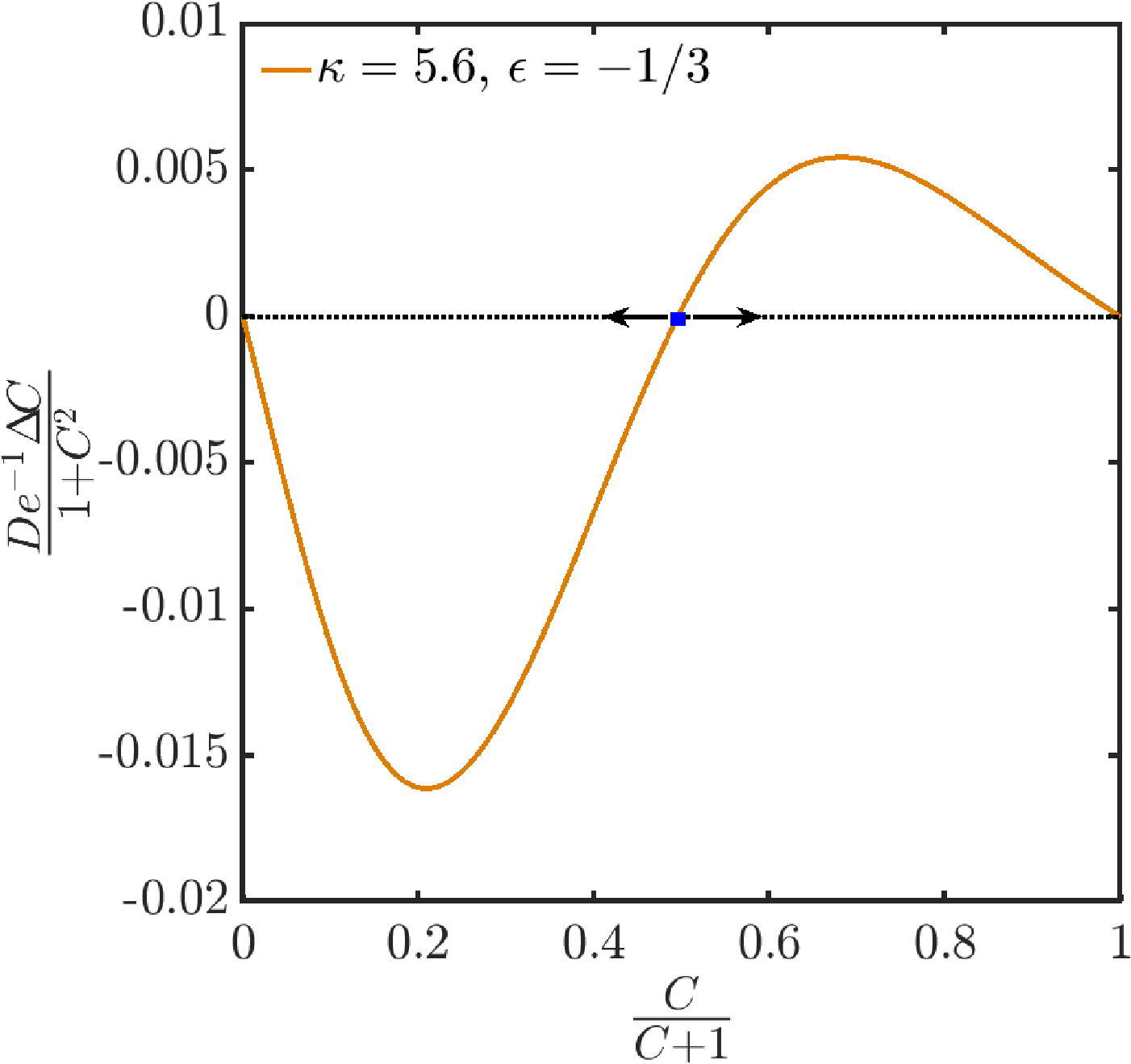}\label{fig4c} }
		\subfigure[]{\includegraphics[scale=0.202]{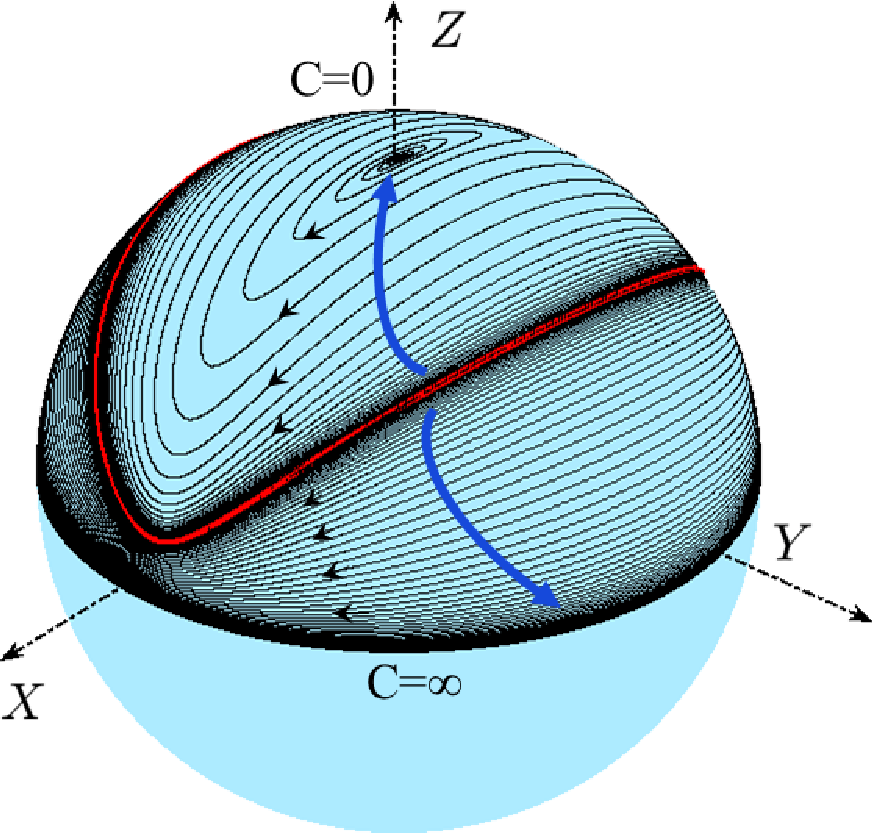}\label{fig4d}}
		\caption{The drift due to elasticity, $De^{-1} \Delta C/(C^2+1)$, as characterized by the normalized change in the orbit constant in a single Jeffery period, plotted as a function of $C/(C + 1)$, for a prolate spheroid with $De=5.0$. (a) $\Delta C$ has a zero crossing with a negative slope, and (c) has a zero crossing with a positive slope, and the corresponding trajectories on the unit sphere are in (b) and (d), respectively. Here, the solid red curve corresponds to an intermediate kayaking orbit and the blue arrow indicates the direction of drift.  In (a) and (c), solid marker represent the position of zero crossing of $\Delta C$.   }\label{fig4}
\end{figure}

Plots of $De^{-1} \Delta C/(C^2+1)$ vs $C/(C+1)$ may now be used to characterize the nature of spiralling trajectories in presence of weak viscoelasticity. As $C/(C+1)$ varies from $0$ to $1$, one moves from the spinning to the tumbling mode; further, $\Delta C$ is normalized by $C^2+1$, so that $\Delta C/(C^2+1) = 0$ for $C = 0$ and for $C \rightarrow \infty$ - as pointed out in \S\ref{intro}, in the context of the theoretical effort of \citet{brunn1977}, symmetry dictates that spinning and tumbling modes are always allowed equilibria, and it is only their stability that depends on the relevant physical parameters. Our detailed calculations for different $\xi_0$\,(or $\kappa$) and $\epsilon$ reveal four different types of $De^{-1} \Delta C/(C^2+1)$ vs $C/(C+1)$ plots, examples of which are illustrated in figures \ref{fig3} and \ref{fig4}. In figure \ref{fig3a} corresponding to $(\kappa,\epsilon) \equiv (5,-2)$, $\Delta C$ is seen to be negative for all finite non-zero $C$, implying that the spheroid drifts towards the spinning mode starting from any initial orientation; the associated unit-sphere trajectory in figure \ref{fig3b} spirals in towards the vorticity axis. In contrast, in figure \ref{fig3c}, corresponding to $(\kappa,\epsilon) \equiv (5,2)$, $\Delta C$ is positive for all $C$, implying that a spheroid for these parameters drifts towards the tumbling mode starting from any initial orientation; the associated unit-sphere trajectory in  figure \ref{fig3d} spirals out towards the flow-gradient plane. The $\Delta C$ vs $C$ curves in figures \ref{fig4a} and \ref{fig4c} exhibit a zero crossing, corresponding to a limit cycle equilibrium intermediate between the spinning and tumbling modes; for small $De$, the limit cycle closely approximates a finite-$C$ Jeffery orbit. When the slope at the zero-crossing is negative, as in figure \ref{fig4a}, the limit cycle is stable, and almost all initial orientations\,(except those corresponding to the tumbling and spinning modes) approach a finite-$C$ kayaking motion for sufficiently long times - figure \ref{fig4b} shows a pair of spiralling trajectories approaching the limit cycle\,(red curve) for long times. When the slope at the zero-crossing is positive, as in figure \ref{fig4c}, the limit cycle is an unstable equilibrium, and acts to divide each unit hemisphere into distinct basins of attraction for the tumbling and spinning modes. All initial orientations in one basin approach the spinning mode, while the ones in the other approach the tumbling mode. Figure \ref{fig4d} depicts a pair of trajectories starting from initial orientations on either side of the unstable limit cycle\,(red curve), and spiralling away from it. For future reference, the scenarios in figures \ref{fig3a}, \ref{fig3c}, \ref{fig4a} and \ref{fig4c} will be referred to as Regimes $1$, $2$, $3$ and $4$, respectively.


\subsection{\bf Viscoelastic-orientation dynamics on the $\kappa$-$\epsilon$ plane}\label{Results:epsilon_kappa}

In this section, we will show $De^{-1} \Delta C/(C^2+1)$ vs $C/(C+1)$ plots over a wide range of spheroid aspect ratios, in order to illustrate transitions in the orientation dynamics that occur with changing $\kappa$, at a fixed $\epsilon$. This leads naturally to an organization of the different orientation-dynamics regimes, identified in figures \ref{fig3} and \ref{fig4}, on the $\kappa-\epsilon$ plane. To do the above, there is the need to scale $De^{-1} \Delta C/(C^2+1)$ appropriately, so it remains finite both in the near-sphere limit, and for extreme aspect ratios\,(slender fibers and flat disks). Barring exceptional values of $\epsilon$, $De^{-1} \Delta C$ scales $O(\xi_{0}^{-2})$ in the near-sphere limit, with the sign of this contribution depending on whether one approaches the sphere from the prolate or the oblate side. $De^{-1} \Delta C$ is $O(1/\sqrt{\xi_0-1})$ in the slender-fiber and flat-disk limits. The divergence for $\xi_0 \rightarrow 1$ arises because the near-meridional nature of the Jeffery orbits, in the said limits, leads to them being squeezed together in an increasingly small neighborhood of the flow or gradient axis. As a result, a fixed $O(De)$ angular displacement in these neighborhoods amounts to an increasingly large drift when measured in units of $C$. The divergent scaling above already points to a breakdown of the weak-drift approximation when $\Delta C \sim De/\sqrt{\xi_0-1} \sim O(1)$; as already mentioned, this breakdown and the associated rotation arrest is examined in \S\ref{Rotation_arrest}. It is worth adding here that, for the inertial case, $\Delta C \sim Re.O(1/\sqrt{\xi_0-1}\ln (\xi_0-1))$ in the slender-fiber limit\,\citep{ganesh2005,vivek2016}, with the logarithmic smallness arising due to different asymptotic regions that contribute dominantly to fiber rotation in the inertial\,(outer region with $r \sim O(1)$ in units of the fiber length) and viscoelastic cases\,(matching region with $\sqrt{\xi_0-1} \ll r \ll 1$) - these aspects are clarified by the analysis in \S\ref{appB}. With the above scalings in mind, we replace $De^{-1} \Delta C /(C^2+1)$ by $De^{-1} \Delta C \xi_0^{\frac{3}{2}}\sqrt{\xi_0-1}/(C^2+1)$ in the orbital drift plots below.

While the generic $\xi_0$-scalings of $\Delta C$ have been clarified above, the exceptional scenarios where these scalings do not hold are also important, since they correspond to boundaries demarcating the different orientation dynamics regimes. To examine these instances, we consider the different limiting forms of $\Delta C$, as defined by (\ref{eq3p8}). In the near-sphere limit, one finds $\lim_{\xi_0\rightarrow \infty} De^{-1} \Delta C = \pm \pi C (8 \epsilon +3)/(14 \xi_{0}^{2})$, with the plus\,(minus) sign corresponding to a prolate\,(oblate) spheroid. This identifies a critical value of $\epsilon\,(=-3/8)$ at which $De^{-1} \Delta C \sim O(\xi_0^{-4})$ in the near-sphere limit, and that marks a bifurcation in the orientation dynamics for $\kappa \rightarrow 1^\pm$ - this bifurcation mediates a transition from Regime $2$ to $1$\,($1$ to $2$) as $\epsilon$ decreases below $-3/8$ on the prolate\,(oblate) side. In the slender-fiber limit, one finds $\lim_{\xi_0\rightarrow 1} De^{-1} \Delta C = \frac{\pi  C (2 \epsilon +1)}{2 \sqrt{2} \sqrt{\xi_{0}-1}}$, implying that the point $(\kappa,\epsilon) \equiv (\infty,-1/2)$ now marks a bifurcation in the orientation dynamics, with $\lim_{\xi_0\rightarrow 1} De^{-1} \Delta C \sim O(1/\sqrt{\xi_0-1}\ln (\xi_0-1))$ at $\epsilon=-1/2$. Inferring the change in the orientation dynamics regime in the slender-fiber limit is more subtle, however, since the limiting expression for $\Delta C$ above is not uniformly valid. A separate asymptote exists for small $C$, and is given by $De^{-1} \Delta C = -\sqrt{2}\pi C (1+\epsilon) \sqrt{\xi_0-1}$. For $\epsilon > -1/2$, this small-$C$ asymptote has a sign\,(negative) opposite to the asymptote for $C \sim O(1)$ above, implying a zero crossing with a positive slope in between, with a positive slope, and therefore, an unstable intermediate limit cycle. For $-1 < \epsilon < -1/2$, both asymptotes have the same (negative)\,sign, and the spinning mode alone is stable. Finally, for $\epsilon < -1$, the small-$C$ asymptote reverses sign, and one must then have a stable intermediate limit cycle. Thus, in the limit $\kappa \rightarrow \infty$, one must transition from Regime $4$, through Regime $1$, to Regime $3$, with decreasing $\epsilon$. In the flat-disk limit, one finds $\lim_{\xi_0\rightarrow 1} De^{-1} \Delta C \sim -\frac{\pi  \left[2 \left(\sqrt{C^2+1}-1\right) (6 \epsilon +1)+C^2\right]}{6 \sqrt{2} C \sqrt{\xi_0-1}}$ which is uniformly valid for all $C$. For $C \rightarrow 0$, this asymptote reduces to $-\frac{\pi}{\sqrt{2}\sqrt{\xi_0-1}}[C(\epsilon+1/3) -\frac{C^3}{4}(\epsilon + 1/6)]+ O(C^5)$, which shows that the point $(\kappa,\epsilon) \equiv (0,-1/3)$ corresponds to a bifurcation in the orientation dynamics. For $\epsilon > -1/3$, the spinning mode alone is stable, while for $\epsilon$ less than but close to $-1/3$, one obtains $\Delta C = 0$ for $C^* \approx 2[-6(\epsilon+1/3)]^{\frac{1}{2}}$ from the small-$C$ asymptote, which corresponds to a stable limit cycle emerging from the vorticity axis; one may, in fact, obtain the location of the stable limit cycle for arbitrary $\epsilon$, $C^* = 2\sqrt{2}[(3 \epsilon +1) (6 \epsilon +1)]^{\frac{1}{2}}$.  using the aforementioned flat-disk asymptote. Thus, in the limit $\kappa \rightarrow 0$, one must transition from Regime $1$ to Regime $3$ as $\epsilon$ decreases across $-1/3$. Consideration of the next term in the flat-disk asymptote for $\Delta C$ shows that, for any small but finite $\kappa$, there must be a second transition from Regime $3$ to Regime $2$ across a large but negative $\epsilon$ of $O(1/\sqrt{\xi_0-1})$.

The above analysis helps identify the limiting forms of the bifurcation loci, that separate the different orientation-dynamics regimes on the $\kappa-\epsilon$ plane. A pair of loci on the prolate side must asymptote to the vertical lines $\epsilon = -1$ and $\epsilon = -1/2$ in the limit $\kappa \rightarrow \infty$. On the oblate side, one of the loci must asymptote to the line $\epsilon = -1/3$, with the other assuming the limiting form $\epsilon \propto -\kappa^{-1}$, in the limit $\kappa \rightarrow 0$. Further, a critical point $(\kappa,\epsilon) \equiv (1,-3/8)$ on the sphere-line is expected to act as an anchor point for one or more of the aforementioned curves, and control the bifurcations in the near-sphere limit. An additional benefit of the arguments above is a broad identification of $\epsilon$-intervals corresponding to distinct bifurcation sequences for the spheroid orientation dynamics. These intervals are $-\infty < \epsilon < -1$, $-1 <\epsilon < -1/2$ and $-1/2 < \epsilon < \infty$ on the prolate side. Neglecting the locus that diverges to infinity for $\kappa \rightarrow 0$, the intervals are $-\infty < \epsilon < -1/3$ and $-1/3 < \epsilon < \infty$ on the oblate side. Figures \ref{fig6} and \ref{fig7} below show the $De^{-1} \Delta C\xi_0^{\frac{3}{2}}\sqrt{\xi_0-1}/(C^2+1)$ vs $C/(C+1)$ plots for $\epsilon$ values in each of the aforesaid intervals.

\begin{figure}
		\centering
          \subfigure[]{\includegraphics[scale=0.242]{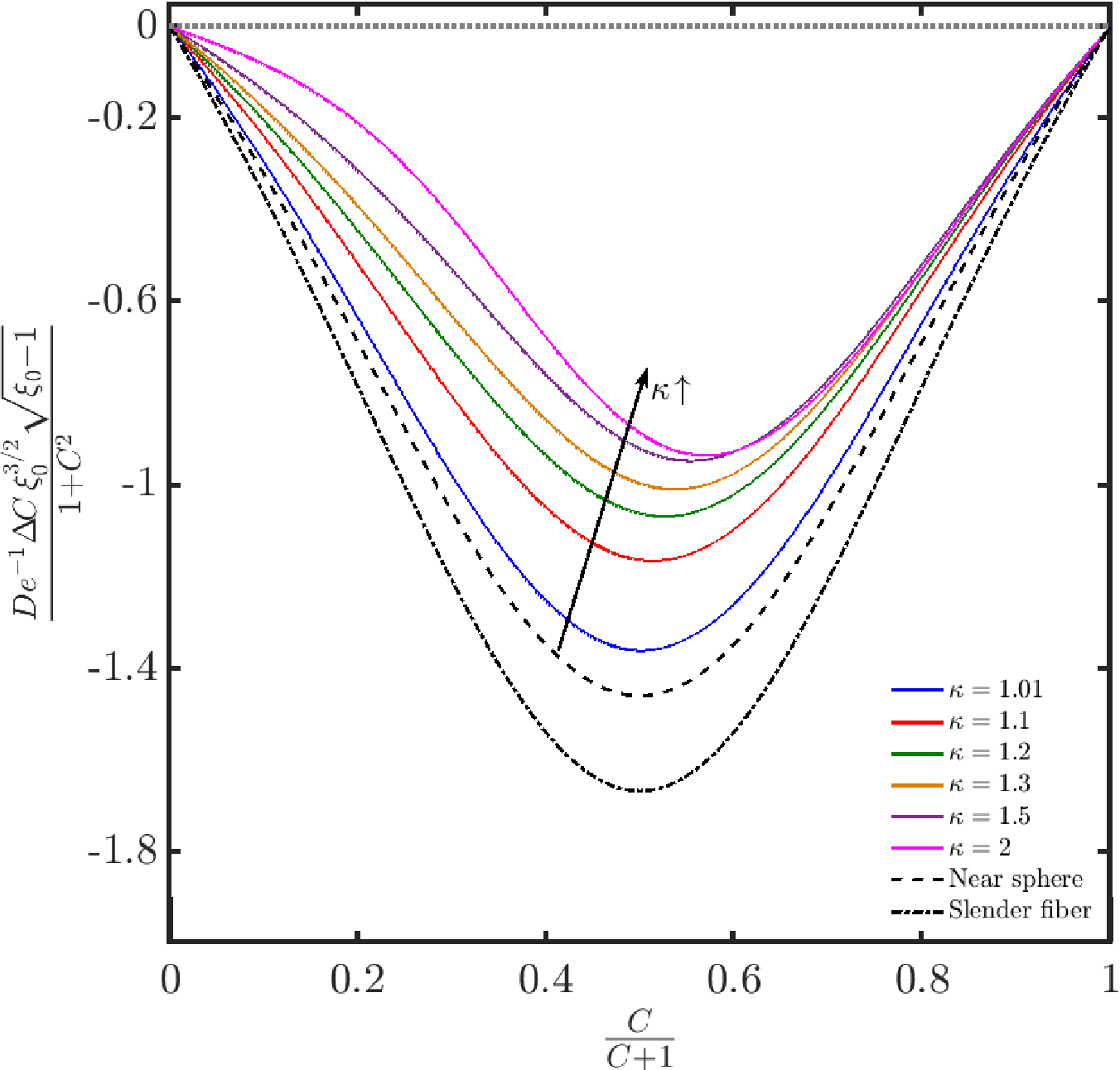}\label{fig6a}}
		\subfigure[]{\includegraphics[scale=0.28]{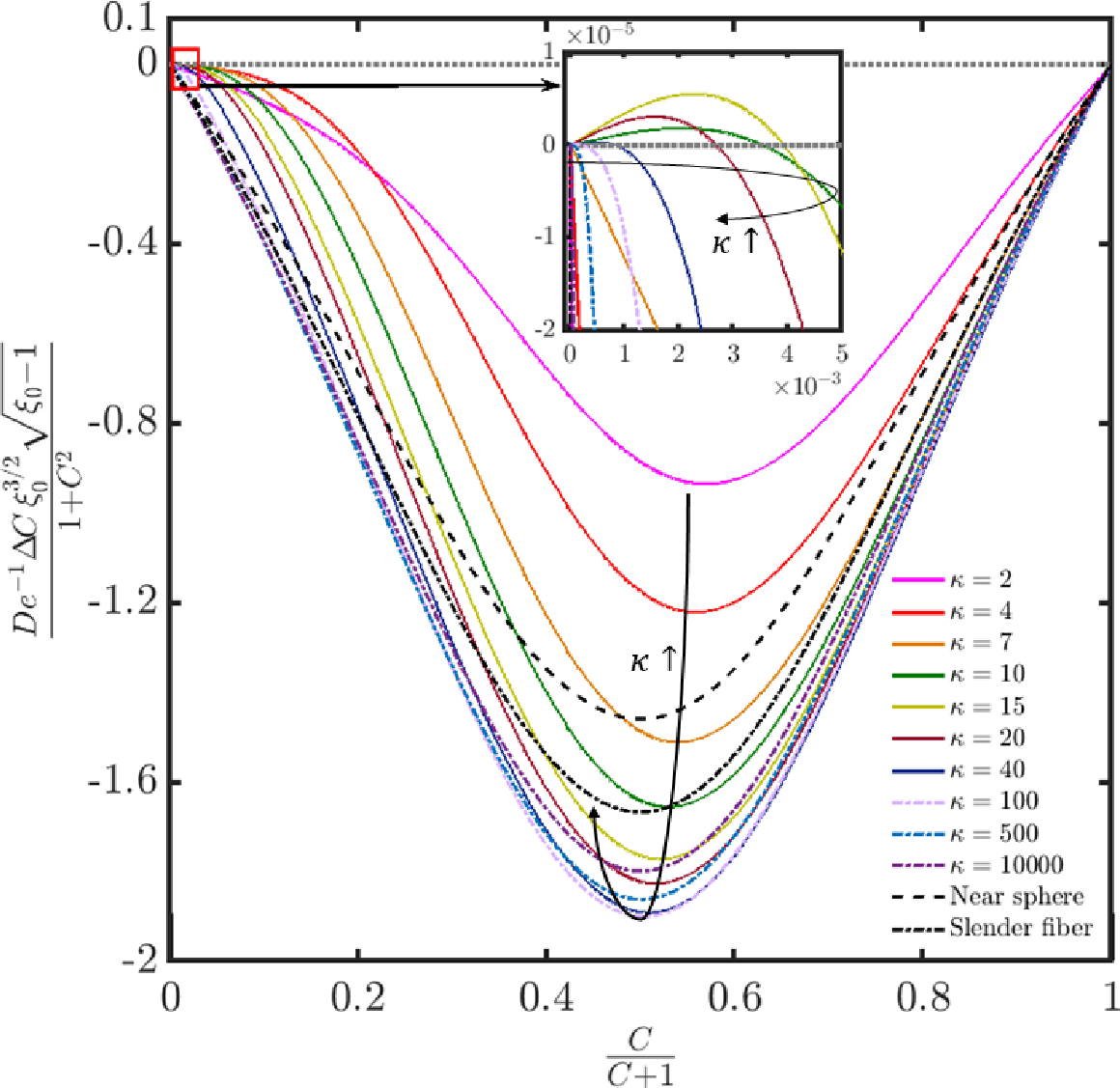}\label{fig6b}}
        \subfigure[]{\includegraphics[scale=0.26]{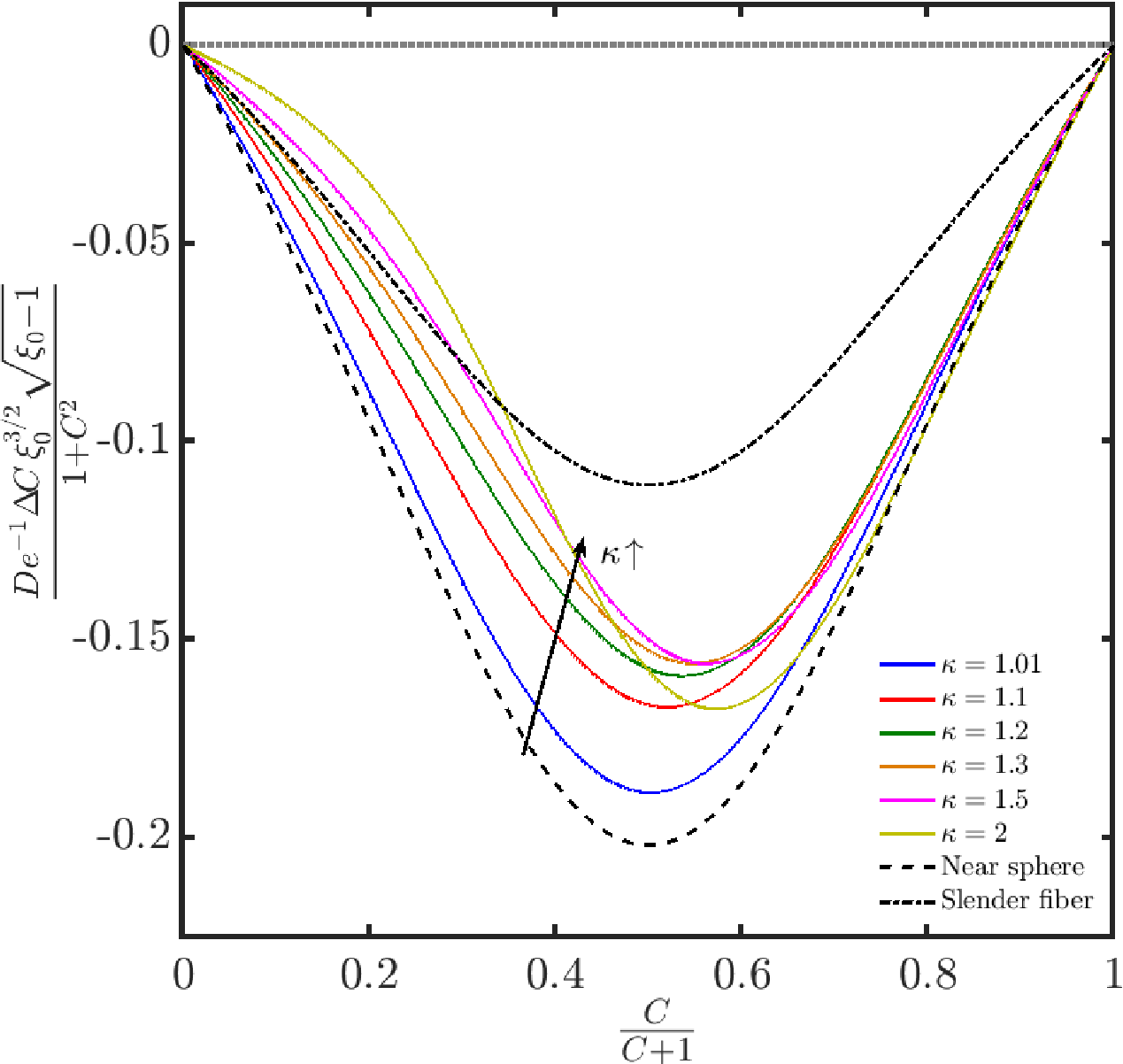}\label{fig6c}  }
		\subfigure[]{\includegraphics[scale=0.28]{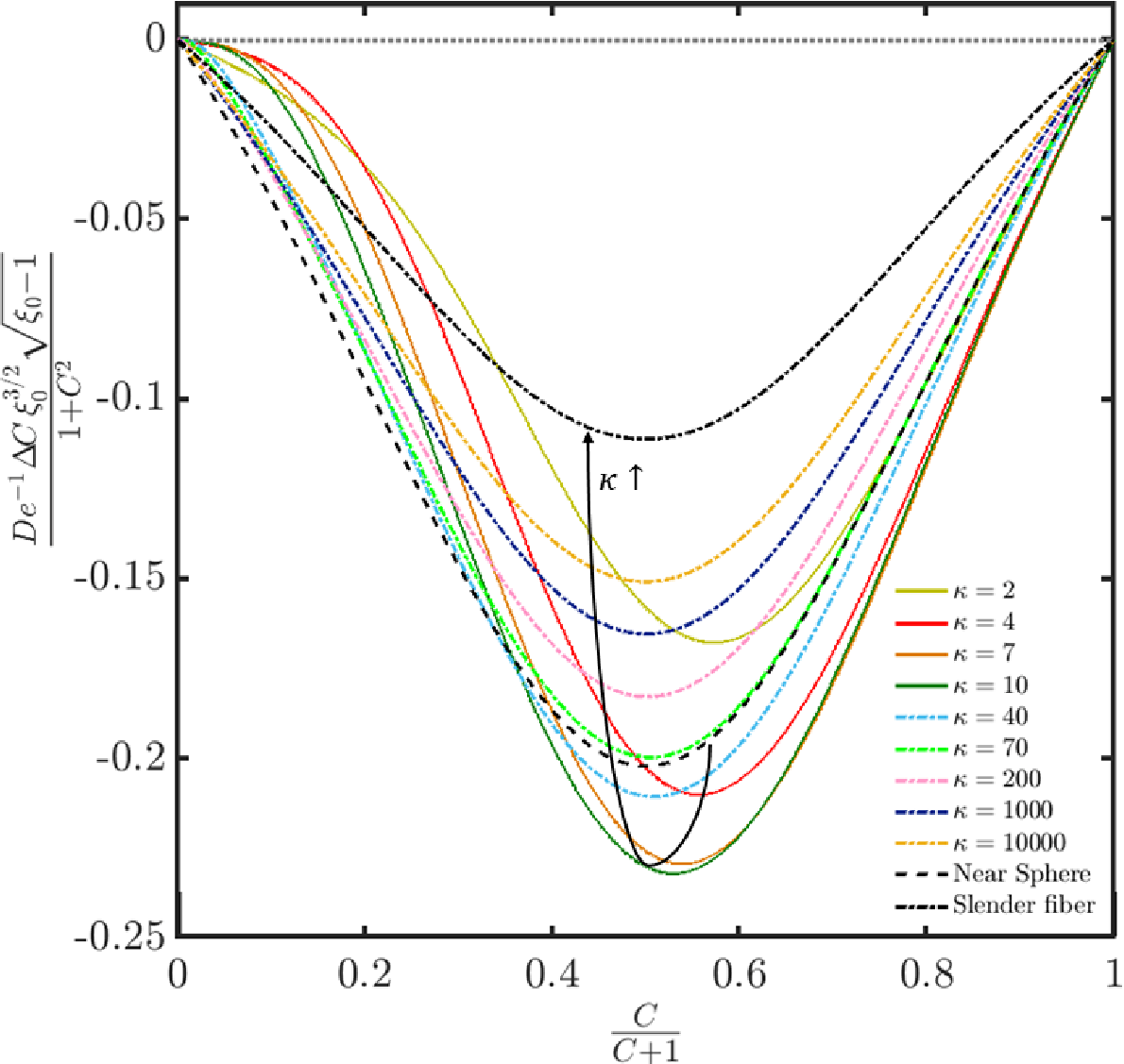}\label{fig6d}}
        \subfigure[]{\includegraphics[scale=0.242]{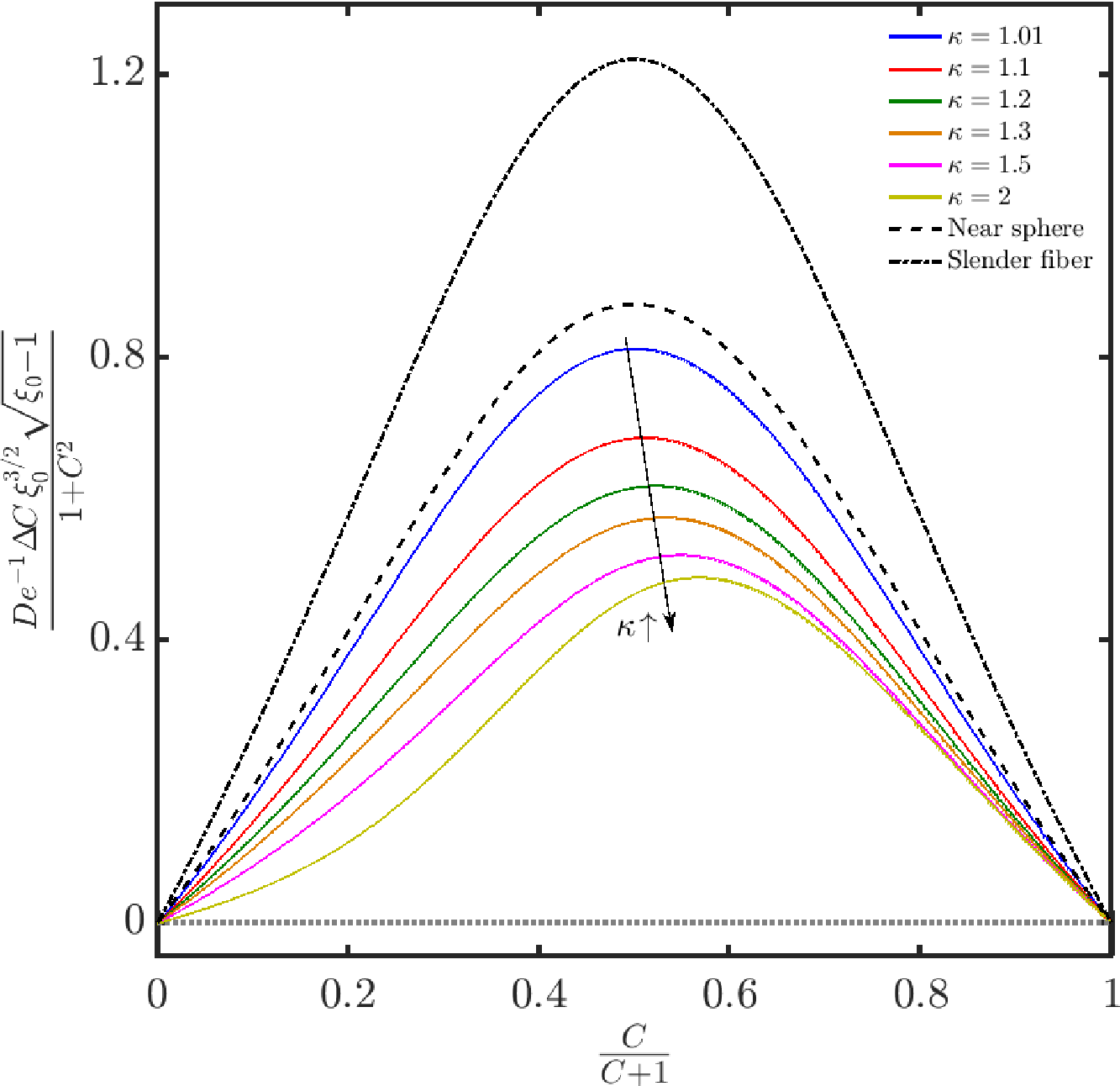}\label{fig6e}}
          \subfigure[]{\includegraphics[scale=0.28]{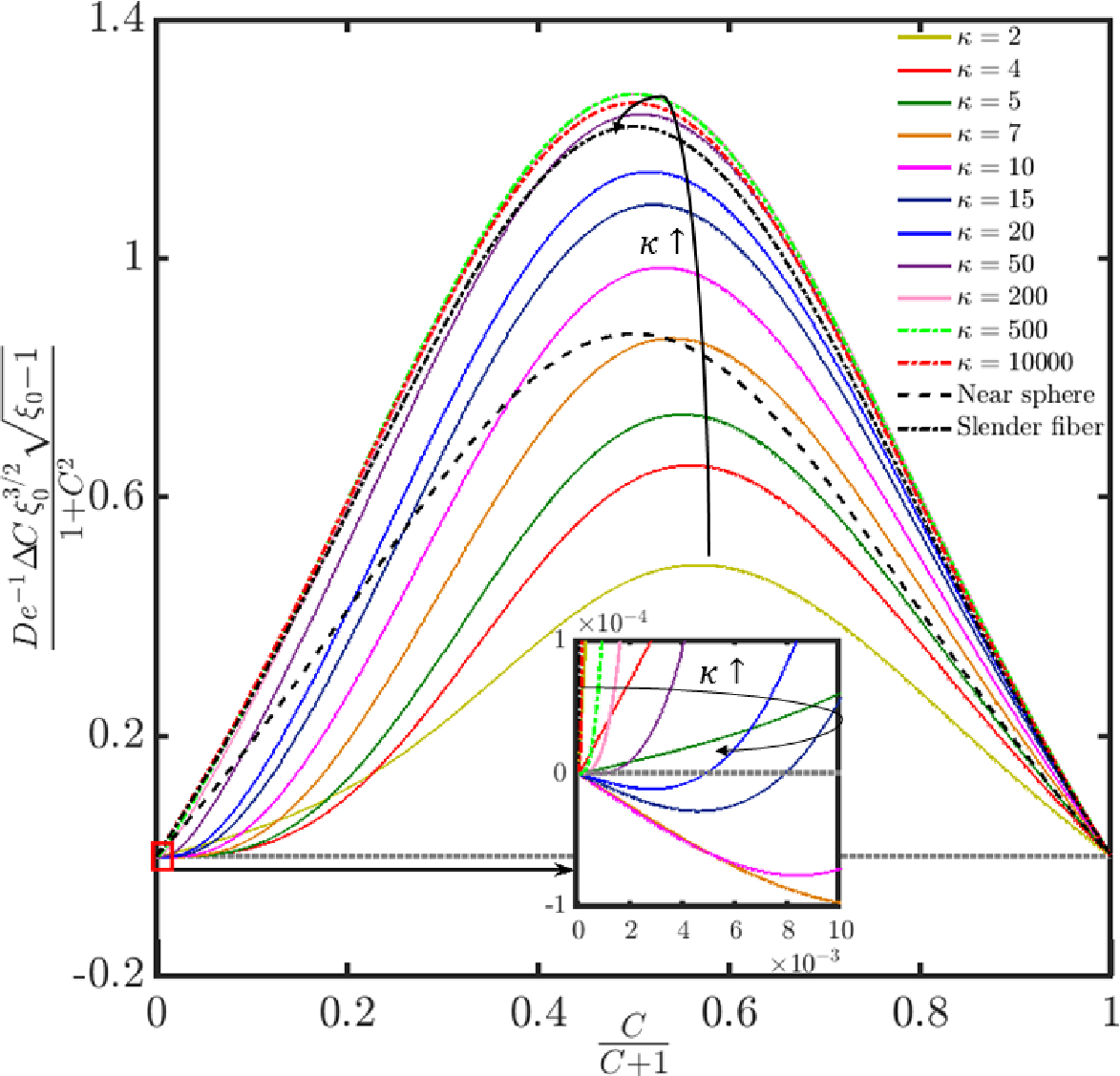}\label{fig6f}}
        \caption{The scaled orbital drift, $De^{-1} \Delta C \xi_{0}^{3/2}\sqrt{\xi_{0}-1}/(C^2+1)$, plotted as a function of $C/(C + 1)$, for prolate spheroids of a varying aspect-ratio $\kappa$ with $\epsilon$ fixed; the black dashed and dash-dotted curves denote the near-sphere and slender-fiber asymptotes, respectively. (a)-(b) $\epsilon=-2$, (c)-(d) $\epsilon=-0.6$, and (e)-(f) $\epsilon=0.6$. The inset views in (b) and (f) highlight the emergence of a zero-crossing from the spinning end\,($C=0$) above a threshold $\kappa$; the zero-crossing corresponds to a stable and unstable limit cycle in the respective cases. Black arrows, either straight or curved, indicate the movement of the drift curves with increasing $\kappa$.
        }\label{fig6}
\end{figure}
Figures \ref{fig6a}-\ref{fig6f} illustrate the orbital drift curves for a prolate spheroid, over the entire range of aspect ratios, for $\epsilon = -2$\,(figures \ref{fig6a} and \ref{fig6b}), $-0.6$\,(figures \ref{fig6c} and \ref{fig6d}) and $0.6$\,(figures \ref{fig6e} and \ref{fig6f}); note that $\epsilon = -0.6$ falls in the range typical of polymer solutions. The black dashed and dash-dotted curves in each of these figures denote the limiting near-sphere and slender-fiber asymptotes, respectively. The drift curves in figure \ref{fig6a} conform to the stable spinning regime. The scaled orbital drift decreases in magnitude starting from the near-sphere asymptote, attaining its minimal limit for $\kappa \approx 2$. The movement of the drift curve reverses thereafter. As shown in figure \ref{fig6b}, the drift increases in magnitude for $\kappa$ larger than $2$, overshooting the slender-fiber asymptote for $\kappa \gtrsim 15$, before turning around and converging back to this asymptote in the limit $\kappa \rightarrow \infty$. The convergence is only logarithmic, with the difference between the asymptote and the large-$\kappa$ curves decreasing as $O(\ln \kappa)^{-1}$, as also evident from the perceptible difference between the two for $\kappa = 10^4$. The inset in figure \ref{fig6b} shows a magnified view of the region near $C = 0$, highlighting the emergence of a negative-slope zero-crossing from $C = 0$ at $\kappa \approx 9.324$. The zero-crossing moves towards larger $C$ with increasing $\kappa$, but turns around thereafter, heading back towards $C = 0$, and eventually coalescing with it in the limit $\kappa = \infty$. Thus, for $\epsilon = -2$, a prolate spheroid transitions from Regime $1$ to Regime $3$ with increasing $\kappa$. This transition is marked by a stable limit cycle emerging from the vorticity axis at $\kappa \approx 9.324$, attaining a maximum size at $\kappa \approx 1000$, and collapsing back onto the vorticity axis for $\kappa \rightarrow \infty$.

Figures \ref{fig6c} and \ref{fig6d} depict the orbital drift for $\epsilon=-0.6$. In this case, the prolate spheroid drifts to the spinning mode regardless of $\kappa$, with there being no bifurcations. Interestingly, the movement of the scaled drift-curve from the near-sphere to the slender-fiber asymptote continues to have a non-monotonic character. With increasing $\kappa$, the drift first decreases away from the near-sphere asymptote until $\kappa \approx 2$, then increases and overshoots the near-sphere asymptote for $\kappa \approx 10$, before turning around and converging at a logarithmic rate towards the slender-fiber asymptote. Figures \ref{fig6e} and \ref{fig6f} illustrate the drift of a prolate spheroid for $\epsilon=0.6$. The drift curves are now positive for all $C$ to begin with. Mirroring the scenario for $\epsilon = -2$, the drift curve in figure \ref{fig6e} decreases relative to the near-sphere asymptote, attaining a minimum for $\kappa \approx 2$. It then increases, again overshooting the slender-fiber asymptote, before converging back to it. The inset view in figure \ref{fig6f} shows the emergence of a positive-slope zero-crossing\,(an unstable limit cycle) from $C = 0$, at $\kappa \approx 5.264$, that eventually collapses back into $C = 0$ for $\kappa \rightarrow \infty$. Thus, a prolate spheroid, for $\epsilon = 0.6$, transitions from Regime $2$ to Regime $4$ with increasing $\kappa$. 

\begin{figure}
		\centering
	\subfigure[]{\includegraphics[scale=0.31]{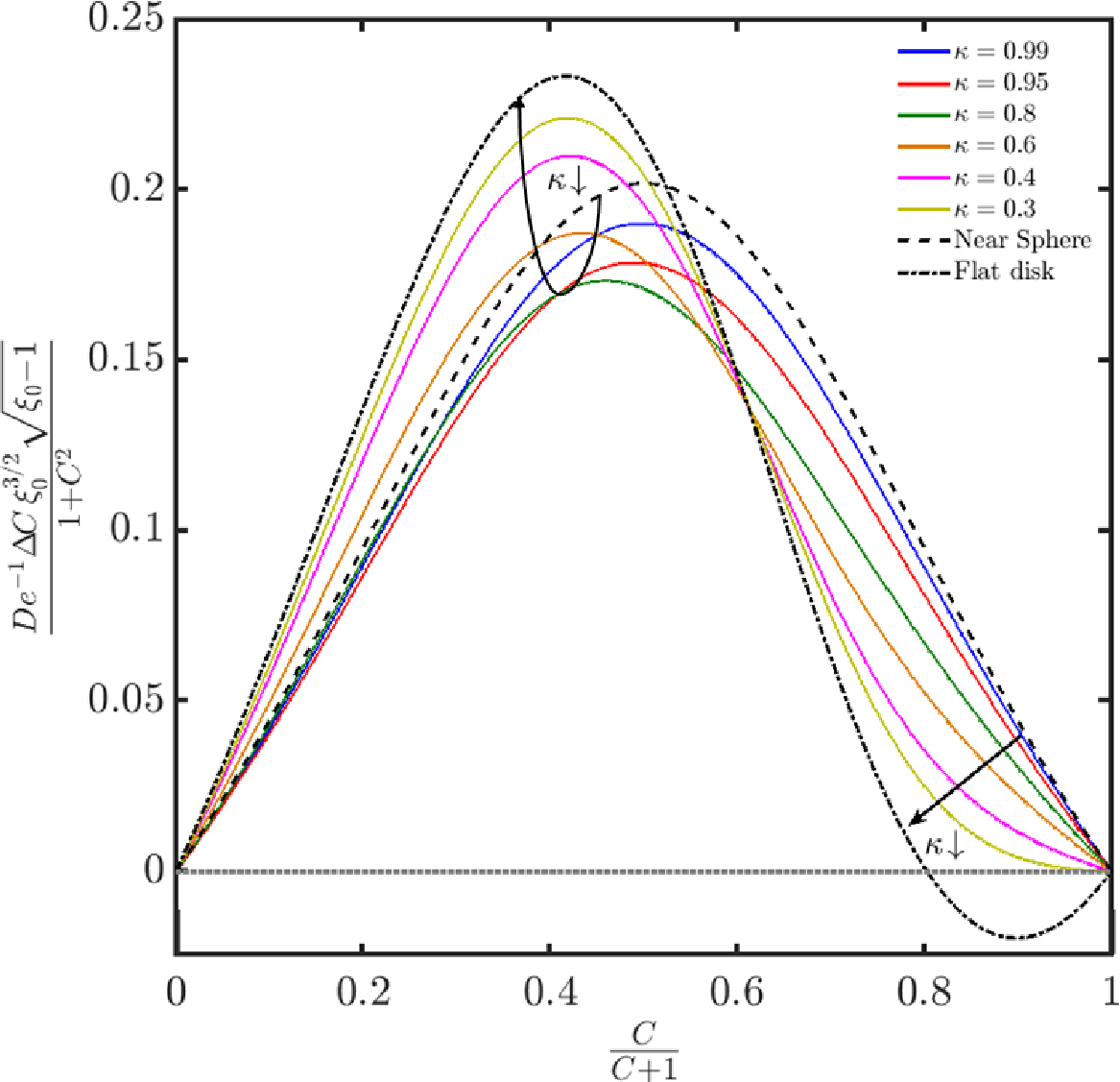}\label{fig7a}}
    \subfigure[]{\includegraphics[scale=0.27]{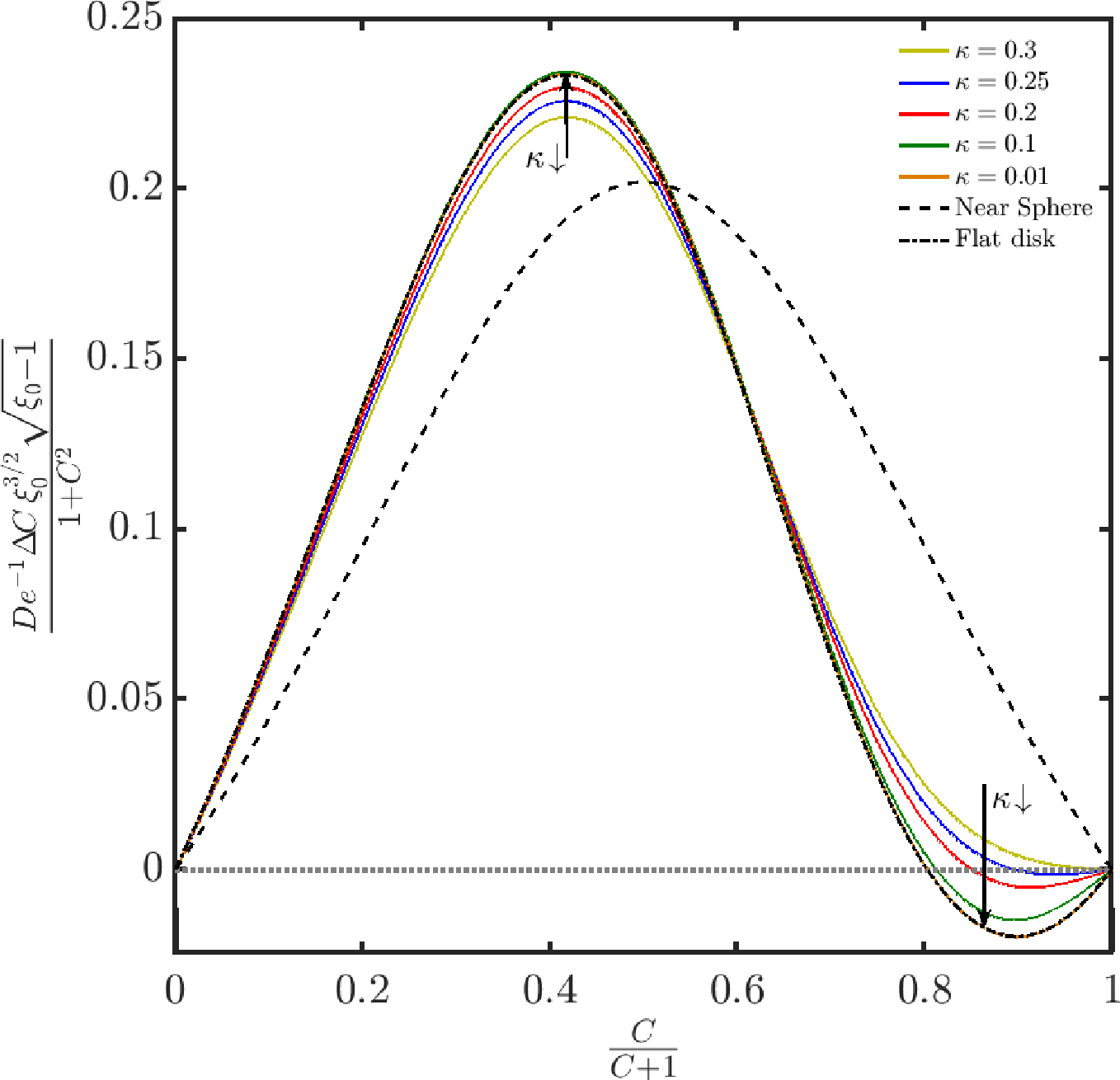}\label{fig7b}}
    \subfigure[]{\includegraphics[scale=0.27]{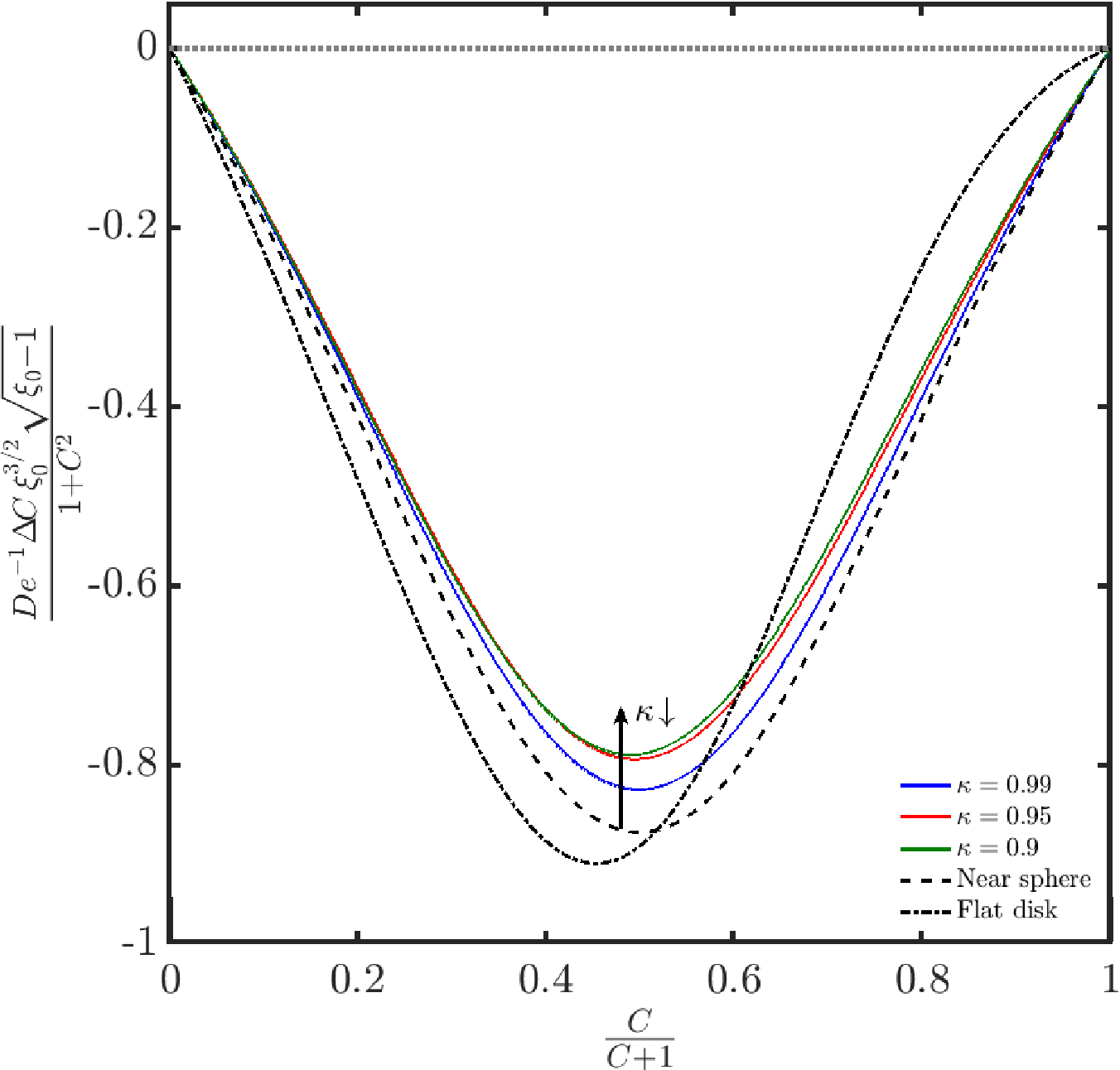}\label{fig7c}  }
     \subfigure[]{\includegraphics[scale=0.31]{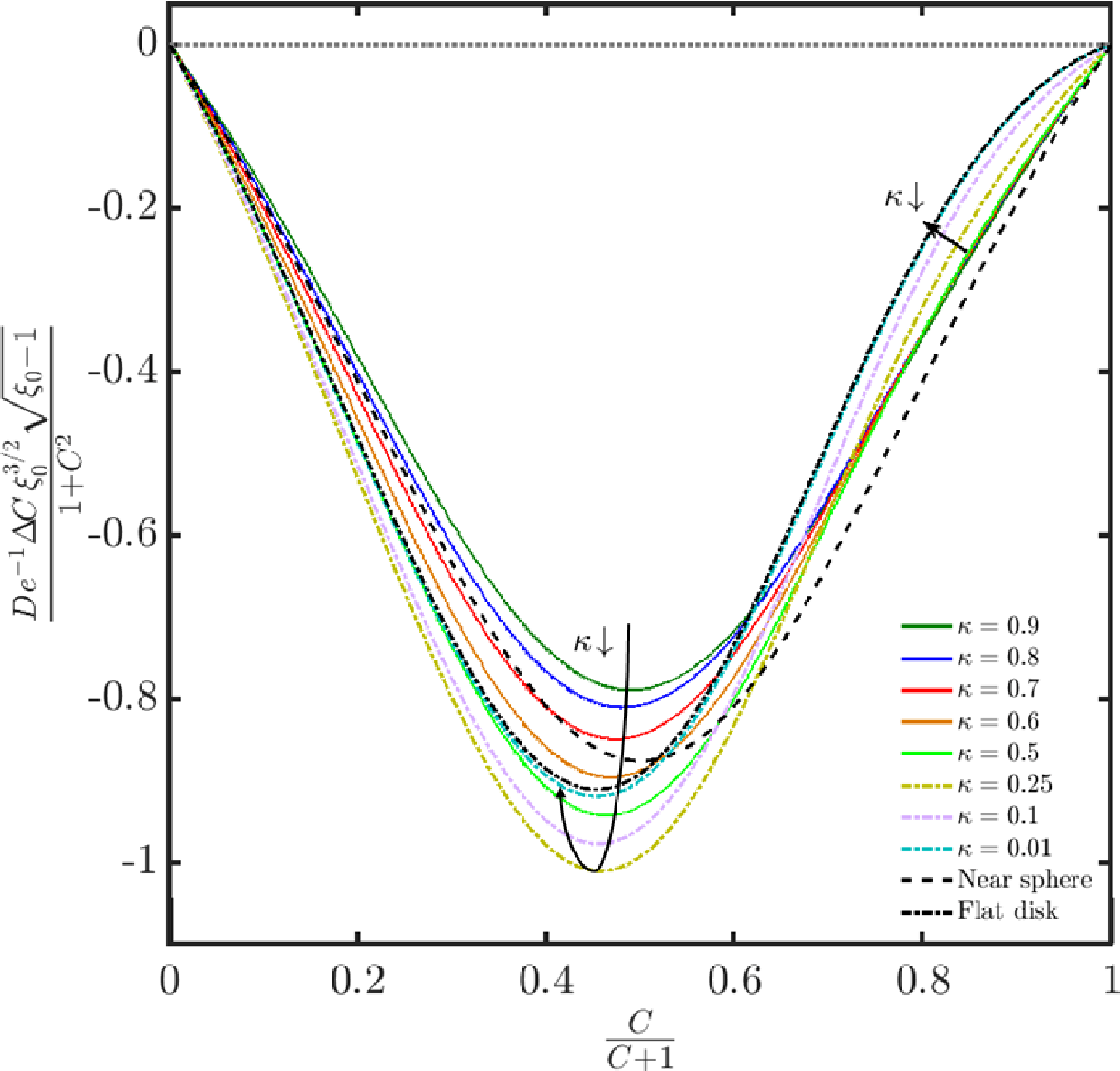}\label{fig7d}}
  	\caption{The scaled orbital drift, $De^{-1} \Delta C \xi_{0}^{3/2}\sqrt{\xi_{0}-1}/(C^2+1)$, plotted as a function of $C/(C + 1)$, for oblate spheroids of a varying aspect-ratio $\kappa$ with $\epsilon$ fixed. The black dashed and dash-dotted curves denote the near-sphere and flat-disk asymptotes, respectively; the latter exhibits a zero-crossing at $C^* = 2\sqrt{2}[(3 \epsilon +1) (6 \epsilon +1)]^{\frac{1}{2}}$ for $\epsilon < -1/3$. (a)-(b) $\epsilon=-0.6$ and (c)-(d) $\epsilon=0.6$. Black arrows, either straight or curved, indicate the movement of the drift curves with decreasing $\kappa$.}\label{fig7}
\end{figure}
Figures \ref{fig7a}-\ref{fig7d} illustrate the orbital drift curves for an oblate spheroid, over the entire range of aspect ratios, for $\epsilon = -0.6$\,(figures \ref{fig7a} and \ref{fig7b}) and $0.6$\,(figures \ref{fig7c} and \ref{fig7d}). In figure \ref{fig7a}, the drift curves conform to the stable tumbling regime to begin with. The portion of the drift curve for $C/(C+1) \lesssim 0.6$ again exhibits a non-monotonic movement, decreasing from the near-sphere asymptote to begin with, but heading back up towards the flat-disk asymptote for $\kappa \lesssim 0.8$. While the movement of the drift curve from the near-sphere to the flat-disk asymptote has a monotonic character for larger $C/(C+1)$, there is nevertheless a bifurcation at $\kappa \approx 0.28$ due to the emergence of a negative-slope zero-crossing from the tumbling end\,($C= \infty$), and that asymptotes to a finite $C\,(=C^*)$ in the flat-disk limit. The expression for $C^*$, for $\kappa \rightarrow 0$, was obtained above, and equals $4$ for the chosen $\epsilon$. Thus, an oblate spheroid, for $\epsilon = -0.6$, transitions from Regime $2$ to Regime $3$ with decreasing $\kappa$. Figures \ref{fig7c} and \ref{fig7d} depict the drift curves for $\epsilon=0.6$. While all of the drift curves conform to the stable spinning regime in this case, the movement with decreasing $\kappa$ has a non-monotonic character, with the drift attaining a maximum at $\kappa \approx 0.25$.

\begin{figure}
	\centering
	 \includegraphics[scale=0.3]{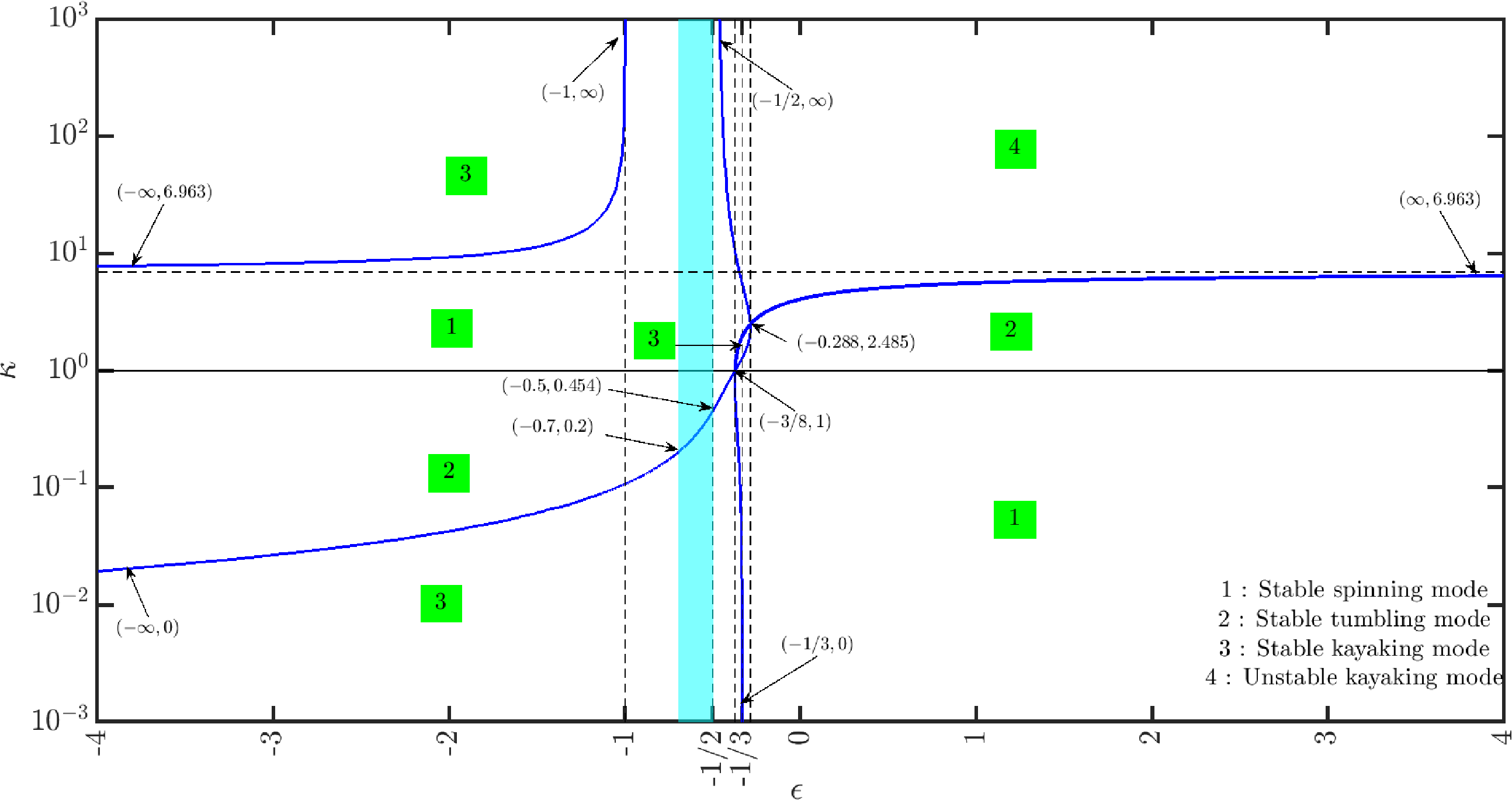}
	\caption{The orientation dynamics regimes on the $\kappa$-$\epsilon$ plane, for small but finite $De$; the four different regimes are numbered as follows: $1$ - stable spinning; $2$ - stable tumbling; $3$ - stable kayaking; $4$ -unstable kayaking sandwiched between stable spinning and tumbling modes. The horizontal black line is $\kappa=1$, with regions above and below corresponding to prolate and oblate spheroids, respectively; the horizontal dashed line, $\kappa = 6.963$, is the infinite-$\epsilon$ asymptote. The cyan vertical band denotes the range $\epsilon \in (-0.7,-0.5)$ relevant to polymer solutions. Vertical dashed lines serve to demarcate the $\epsilon$-intervals corresponding to distinct bifurcation sequences for the orientation dynamics, as a function of $\kappa$.}\label{fig_kappa_epsilon} 
\end{figure}

\begin{figure}
		\centering
         \subfigure[]{\includegraphics[scale=0.245]{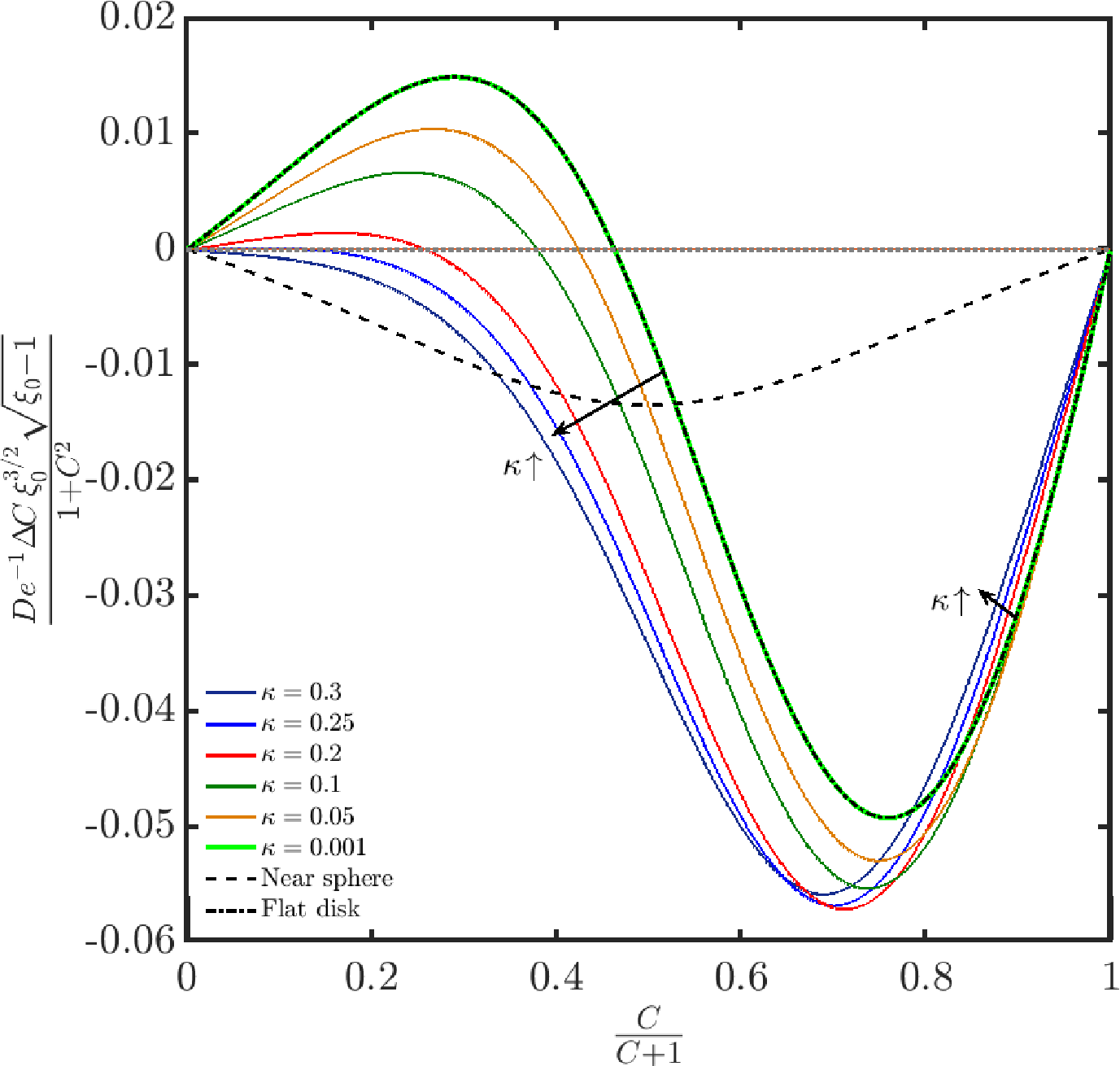}\label{fig9a}}
         \subfigure[]{\includegraphics[scale=0.245]{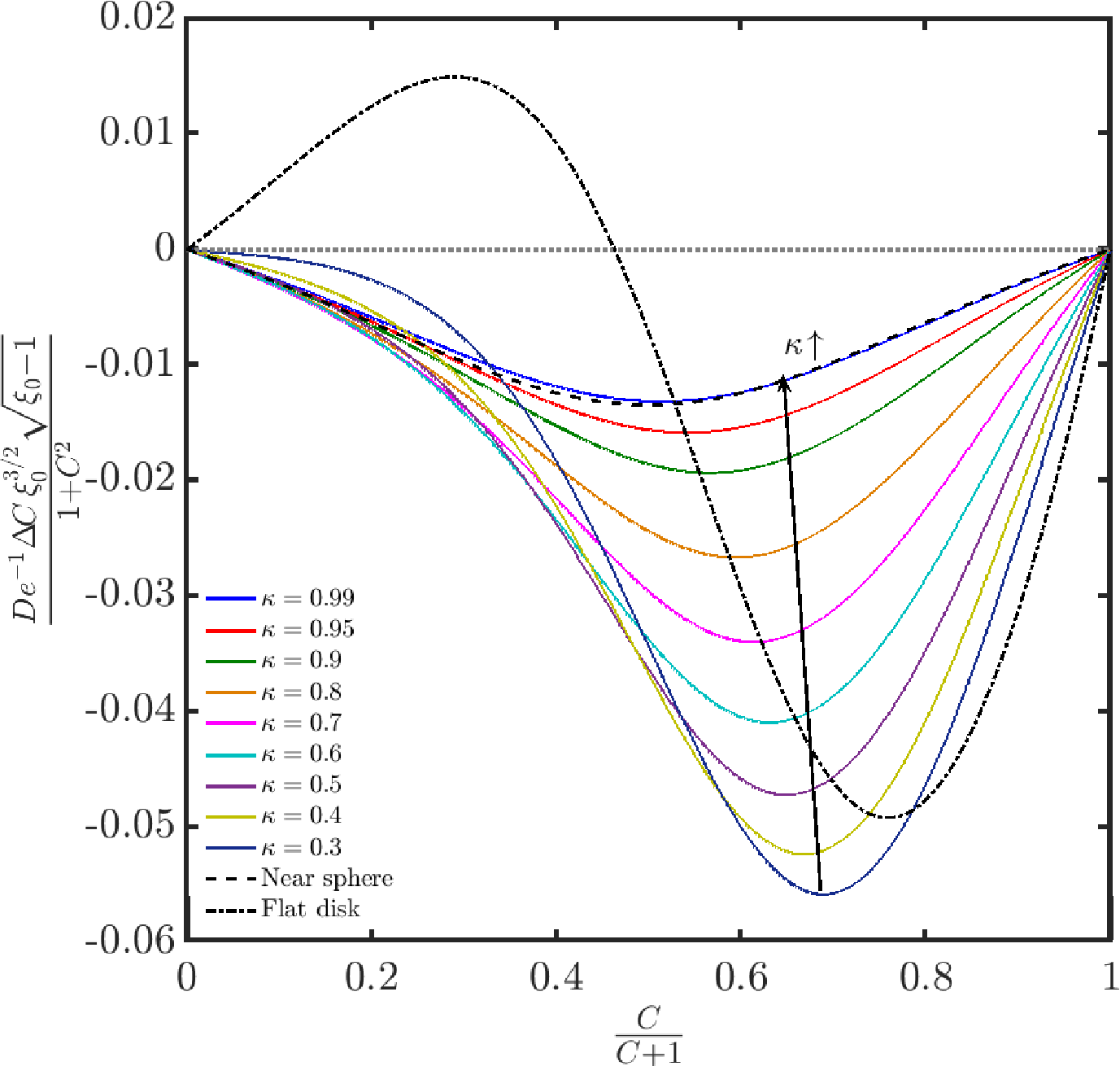}\label{fig9b}}
         \subfigure[]{\includegraphics[scale=0.245]{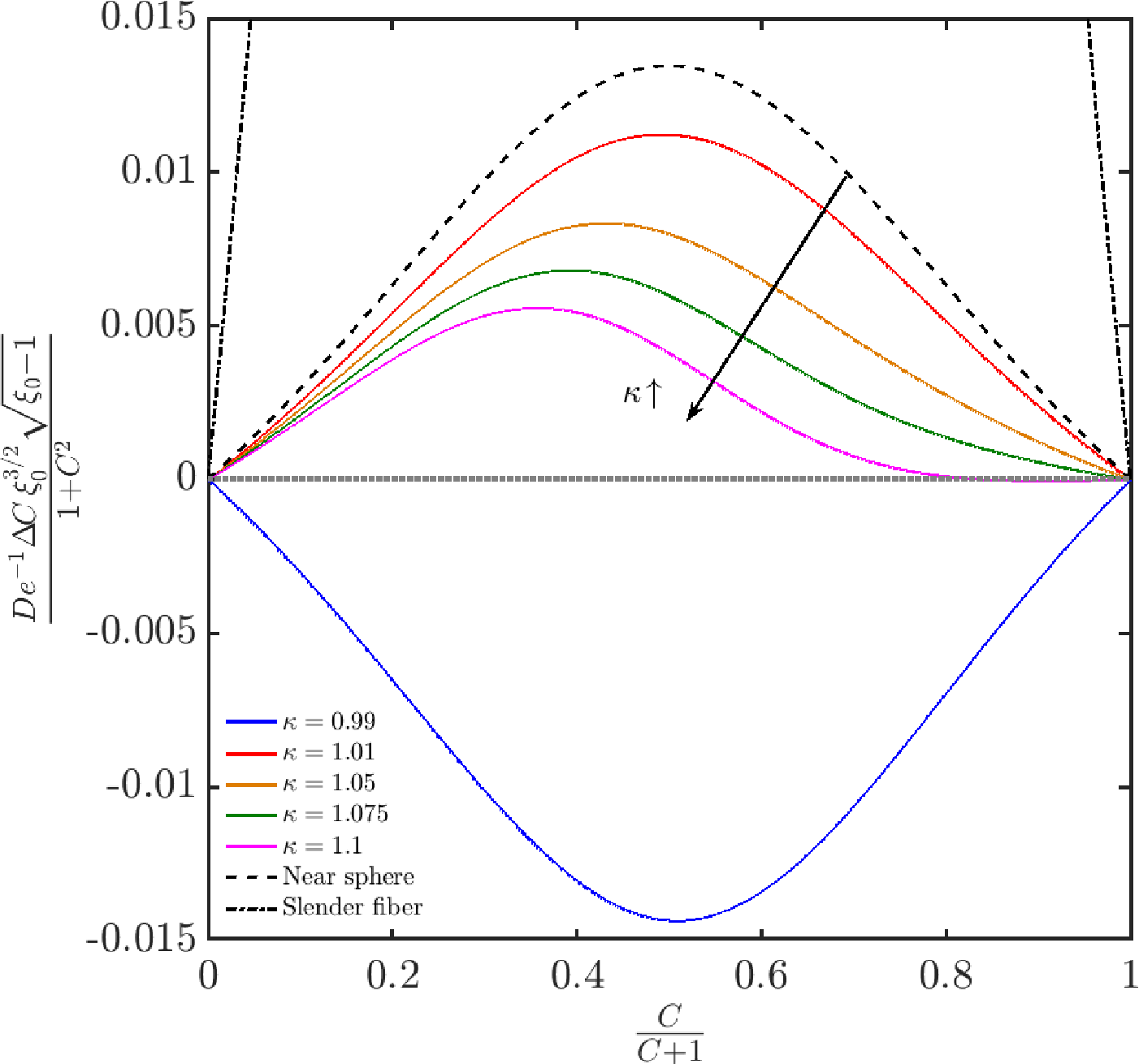}\label{fig9c}}
         \subfigure[]{\includegraphics[scale=0.24]{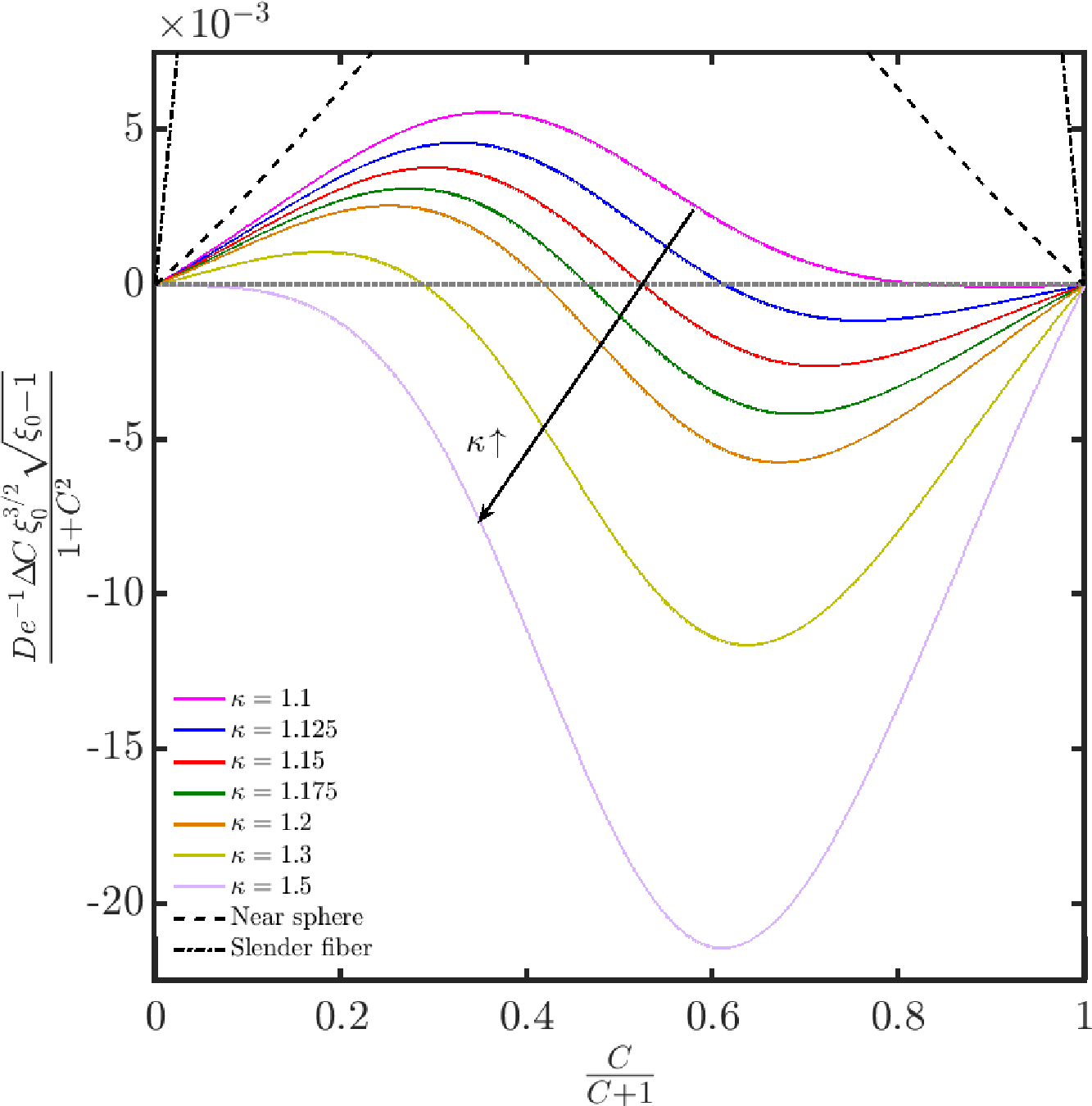}\label{fig9d}}
         \subfigure[]{\includegraphics[scale=0.28]{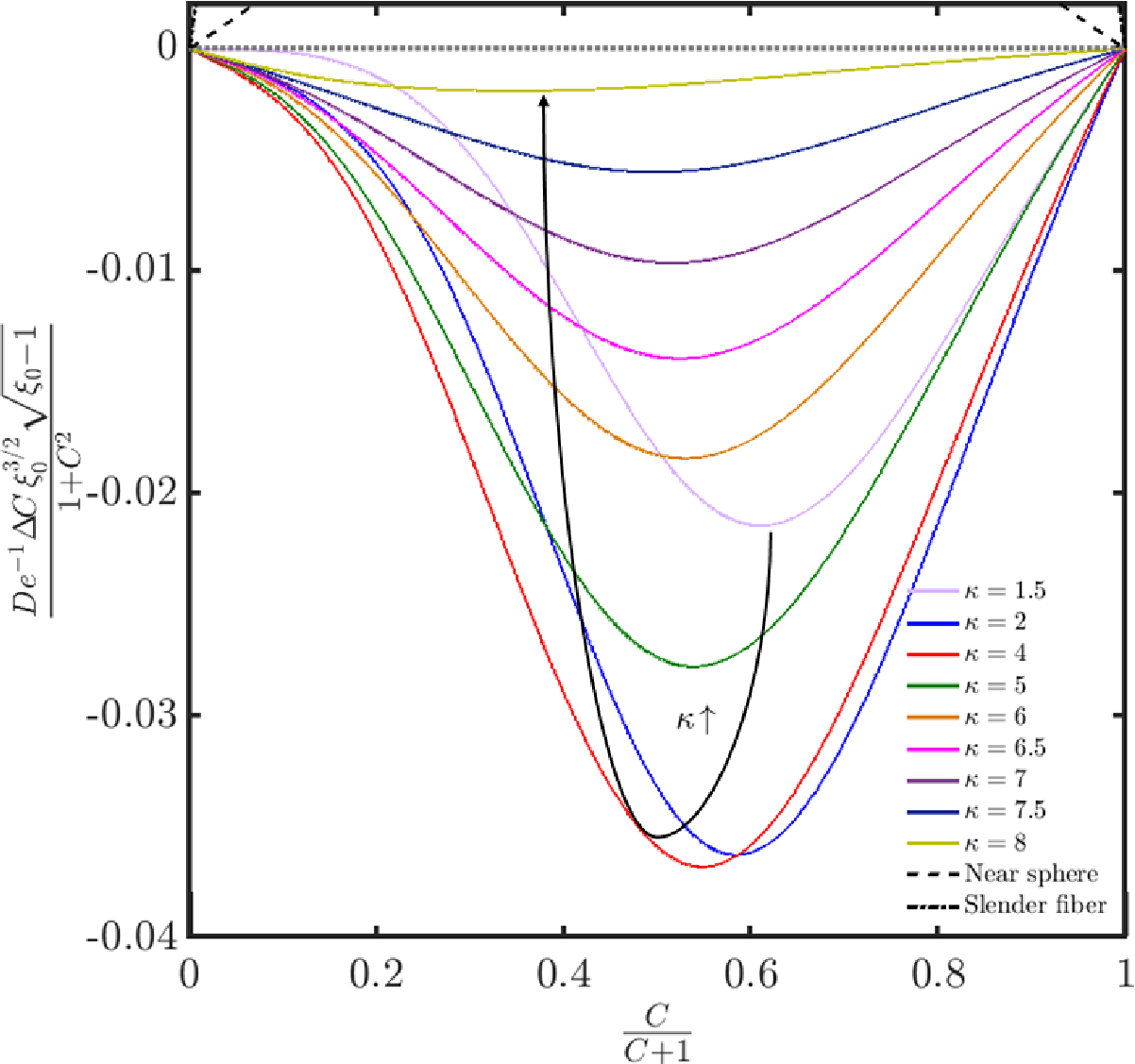}\label{fig9e}}
         \subfigure[]{\includegraphics[scale=0.245]{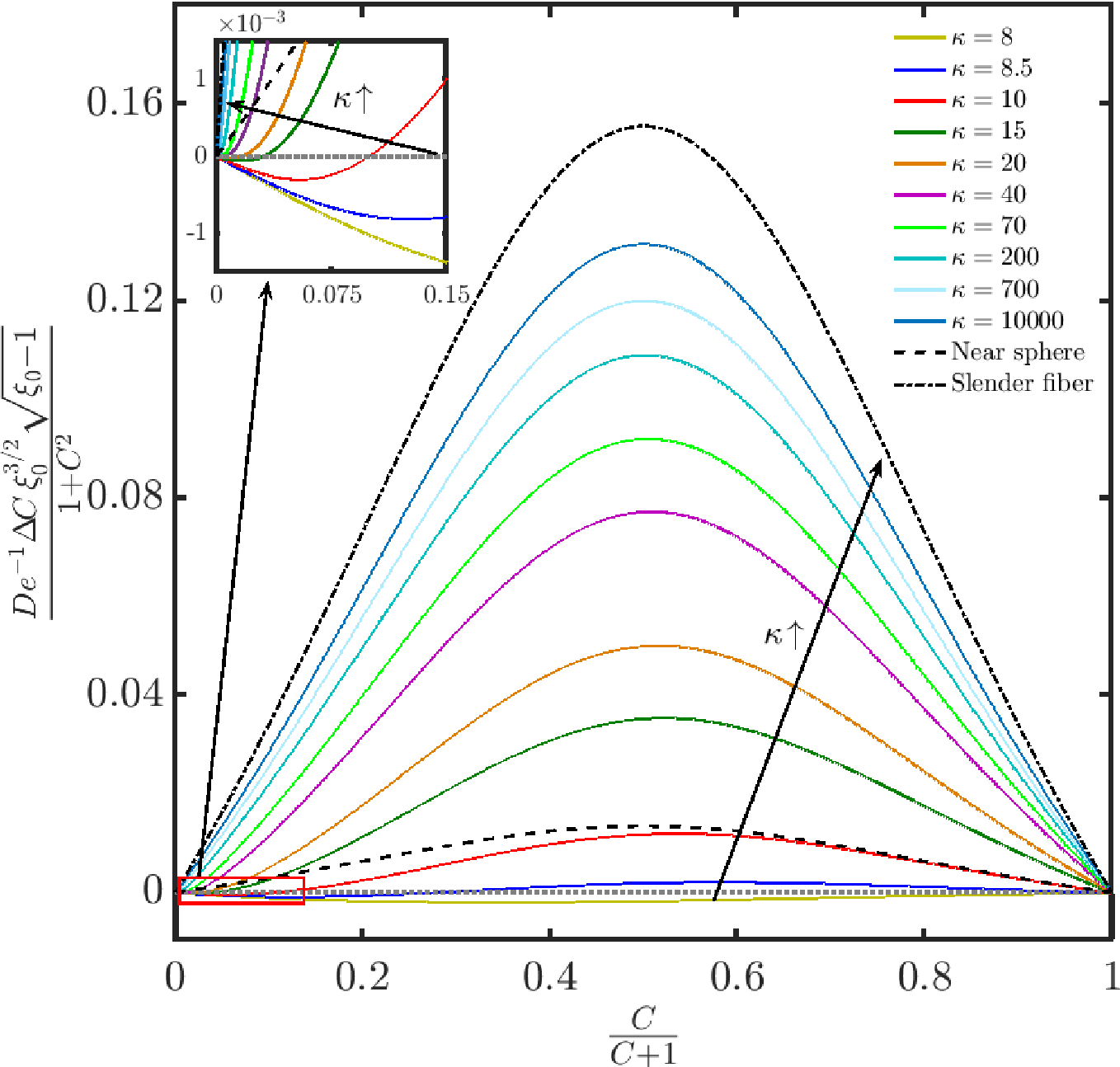}\label{fig9f}}
        \caption{The scaled orbital drift, $De^{-1} \Delta C \xi_{0}^{3/2}\sqrt{\xi_{0}-1}/(C^2+1)$, plotted as a function of $C/(C + 1)$, for spheroids with $\kappa$ increasing from zero to infinity; $\epsilon=-0.36$. Figures (a) and (b) pertain to oblate spheroids, and (d)-(f) to prolate spheroids; (c) shows the transition of the drift curve from the oblate to the prolate side. Black dashed and dash-dotted curves denote the near-sphere and flat-disk asymptotes in (a)-(b), and the near-sphere and slender-fiber asymptotes in (c)-(f). Black arrows in all figures denote the direction of movement of the drift curve with increasing $\kappa$.}\label{fig9} 
\end{figure}

Having characterized the orientation dynamics bifurcations as a function of varying $\kappa$, for specific $\epsilon$ values, figure \ref{fig_kappa_epsilon} shows the different small-$De$ orientation dynamics regimes on the $\kappa-\epsilon$ plane, for an arbitrary aspect ratio spheroid, in an ambient simple shear flow. It is worth reiterating that $De$ here is not fixed; instead, in order for the orbital-drift interpretation to remain valid, $De$ is constrained to be increasingly small for $\kappa \rightarrow 0$\,($De \ll \kappa$) and $\infty$\,($De \ll \kappa^{-1}$). The boundaries, that demarcate the different regimes in figure \ref{fig_kappa_epsilon}, have a more complicated form than that inferred using arguments based on the limiting forms of $\Delta C$ above. One of the prolate-side boundaries starts from its large-$\kappa$ asymptotic form, $\epsilon = -1$, but turns around for finite $\kappa$, eventually heading to negative infinity parallel to the $\epsilon$-axis, with $\kappa \rightarrow 6.963$ in his limit. The second prolate-side boundary continues down to the sphere line\,($\kappa =1$), intersecting it at $\epsilon = -3/8$. It then continues further into the oblate side, eventually asymptoting to the curve $\epsilon = (2 \pi  \kappa^{-1})/\left(-1168-24 \pi +117 \pi ^2 \right)$ in the limit $\kappa \rightarrow 0$. The oblate-side boundary that starts off along $\epsilon = -1/3$ in the limit $\kappa \rightarrow 0$, continues up and intersects the sphere-line at $\epsilon = -3/8$. It then turns around on the prolate side, heading off to positive infinity parallel to the $\epsilon$-axis, with $\kappa$ again approaching $6.963$ in this limit. The full form of the equations governing the two bounding curves in figure \ref{fig_kappa_epsilon}, in the form $\epsilon \equiv f(\xi_0)$, is given in \S\ref{appC}. The two bounding curves, that intersect the sphere line at $\epsilon = -3/8$, also intersect in a second point, $(\epsilon,\kappa) \equiv (-0.288,2.485)$, on the prolate side. The unit-sphere exhibits a closed Jeffery-orbit topology at the latter point, even for small but finite $De$; by symmetry, the orbits are (trivially)\,circles at $(-3/8,1)$. The portions of the boundaries between $(-3/8,1)$ and $(-0.288,2.485)$ enclose an island where the orientation dynamics conforms to Regime $3$. Accounting for this island, the transitions in the small-$De$ orientation dynamics, for $\kappa$ increasing from zero to infinity, may be organized across the following six distinct $\epsilon$-intervals:
\begin{enumerate}
\item $-\infty < \epsilon < -1$: stable kayaking $\rightarrow$ stable tumbling $\rightarrow$ stable spinning $\rightarrow$ stable kayaking. 
\item $-1< \epsilon <-1/2$: stable kayaking $\rightarrow$ stable tumbling $\rightarrow$ stable spinning.
\item $-1/2< \epsilon <-3/8$: stable kayaking $\rightarrow$ stable tumbling $\rightarrow$ stable spinning $\rightarrow$ unstable kayaking sandwiched between stable spinning and tumbling.
\item $-3/8 < \epsilon < -1/3$: stable kayaking $\rightarrow$ stable spinning $\rightarrow$ stable tumbling $\rightarrow$ stable kayaking $\rightarrow$ stable spinning $\rightarrow$ unstable kayaking sandwiched between stable spinning and tumbling.
\item $-1/3 < \epsilon < -0.288$: stable spinning $\rightarrow$ stable tumbling $\rightarrow$ stable kayaking $\rightarrow$ stable spinning $\rightarrow$ unstable limit cycle.
\item $-0.288 < \epsilon < \infty$: stable spinning $\rightarrow$ stable tumbling $\rightarrow$ unstable kayaking sandwiched between stable spinning and tumbling. 
\end{enumerate}
The Regime $3$ island in figure \ref{fig_kappa_epsilon} implies that a vertical path corresponding to any $\epsilon$ in the interval $(-3/8,-1/3)$ exhibits a more complex sequence of bifurcations than those identified earlier in figures \ref{fig6} and \ref{fig7}, based on the limiting forms of $\Delta C$; this sequence corresponds to item (iv) in the list above. The scaled orbital drift curves for $\epsilon = -0.36$, belonging to the above interval, are shown in figures \ref{fig9a}-\ref{fig9f}. One starts off in figure \ref{fig9a} with the drift curve exhibiting a negative-slope zero crossing at $C^* =0.862 $. The zero-crossing migrates towards the spinning end, coalescing with it at $\kappa \approx 0.265$. The drift in figure \ref{fig9b} decreases in magnitude towards the near-sphere asymptote, while conforming to a stable spinning regime. Figure \ref{fig9c} shows the expected transition to a stable tumbling regime as $\kappa$ increases across unity. A negative-slope zero-crossing emerges from the tumbling end at $\kappa \approx 1.098$, and as shown in figure \ref{fig9d}, moves towards the spinning end with increasing $\kappa$, coalescing with it at $\kappa \approx 1.499$; this movement also leads to $\Delta C$ reversing its sign\,(from positive to negative). Figure \ref{fig9e} shows that, with a further increase in $\kappa$, the drift magnitude exhibits a non-monotonic variation, attaining a maximum at $\kappa \approx 4$. In figure \ref{fig9f}, a positive-slope zero crossing is seen to emerge from the spinning end at $\kappa \approx 8.154$, again coalescing with it in the limit $\kappa \rightarrow \infty$. With increasing $\kappa$, therefore, figures \ref{fig9a}-\ref{fig9f} describe the following sequence of regimes: (i) $0<\kappa<0.265$: stable kayaking, (ii) $0.265<\kappa<1$: stable spinning,  (iii) $1<\kappa<1.098$: stable tumbling, (iv) $1.098<\kappa<1.499$: stable kayaking, (v) $1.499<\kappa<8.154$: stable spinning and (vi) $8.154<\kappa<\infty$: unstable kayaking.
\section{\bf Rotation arrest: $De$ comparable to $De_c(\kappa)$}\label{Rotation_arrest}
\begin{figure}
	\centering
    \subfigure[]{\includegraphics[scale=0.265]{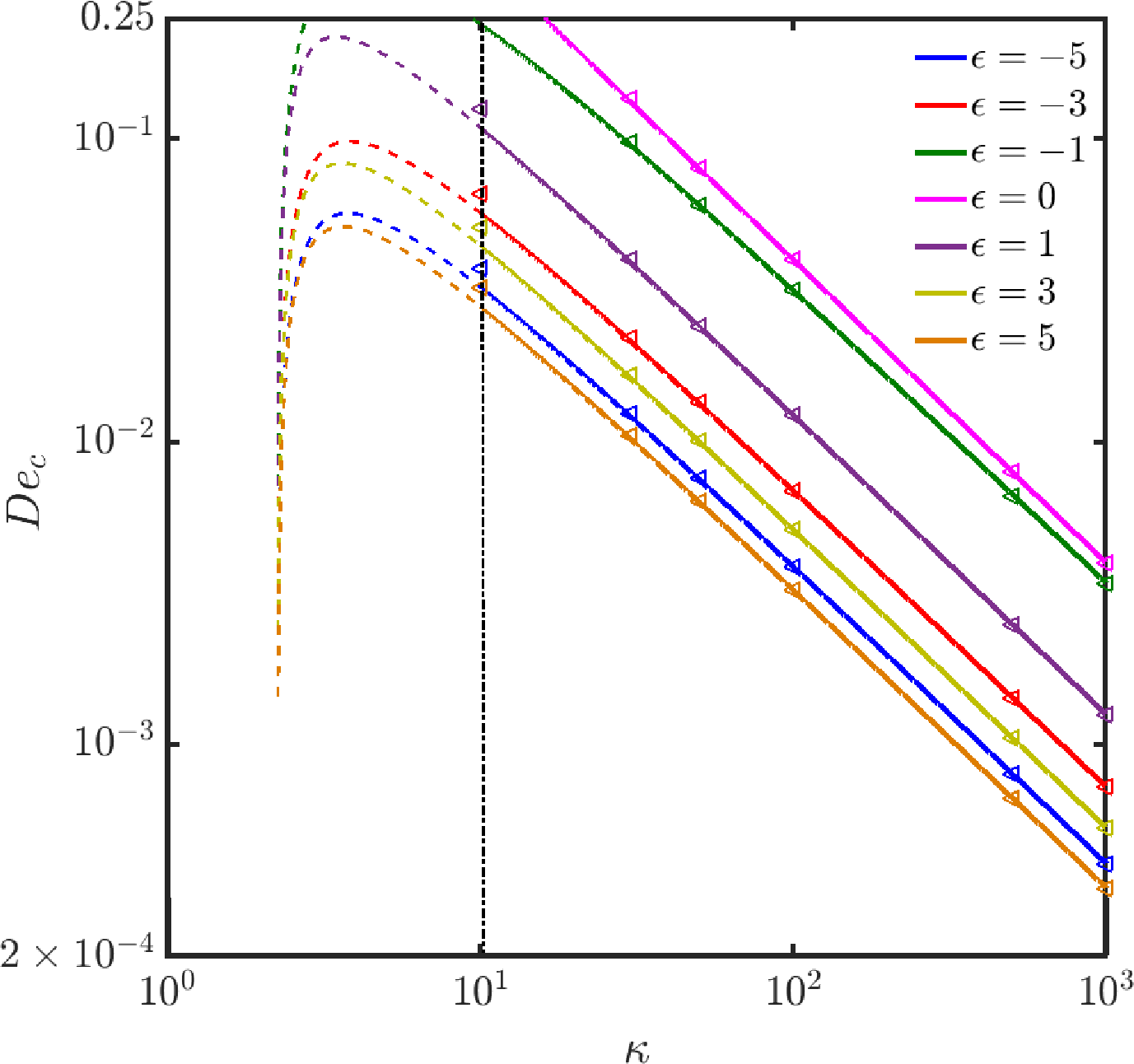}\label{fig_rot_Prolate_a}}\hspace{0.2cm}
 \subfigure[]{\includegraphics[scale=0.265]{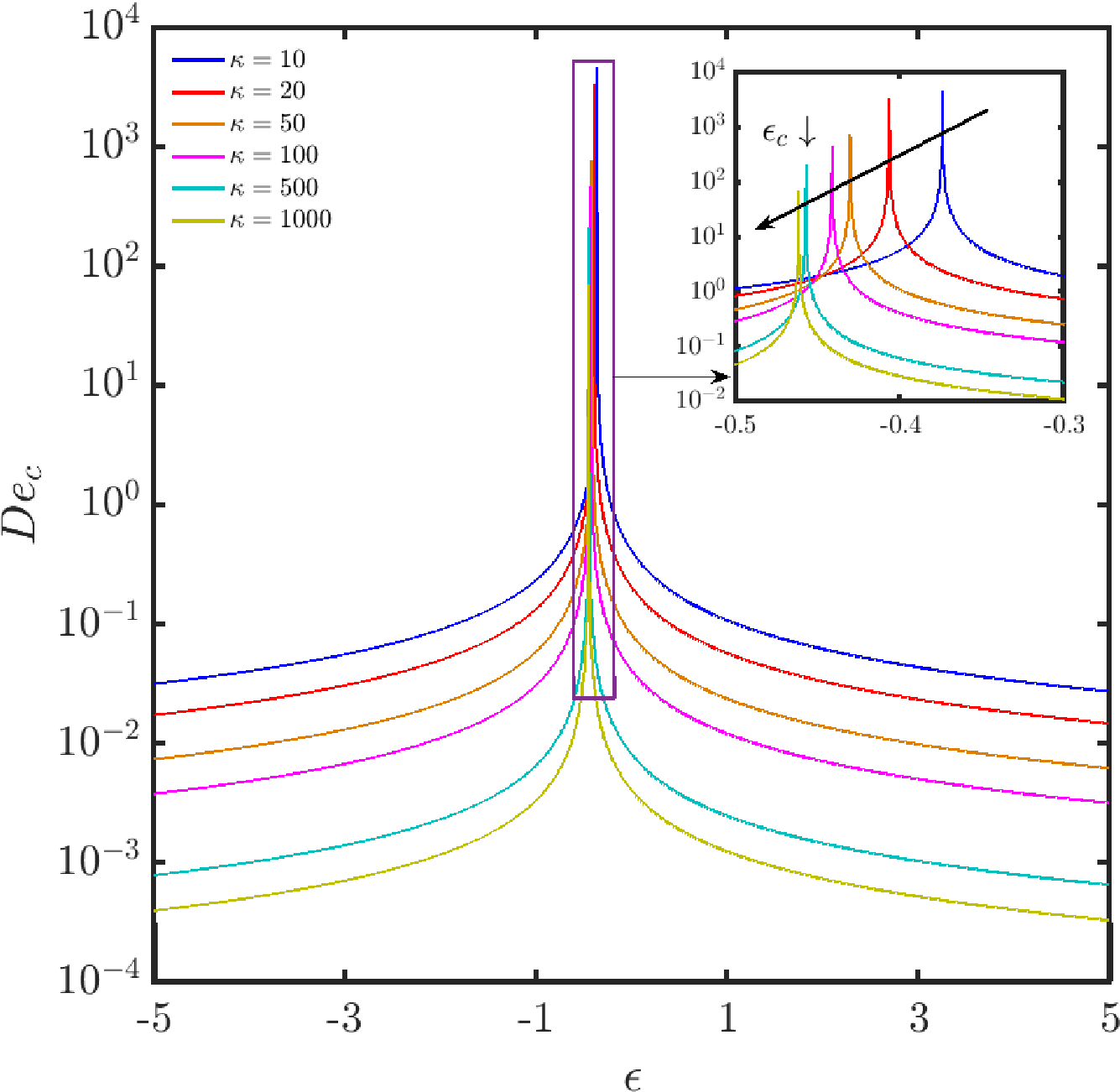}\label{fig_rot_Prolate_b}}
	\caption{The threshold Deborah number, $De_c(\kappa)$ for rotation arrest of a prolate spheroid\,(in the flow-gradient plane) in an ambient simple shear flow: (a) $De_c$ vs $\kappa$ for different $\epsilon$, and (b) $De_c$ vs $\epsilon$ for different $\kappa$. The continuous curves in (a), either solid or dashed, denote the prediction based on (\ref{eq4p2}), with symbols corresponding to a numerical solution of the equilibrium conditions $\dot{\theta}_{j}=0$ with $\dot{\theta}_{j}=\frac{\pi}{2}$; the vertical dash-dotted line in (a) demarcates the physically unrealistic portions of the rotation-arrest curves on its left\,(denoted by dashed lines) that turn around and precipitously dip to zero at a finite $\kappa$. In (b), the movement of divergent location towards $\epsilon = -1/2$\,(the upper end of the polymeric range), with increasing $\kappa$, is illustrated via the inset view.}\label{fig_rot_Prolate} 
\end{figure}

As pointed out at the beginning of \S \ref{Orientation_dynamics}, the multiple scales formalism breaks down for any fixed $De$ for sufficiently slender prolate, and sufficiently flat oblate, spheroids. The Jeffery rotation of such spheroids is highly non-uniform with the slender prolate spheroid staying nearly aligned with the flow axis for most of the $O(\kappa \dot{\gamma}^{-1})$ time period, while rotating through non-aligned orientations in a much shorter time of $O(\dot{\gamma}^{-1})$. An analogous scenario prevails for a flat oblate spheroid, with the aligned phase occurring in the vicinity of the gradient axis and the duration of alignment being $O(\kappa^{-1}\dot{\gamma}^{-1})$. The aligned-phase angular velocity of a slender spheroid is $O(\phi_j^{2})$, with $\phi_j \sim O(\kappa^{-1})$, as may be seen from writing the small-$\phi_j$ large-$\kappa$ form of the equation for $\dot{\phi}_j$, to $O(De)$, obtained from combining (\ref{eq37}) and (\ref{eq41}), and given by:
\begin{equation} \label{eq4p1}
\dot{\phi}_j = -(\phi_{j}^2 + \frac{1}{\kappa^2})+ De\, \phi_{j} \left(\frac{\epsilon }{-3 + \ln 4 + 2 \ln \kappa } + \frac{1 + 2\epsilon }{2}\right) \sin^{2}\!\theta_j. 
\end{equation}
Here, the $O(De)$ viscoelastic contribution is accurate to within algebraically small terms in $\kappa^{-1}$. Since this contribution is proportional to $\phi_j$, it becomes comparable to the leading order Jeffery terms for sufficiently small $\phi_j$, for $De$ and $\theta_j$ fixed, which leads to rotation arrest. The criterion for arrest may be obtained by setting $\dot{\theta}_j$ and $\dot{\phi}_j$ to zero. The former (trivially)\,yields $\theta_j = \frac{\pi}{2}$, while the latter, from (\ref{eq4p1}), leads to a quadratic equation for $\phi_j$. The equation admits real roots, denoting to the pair of rotation-arrested states, when $De$ exceeds a threshold. The threshold corresponds to the said equation having coincident roots, and is given by: 
\begin{equation} \label{eq4p2}
De_{c}(\kappa) = 
\begin{cases} 
\frac{-4}{\kappa\left(1+2\epsilon +\frac{2\epsilon}{(-3+\ln{4}+2 \ln{\kappa})}\right)} & \epsilon <\epsilon_c, \\
\frac{4}{\kappa\left(1+2\epsilon +\frac{2\epsilon}{(-3+\ln{4}+2 \ln{\kappa})}\right)} & \epsilon > \epsilon_c. \\
\end{cases} 
\end{equation} 
Here, $\epsilon_c$ corresponds to the bracketed term in the denominator, in (\ref{eq4p2}), equalling zero, and is given by $\epsilon_c(\kappa) =\left(\frac{3-\ln 4-2  \ln \kappa}{-4+2 \ln 4+4 \ln \kappa} \right)$; $\epsilon_c \rightarrow -1/2 + O(\ln \kappa)^{-1}$ for $\kappa \gg 1$. The viscoelastic contribution is $O(\phi_j^2)$ at $\epsilon = \epsilon_c$, implying that rotation arrest is not possible for small $De$, and accordingly, $De_c$ diverges for $\epsilon \rightarrow \epsilon_c$. The threshold $De$ is plotted as a function of $\kappa$ and $\epsilon$ in figures \ref{fig_rot_Prolate_a} and \ref{fig_rot_Prolate_b}, respectively, with the latter emphasizing the aforementioned divergence at a $\kappa$-dependent $\epsilon_c$ that approaches $-1/2$ for $\kappa \rightarrow \infty$. Since (\ref{eq4p2}) is only valid for large $\kappa$, the moderate-$\kappa\,(< 10)$ portions of the threshold curves in figure \ref{fig_rot_Prolate_a} are shown using dashed lines. The breakdown of the rotation-arrest analysis is also evident from the dashed curves turning around and dropping precipitously to zero at a finite $\kappa$; this behavior is unphysical since the available evidence points to sphere rotation not being arrested at any finite $De$ \citep{hwang2004, dvino2008, snijkers2009,snijkers2011,dvino2015}.

\begin{figure}
	\centering
 \subfigure[]{\includegraphics[scale=0.265]{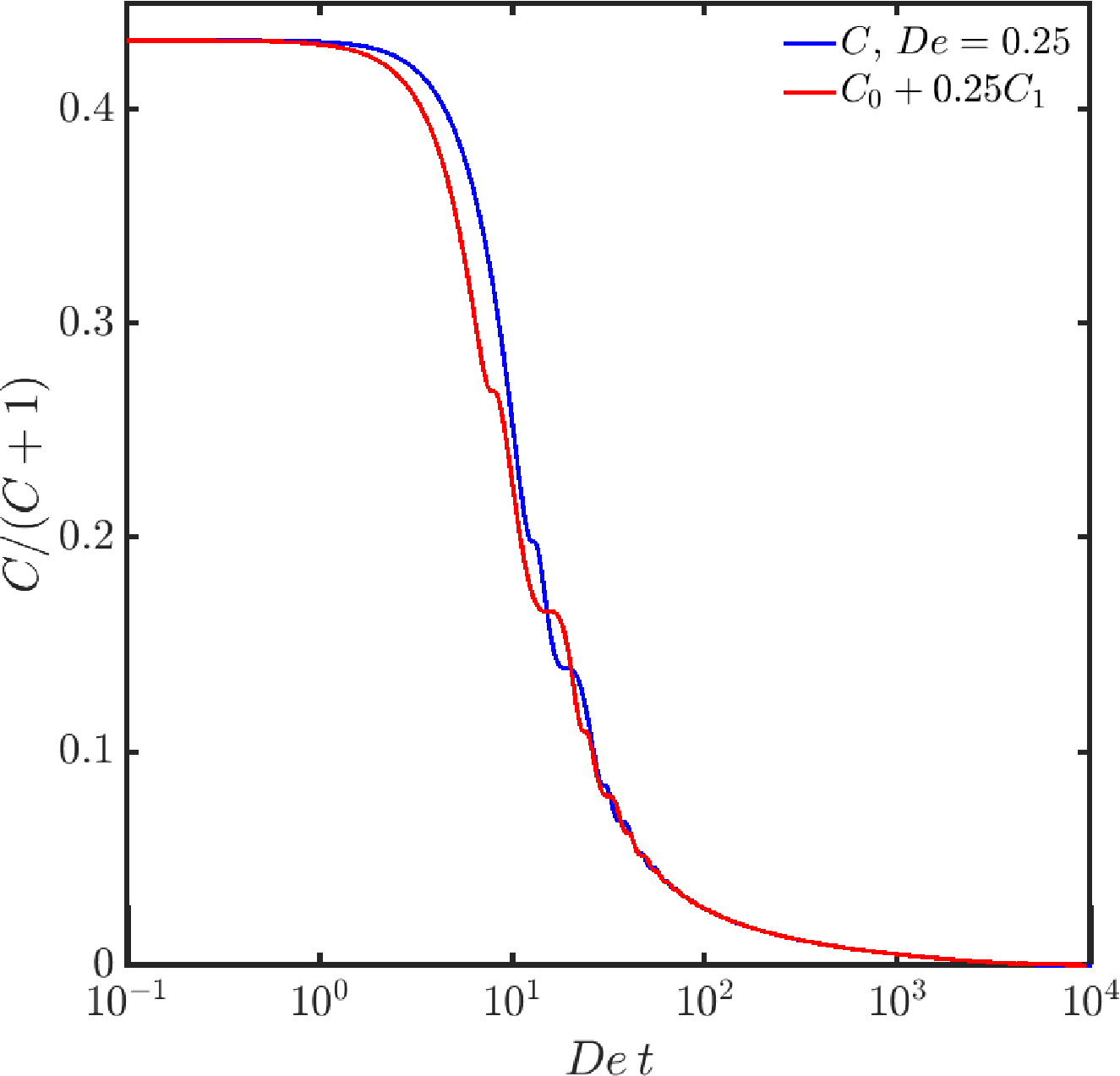}\label{fig_multi_fail_a}}\hspace{0.2cm}
 \subfigure[]{\includegraphics[scale=0.265]{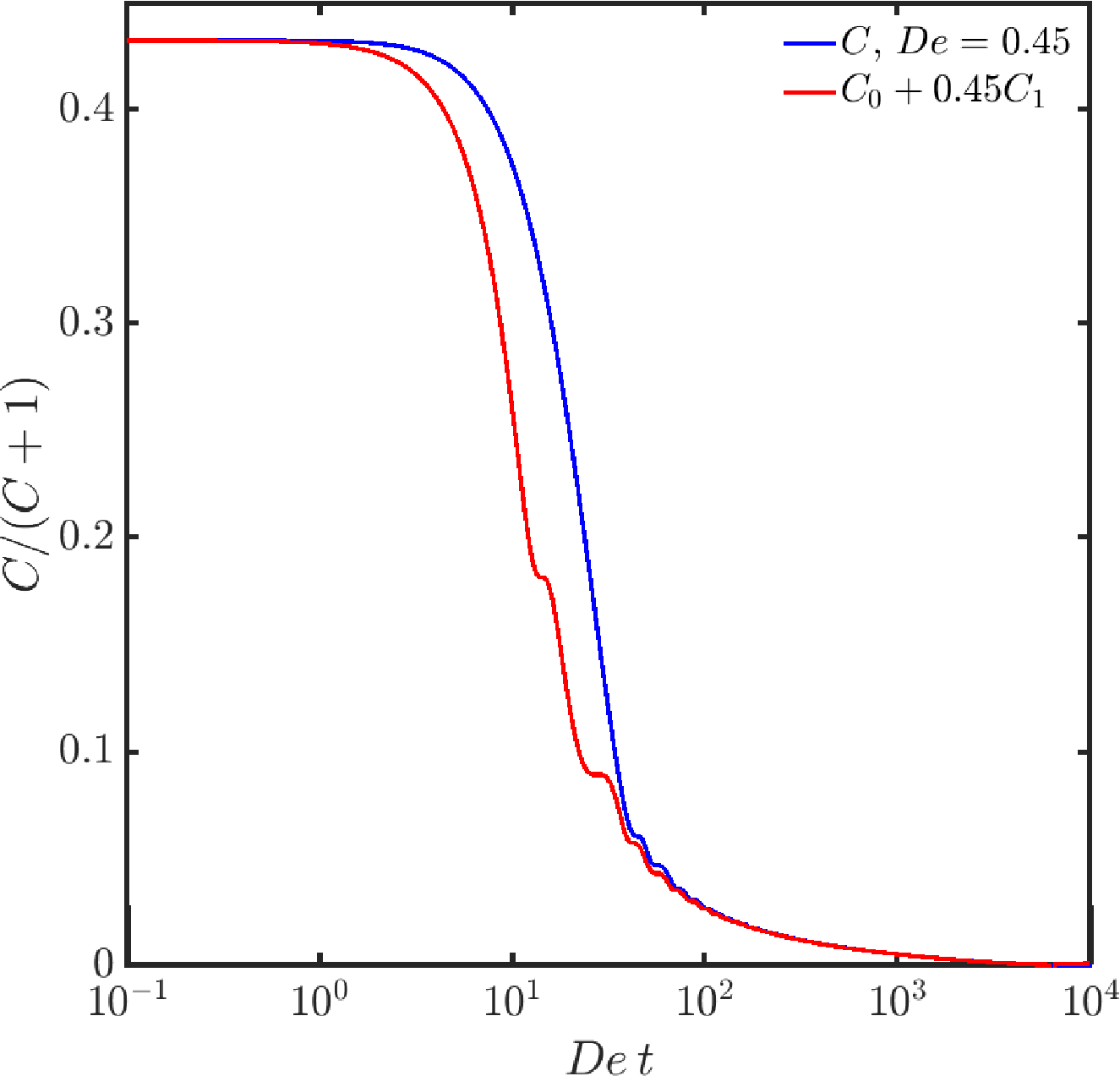}\label{fig_multi_fail_b}}
  \subfigure[]{\includegraphics[scale=0.265]{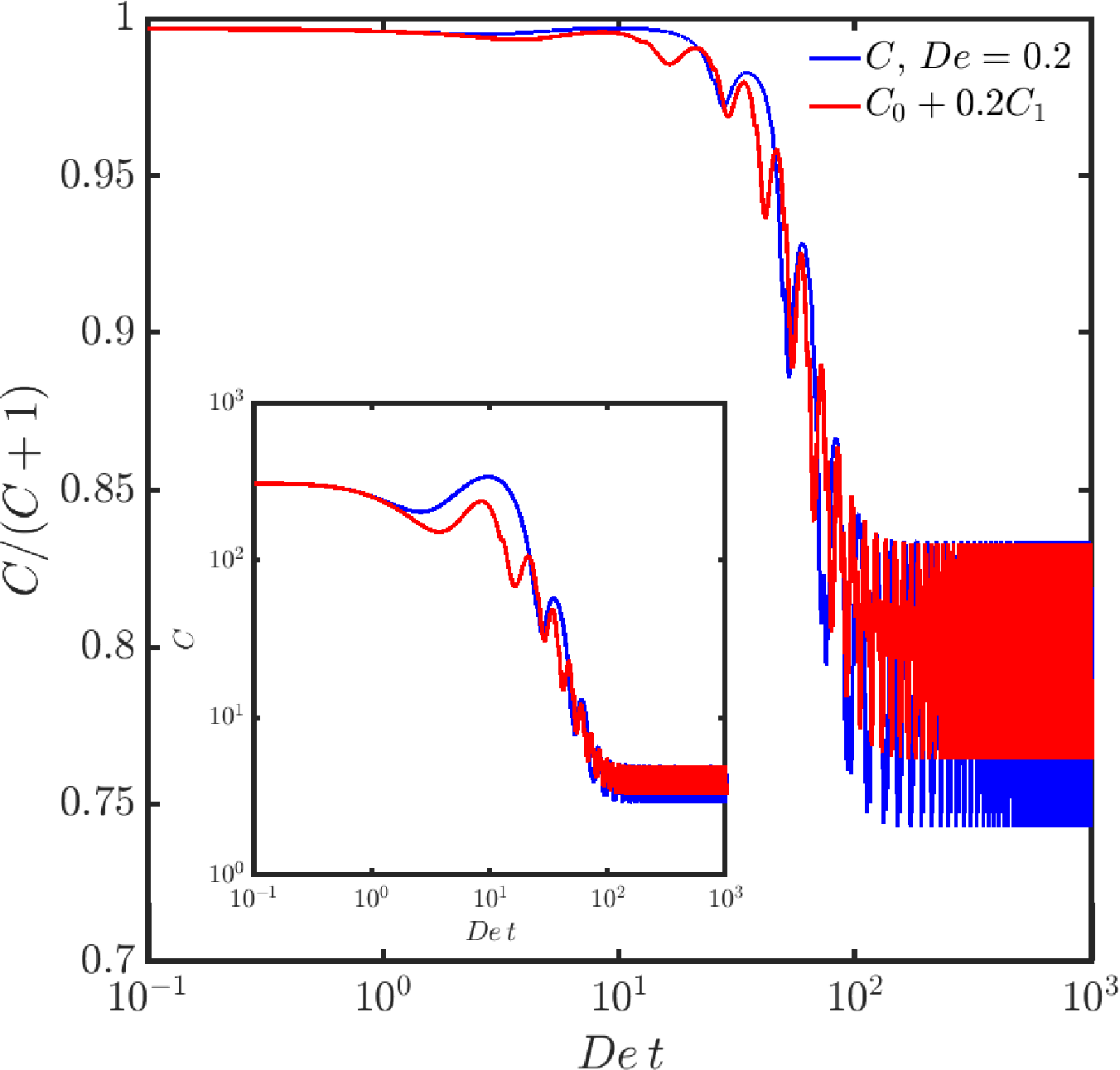}\label{fig_multi_fail_c}}\hspace{0.2cm}
 \subfigure[]{\includegraphics[scale=0.265]{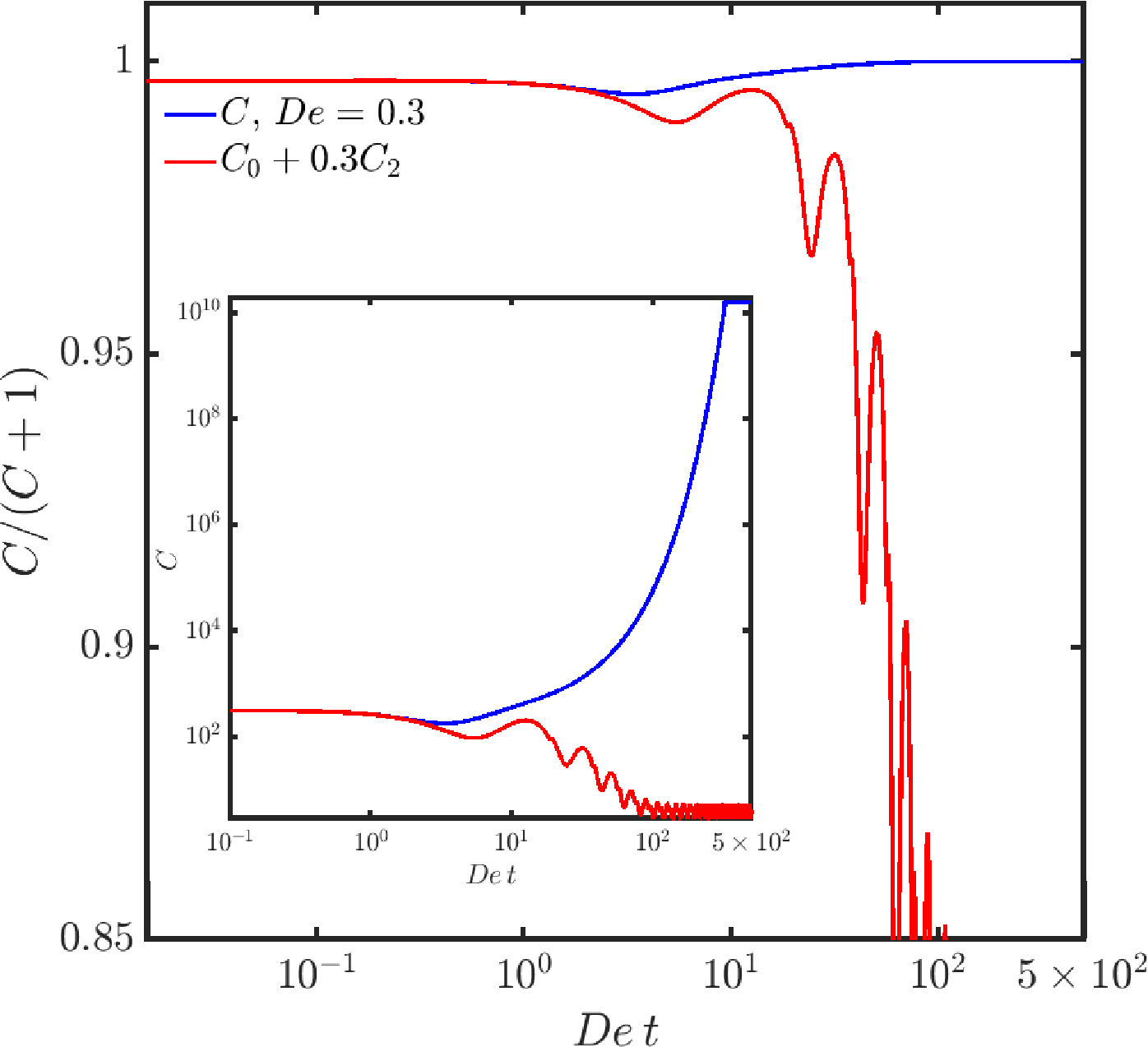}\label{fig_multi_fail_d}}
\caption{The orbit constant $C$ as a function of the slow time $De\,t$, for a spheroid starting from $(\theta_{j},\phi_{j}) \equiv (23 \pi/48,0)$ - with $\kappa=20$, $\epsilon=-0.6$ for (a) $De=0.25$, (b) $De=0.45$, and with $\kappa=0.05$, $\epsilon=-0.6$ for (c) $De=0.2$, (d) $De=0.3$. The blue curves are obtained from numerical integration of the governing equations for $\theta_{j}$ and $\phi_{j}$, with $C$ being calculated thereafter. The red curves denote $C_0 + De\,C_1$, with the expressions for $C_0$ and $C_1$ obtained as part of the multiple scales analysis in \S\ref{Orientation_dynamics:multi}; the insets show the same comparison with the ordinate on a logarithmic scale. The plots highlight the breakdown of the multiple scales analysis for $De > De_c(\kappa)$, with $\kappa$ and $\epsilon$ fixed. Using (\ref{eq4p2}), $De_{c} = 0.422$ for $(\kappa,\epsilon) \equiv (20,-0.6)$; from (\ref{eq4p3}) and (\ref{eq4p4}), $De_{c1} = 0.23$, $De_{c2} =0.25$ for $(\kappa,\epsilon) \equiv (0.05,-0.6)$.}\label{fig_multi_fail} 
\end{figure}

The failure of the multiple scales analysis for $De > De_c$ is illustrated via the orbit constant plots in figures \ref{fig_multi_fail_a}-\ref{fig_multi_fail_d}. Figures \ref{fig_multi_fail_a} and \ref{fig_multi_fail_b} pertain to a prolate spheroid with $\kappa =20$, and figures \ref{fig_multi_fail_c} and \ref{fig_multi_fail_d} to an oblate one with $\kappa =0.05$, both for $\epsilon = -0.6$\,(in the polymeric range). The first figure in each of these pairs corresponds to $De$ below, and the second one to $De$ above, $De_c(\kappa)$. The latter is defined by (\ref{eq4p2}) for the prolate case, and equals $0.422$. As will be shown below, there are multiple threshold $De$'s for the oblate case, with the first threshold\,($De_{c1}$) being defined by (\ref{eq4p3}) below and equalling $0.23$ for the chosen $\kappa$. In figure \ref{fig_multi_fail_a}, the variation of $C$ with the slow time\,($D\!e\,t$), obtained from numerical integration of the $(\theta_j,\phi_j)$-system, compares reasonably well with that predicted by the multiple scales analysis, for $De = 0.25$. In contrast, there is substantial disagreement between the two in figure \ref{fig_multi_fail_b} for $De = 0.45$. Rather surprisingly, even in this case, the two curves agree well for sufficiently long times, which has to do with the nature of slender spheroid rotation. For the chosen $\epsilon$, the orientation dynamics conforms to Regime $1$, and the spheroid spirals in towards the vorticity axis for $De < De_c(\kappa)$. While a spheroid in the flow-gradient plane stops rotating for $De > De_c(\kappa)$, for $\theta_j \neq \frac{\pi}{2}$, only rotation about the ambient vorticity ceases above a threshold $De$ given by $De_c(\kappa)/\sin^2 \theta_j$. For larger $De$, the spheroid continues to drift monotonically towards the vorticity axis, on account of $\dot{\theta}_j$ being non-zero\,(and negative). Since the $\theta_j$-dependent threshold above diverges for $\theta_j \rightarrow  0$, the decrease in $\theta_j$ accompanying the drift towards vorticity leads to $De$ eventually falling below $De_c$, and the spheroid reverting to a spiralling approach towards the limit cycle. This leads to the multiple scales analysis again becoming valid for sufficiently small $\theta_j$, and therefore, explains the long-time agreement in figure \ref{fig_multi_fail_b}. It is worth recalling that the aforementioned dynamical sequence was indeed observed for rod-shaped particles in the experiments of \citet{bartram1975}, discussed in \S\ref{introsec:expt}, and also predicted by \citet{leal1975}. In contrast to the prolate case above, figures \ref{fig_multi_fail_c} and \ref{fig_multi_fail_d} show that the multiple scales analysis breaks down completely for the oblate spheroid above the rotation-arrest threshold. Figure \ref{fig_multi_fail_c} is consistent with Regime $3$ for $De < De_c(\kappa)$, with both the numerical integration and multiple scales analysis predicting an oscillatory approach to a stable limit cycle\,(see inset). In contrast, in figure \ref{fig_multi_fail_d}, while the multiple scales prediction remains virtually unchanged, the numerical integration for $De > De_c(\kappa)$ reveals a monotonic drift towards the flow-gradient plane instead. Since $\theta_j$ increases with time during the course of this drifting motion, unlike the prolate case, $De$ never dips below the threshold for rotation arrest, and the multiple scales analysis and numerical integration remain in disagreement for all time.

\begin{figure}
	\centering
    \subfigure[]{\includegraphics[scale=0.24]{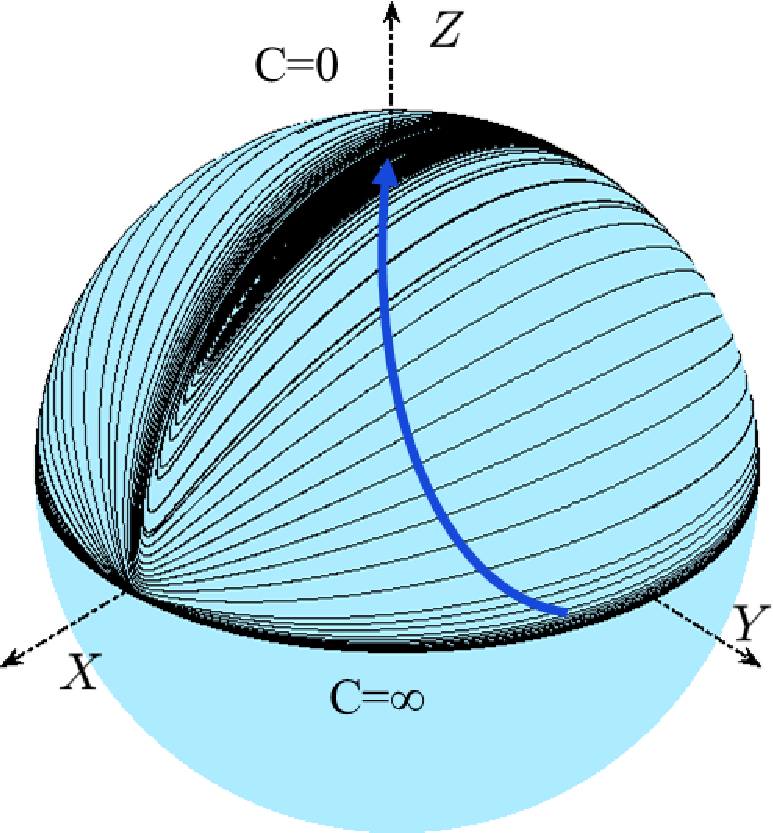}\label{fig_rotation_eps_m0p6_a}} 
    \subfigure[]{\includegraphics[scale=0.14]{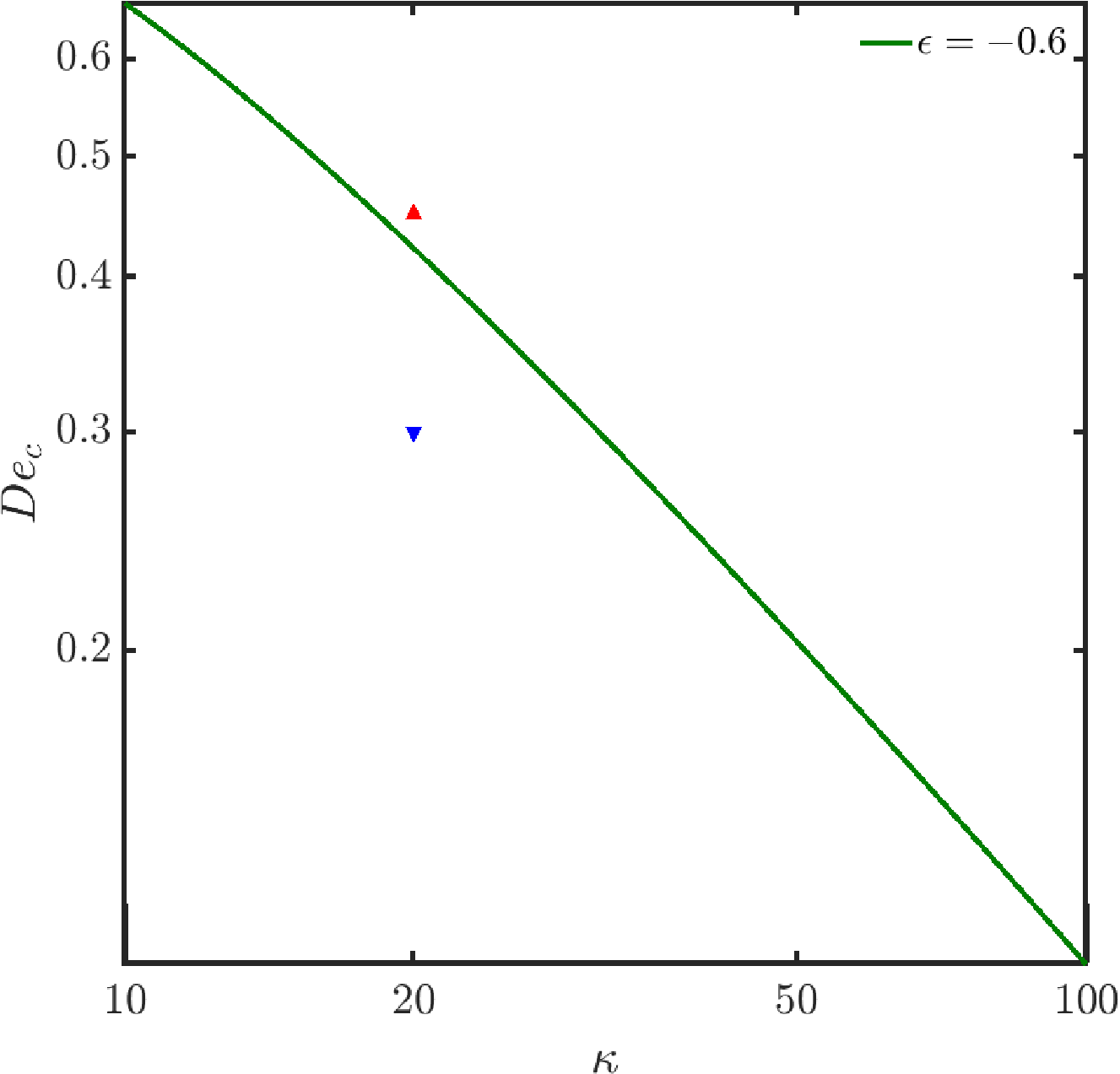}\label{fig_rotation_eps_m0p6_b}} \hspace{0.2cm}
    \subfigure[]{\includegraphics[scale=0.24]{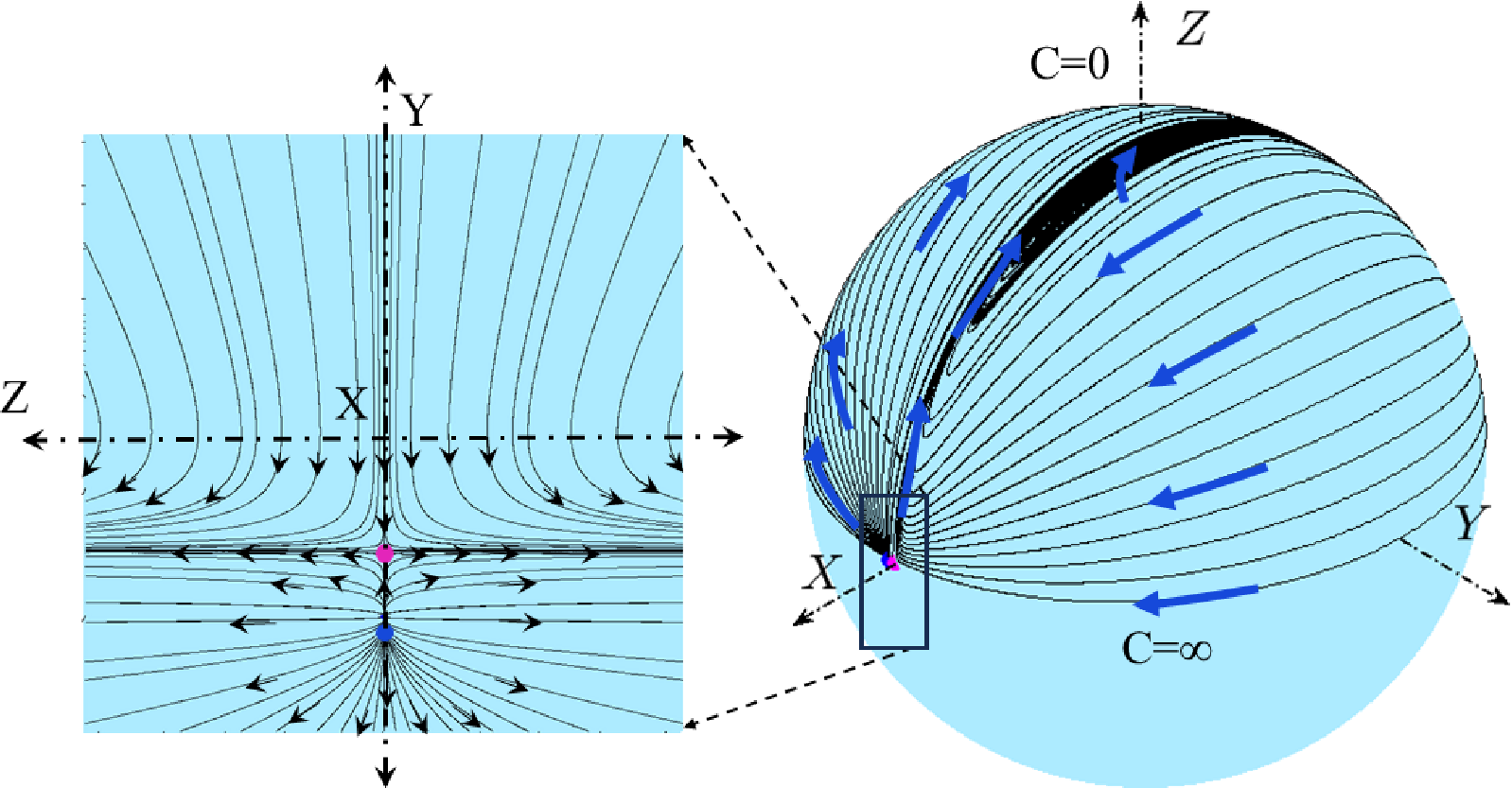}\label{fig_rotation_eps_m0p6_c}}
	\caption{Unit sphere trajectory topologies for $De$ below and above rotation arrest, for a prolate spheroid with $\kappa=20$, for $\epsilon= -0.6$. (a) $De=0.3$ and (c) $De=0.45$; $De_{c}(\kappa) \approx 0.422$. The plot in (b) shows the chosen $De$'s in (a) and (c), in relation to the threshold curve for rotation arrest. The trajectory topology in (a) conforms to Regime $1$, with blue arrows in depicting the direction of viscoelastic drift; the blue arrows in (c) depict the overall direction of the drift arising due to the fixed-point pair that emerges above rotation arrest. The magnified view accompanying (c) shows trajectories in the immediate vicinity of the saddle\,(magenta) and unstable node\,(blue).}\label{fig_rotation_eps_m0p6} 
\end{figure}

From the dynamical systems perspective, the rotation arrest above is a saddle-node bifurcation, which leads to the creation of fixed points in the flow-gradient plane for $De > De_c(\kappa)$ that, for a prolate spheroid, are at an angle of $O(\kappa^{-1})$ to the flow direction. The inversion symmetry of simple shear implies that identical fixed-point pairs emerge in the neighborhoods of both $\phi_j = 0$ and $\pi$. For $\epsilon$ in the polymeric range, as in figure \ref{fig_multi_fail}, calculation of the eigenvalues based on the linearized forms of the equations governing $\theta_j$ and $\phi_j$, shows that the fixed-point pair consists of a saddle and an unstable node. Unit-sphere trajectory topologies below and above $De_c(\kappa)$, for $\epsilon = -0.6$, are depicted in figures \ref{fig_rotation_eps_m0p6_a} and \ref{fig_rotation_eps_m0p6_c}. The latter figure shows the emergence of a pair of fixed points, close to $\phi_j = 0$, just below the flow-vorticity plane; the magnified view accompanying this figure shows trajectories in the immediate vicinity of the fixed-point pair. Since both fixed points are unstable\,(the saddle point being an attractor only for orientations within the flow-gradient plane), a prolate spheroid starting from almost any initial orientation still ends up approaching the spinning mode. Only the manner of approach towards the spinning equilibrium depends on whether $De$ is less or greater than $De_c(\kappa)$ - the approach always having a spiralling character for $De < De_c(\kappa)$\,(as in figure \ref{fig_rotation_eps_m0p6_a}), but having an initially monotonic\,(and eventually spiralling) character for $De > De_c(\kappa)$\,(as in  figure \ref{fig_rotation_eps_m0p6_c}).

It is worth adding that, the location of the pair of degenerate fixed points at $De = De_c(\kappa)$ is independent of $\epsilon$, being given by $\phi_j = -\frac{1}{\kappa}, \pi - \frac{1}{\kappa}$ for a slender spheroid. Saddle-node pairs emerge for larger $De$, with the saddle approaching the flow-vorticty plane and the node moving away, until $De$ is no longer small. The stability of the node depends on the orientation dynamics regime that prevails for $De < De_c(\kappa)$, and thence, on $\epsilon$. From figure \ref{fig_kappa_epsilon}, the orientation dynamics of a slender prolate spheroid is seen to conform to any of Regimes $1$, $3$ or $4$, depending on $\epsilon$. Regime $1$ is the polymeric case described above. For Regime $3$, the diametrically opposite nodes that emerge after bifurcation remain unstable, similar to Regime $1$. In this regime, a prolate spheroid for $De > De_c(\kappa)$, starting from almost any initial orientation, exhibits an initially monotonic approach towards a stable limit cycle\,(as opposed to the vorticity axis). In Regime $4$, corresponding to $\epsilon > -1/2$ for large $\kappa$, the node in each of the saddle-node pairs is a stable one, with each of the saddle-node pairs emerging on the opposite side of the flow-vorticity plane, when compared to Regimes $1$ and $3$. The stability of the nodes implies that a prolate spheroid, starting from any initial orientation outside of the unstable limit cycle, must eventually converge to either of the nodes in the flow-gradient plane. A pair of trajectories that spiral outward from the limit cycle, and are also the stable manifolds of the saddle points\,(in the flow-gradient plane), form the boundaries of the basins of attraction pertaining to the pair of stable nodes - a similar scenario occurs for an oblate spheroid for $\epsilon$ in the polymeric range, and is described in more detail below.

\begin{figure}
	\centering
        \subfigure[]{\includegraphics[scale=0.24]{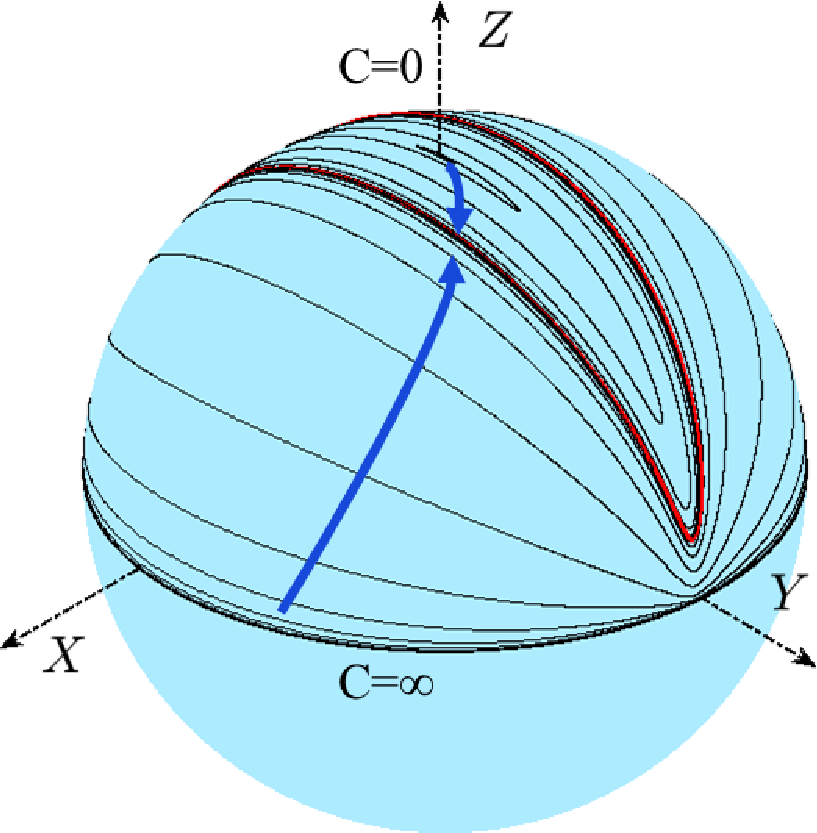}\label{fig_rotation_eps_m2by3_a}}
        \subfigure[]{\includegraphics[scale=0.14]{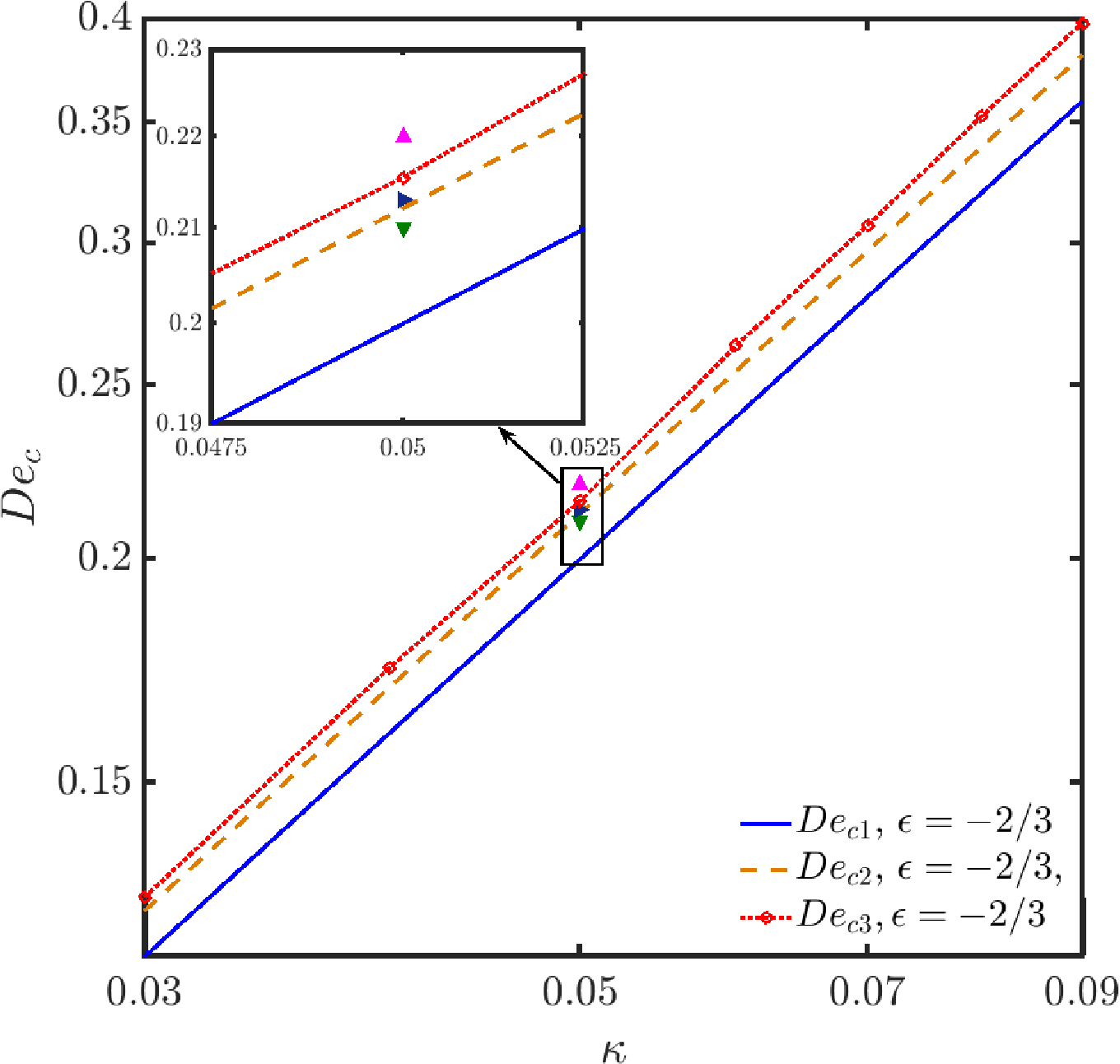}\label{fig_rotation_eps_m2by3_b}}
          	\subfigure[]{\includegraphics[scale=0.24]{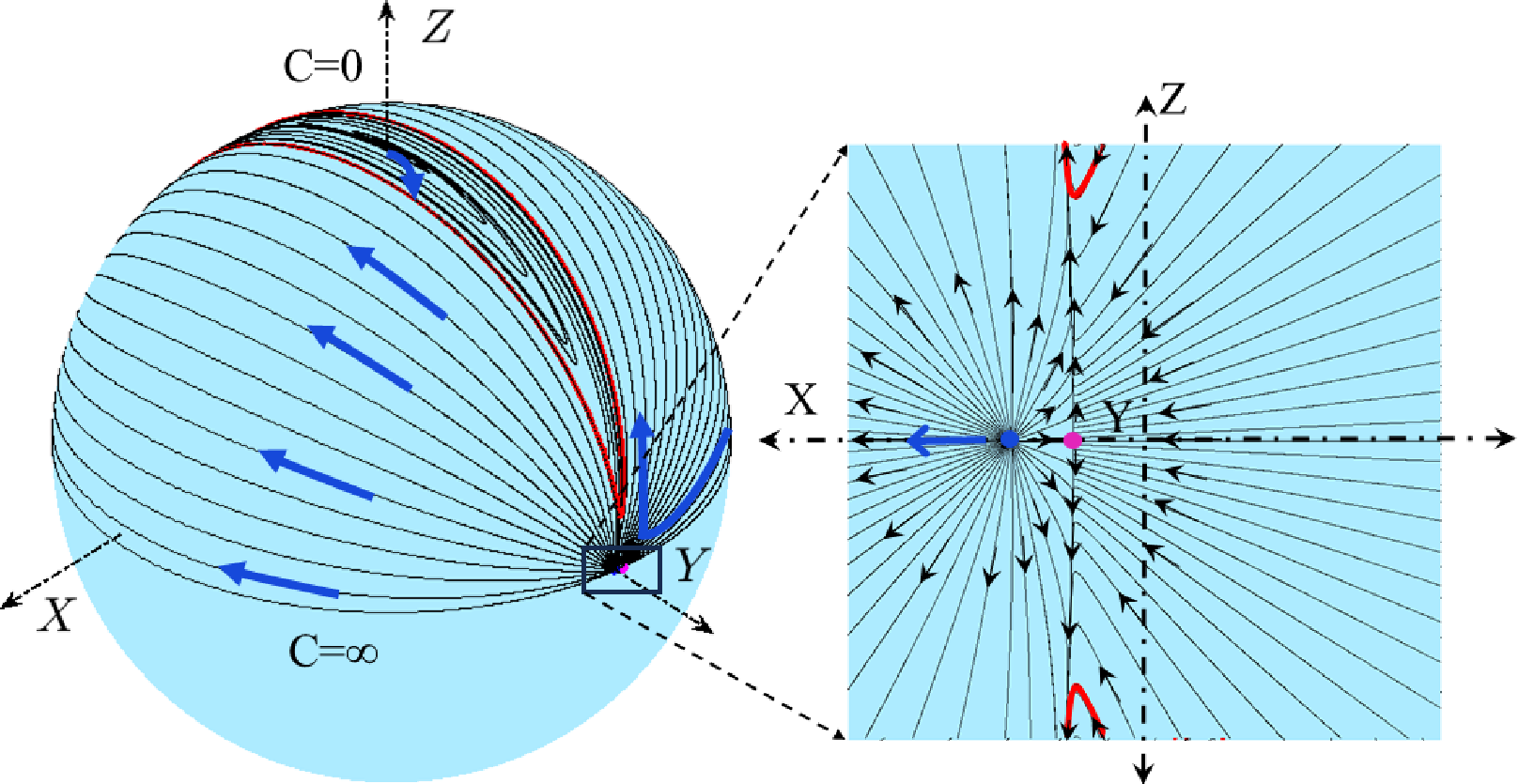}\label{fig_rotation_eps_m2by3_c}}\\
            \subfigure[]{\includegraphics[scale=0.24]{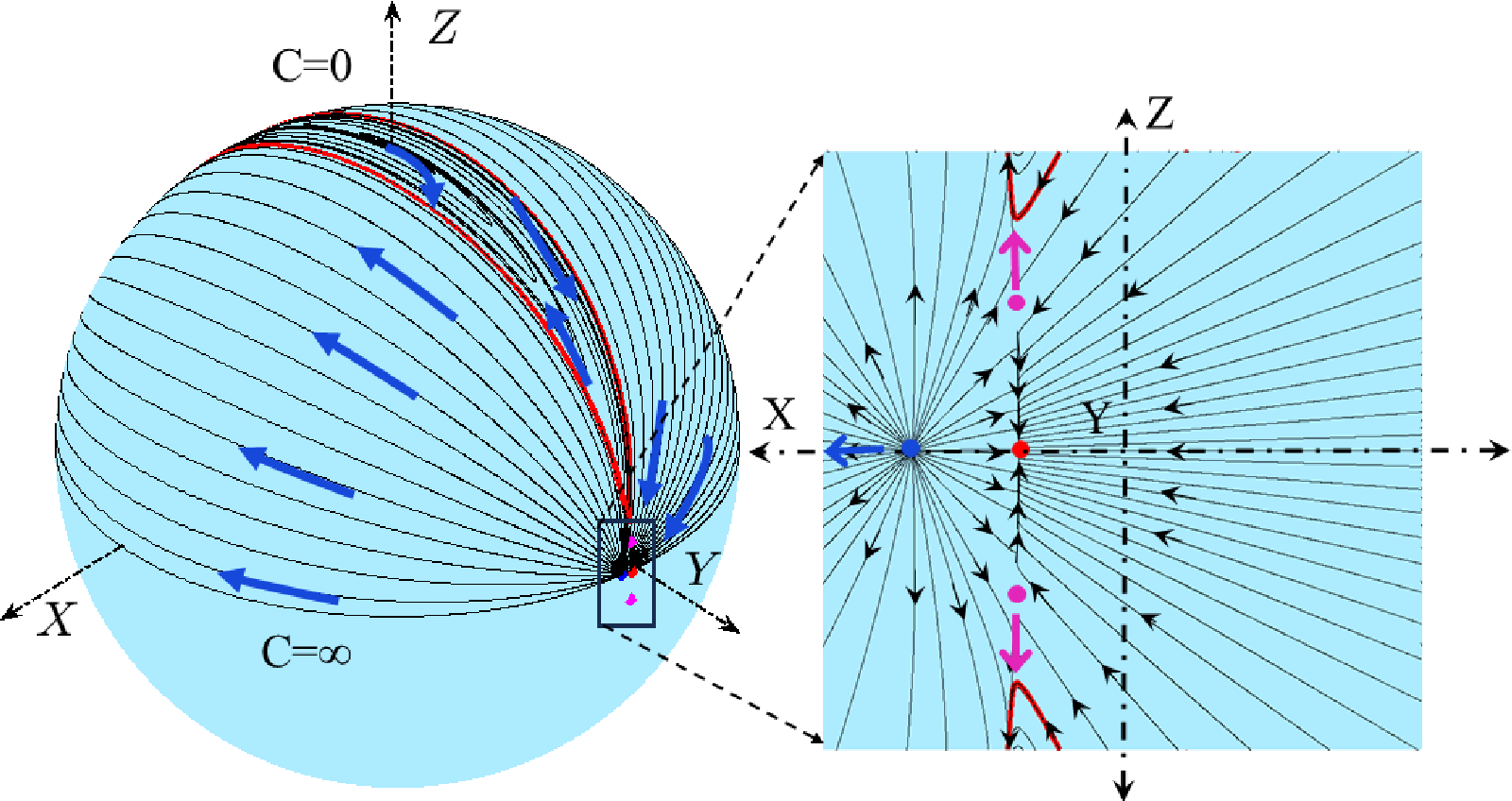}\label{fig_rotation_eps_m2by3_d}} \hspace{0.1cm}
          \subfigure[]{\includegraphics[scale=0.24]{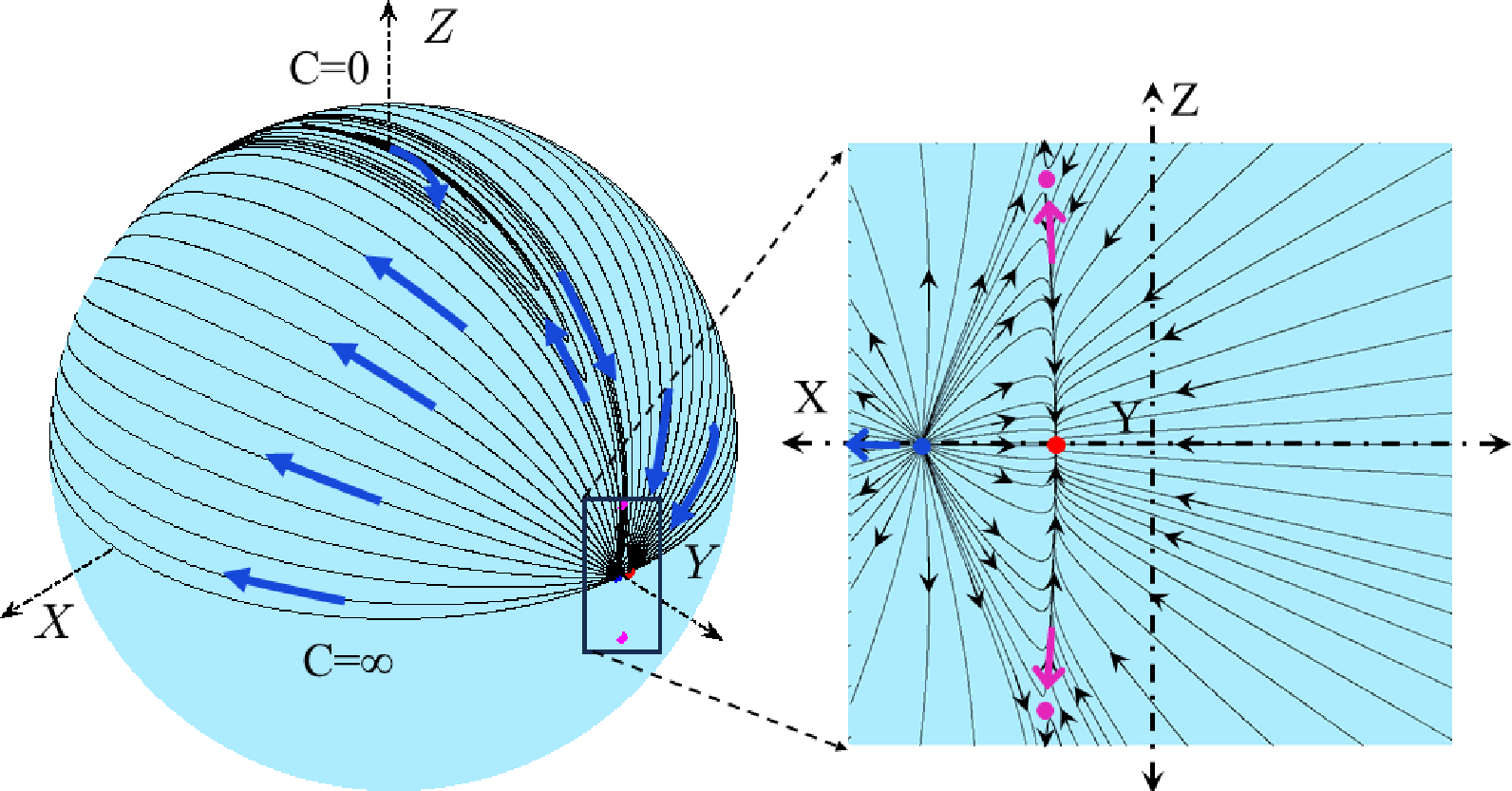}\label{fig_rotation_eps_m2by3_e}}
	\caption{Unit sphere trajectory topologies before and after rotation arrest for an oblate spheroid with $\kappa=0.05$, for $\epsilon=-2/3$, for three different Deborah numbers: (a) $De=0.1$, (c) $De=0.21$, (d) $De=0.213$ and (e) $De=0.22$. The stable limit cycle appears as a red curve in (a), (c) and (d). The three  critical Deborah numbers are plotted as a function of $\kappa$ in (b); for the chosen parameters, corresponding to the points shown in (b), $De_{c1}=0.2$, $De_{c2}=0.212$ and $De_{c3} = 0.215$. (a) corresponds to the trajectory topology before rotation arrest. (c) corresponds to the trajectory topology after the primary bifurcation, with a saddle\,(magenta) and unstable node\,(blue) emerging in the gradient-vorticity plane. (d) corresponds to the trajectory topology after the secondary bifurcation which leads to a stable node\,(red) and pair of saddle interior saddle points\,(magenta) moving towards the pair of limit cycles. (e) corresponds to the trajectory topology after the tertiary bifurcation with the limit cycles having been destroyed. Magnified views accompanying (c), (d) and (e) illustrate the trajectory directions and highlight the behavior of the fixed points. Blue and magenta arrows depict the direction of movement of the fixed points.}\label{fig_rotation_eps_m2by3} 
\end{figure}

The rotation of a thin oblate spheroid is analogous to a slender prolate spheroid, the difference being that the aligned phase occurs in the vicinity of the gradient axis, with $\frac{\pi}{2} -\phi_j$ rather than $\phi_j$, being small.
An analysis along lines similar to the above leads to the threshold Deborah number, for rotation arrest, being given by: 
\begin{equation}\label{eq4p3}
     De_{c1}=\begin{cases}
\frac{-12 \kappa}{(1+6\epsilon)} & \epsilon < -1/6,  \\
\frac{12 \kappa}{(1+6\epsilon)} & \epsilon > -1/6, \\
\end{cases}
    \end{equation}
with the arrest first occurring at $\phi_j = \frac{\pi}{2} -\kappa, \frac{3\pi}{2} -\kappa$. $De_{c1}(\kappa)$ now diverges at a $\kappa$-independent $\epsilon\,( = -1/6)$. Figures \ref{fig_rotation_eps_m2by3_a} and \ref{fig_rotation_eps_m2by3_c} correspond to $De = 0.1$ and $0.21$, respectively, and depict unit-sphere trajectory topologies below, and just above, rotation arrest for an oblate spheroid with $\kappa = 0.05$, and for $\epsilon=-2/3$. For the chosen $\kappa$ and $\epsilon$, rotation arrest occurs at $De_{c1}(\kappa) = 0.2$, and the orientation dynamics for $De < De_{c1}$ conforms to Regime $3$, as evident from the stable limit cycle in figure \ref{fig_rotation_eps_m2by3_a}. As for the prolate spheroid, rotation arrest for $De > De_{c1}$ is a saddle-node bifurcation, that leads to the emergence of a saddle-unstable-node pair in the flow-gradient plane, but now in the neighborhood of the gradient axis; the saddle being closer to the gradient axis. However, unlike the prolate spheroid, the stable limit cycle around the vorticity axis is quite large, extending almost right until the points at which the gradient axis intersects the unit sphere. The vicinity of the limit cycle to the fixed-point pair created after bifurcation has the consequence that the trajectory topology shown in figure \ref{fig_rotation_eps_m2by3_c} only persists over a small $De$-interval, as described in the following paragraph. From figure \ref{fig_kappa_epsilon}, the orientation dynamics of an oblate spheroid is also seen to conform to Regime $2$, if $\kappa$ were greater than an $\epsilon$-dependent threshold - this threshold varies from $0.2$ to $0.454$ for $\epsilon\in[-0.7,-0.5]$. While rotation arrest in this regime would lead to a stable, rather than unstable, node in the flow-gradient plane, the associated threshold $De$ values are found to be of order unity, and therefore, unlikely to be amenable to an ordered-fluid description.

The rotation arrest above is only the first of multiple bifurcations in the oblate case. The saddle-node bifurcation above is followed by a second bifurcation where the saddle point in the flow-gradient plane, shown in figure \ref{fig_rotation_eps_m2by3_c}, and that is now closer to the gradient axis, tri-furcates into a stable node, with a pair of new `interior' saddle points originating from it on either side of the flow-gradient plane, and moving away from it in a symmetrical manner. Now, the transformation of the original saddle into a stable mode requires that the associated stable eigenvalue change sign. The threshold $De$ for this change of sign may be obtained from analyzing the equation for $\dot{\theta}_{j}$, linearized about the saddle point, which leads to the following threshold:
\begin{equation}\label{eq4p4}
    \begin{aligned}
   De_{c2}= \frac{2\sqrt{3} \kappa }{\sqrt{\epsilon(1+3\epsilon)}} \quad  \text{for}\quad \epsilon \in(-\infty,-1/3) \cup (0,\infty),
    \end{aligned}
\end{equation}
with there being no secondary bifurcation in the interval $\epsilon\,\in\,(-1/3,0)$. Note that the emergence of the pair of interior saddle points preserves the Euler characteristic, as is required from topological invariance - the phase space for the spheroid orientation vector, the surface of the unit sphere, has an Euler characteristic of two, and the Euler characteristic must equal the sum of all fixed-point indices\,(saddles have an index of $-1$, while nodes and foci have an index of $1$). Figure \ref{fig_rotation_eps_m2by3_d}, corresponding to $De = 0.213$, illustrates the unit-sphere trajectory topology just after the secondary bifurcation\,($De_{c2}(\kappa) = 0.212$ in this case). A comparison of the magnified views associated with figures \ref{fig_rotation_eps_m2by3_c} and \ref{fig_rotation_eps_m2by3_d}, highlights the tri-furcation of the flow-gradient saddle into a stable node and a pair of interior saddles. It is worth emphasizing that $De_{c2}$, like $De_{c1}$ is $O(\kappa)$, and therefore, small for flat oblate spheroids. In fact, as evident from the plot of the various threshold Deborah numbers against aspect ratio in \ref{fig_rotation_eps_m2by3_b}, $De_{c1}$ and $De_{c2}$ are very close to each other. An analogous secondary bifurcation also occurs for a slender prolate spheroid, but the associated Deborah number is of order unity, and therefore, outside the purview of the ordered-fluid paradigm.

With a further increase in $De$, the interior saddle points move further away from the flow-gradient plane, and towards the pair of stable limit cycles, colliding with them at a tertiary threshold $De$. This collision destroys the limit cycles, with only fixed points remaining on the unit sphere for all higher $De$ - these comprise four interior saddle points, a pair each of unstable and stable nodes in the flow-gradient plane, and a pair of unstable foci at the vorticity axis intersections with the unit sphere. For the oblate spheroid chosen above, with $\epsilon = -2/3$, this collision occurs at $De_{c3} \approx 0.215$. Figure \ref{fig_rotation_eps_m2by3_e} illustrates the unit-sphere trajectory topology for a slightly higher $De=0.22$. A comparison of the magnified views in figures \ref{fig_rotation_eps_m2by3_d} and \ref{fig_rotation_eps_m2by3_e} shows how the pair of red curves\,(corresponding to the stable limit cycles) is destroyed on collision with the saddle points; the latter continue to migrate towards the vorticity axis. 

\begin{figure}
	\centering
        \subfigure[]{\includegraphics[scale=0.35]{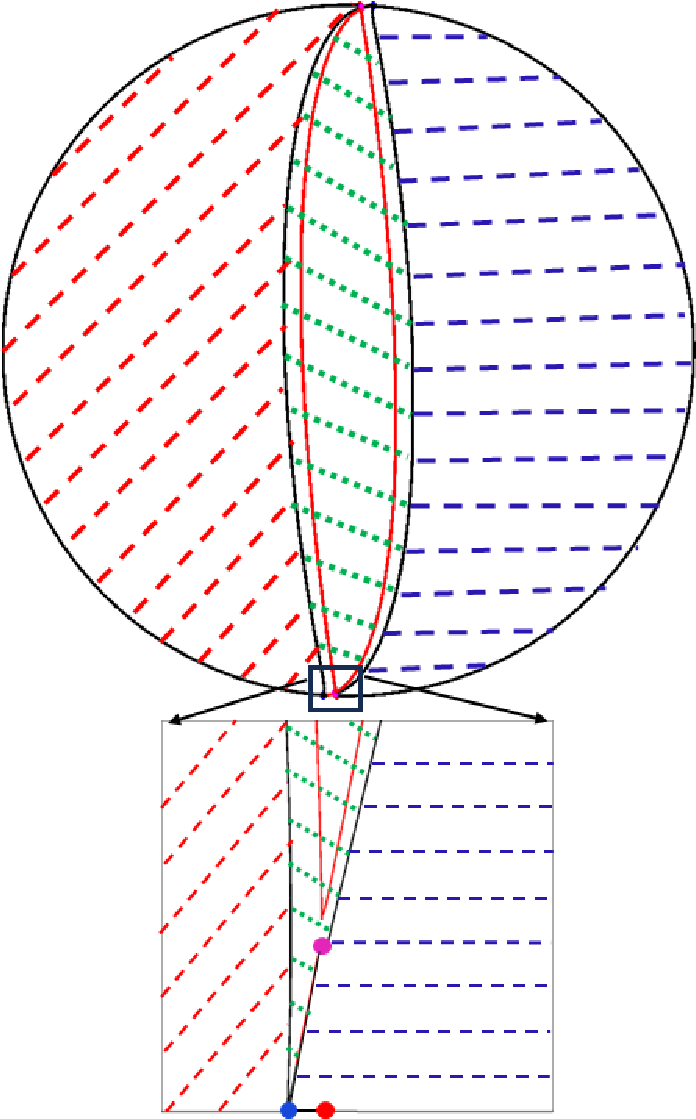}\label{fig_arrest_sketch_a}}\hspace{0.25cm}
        \subfigure[]{\includegraphics[scale=0.35]{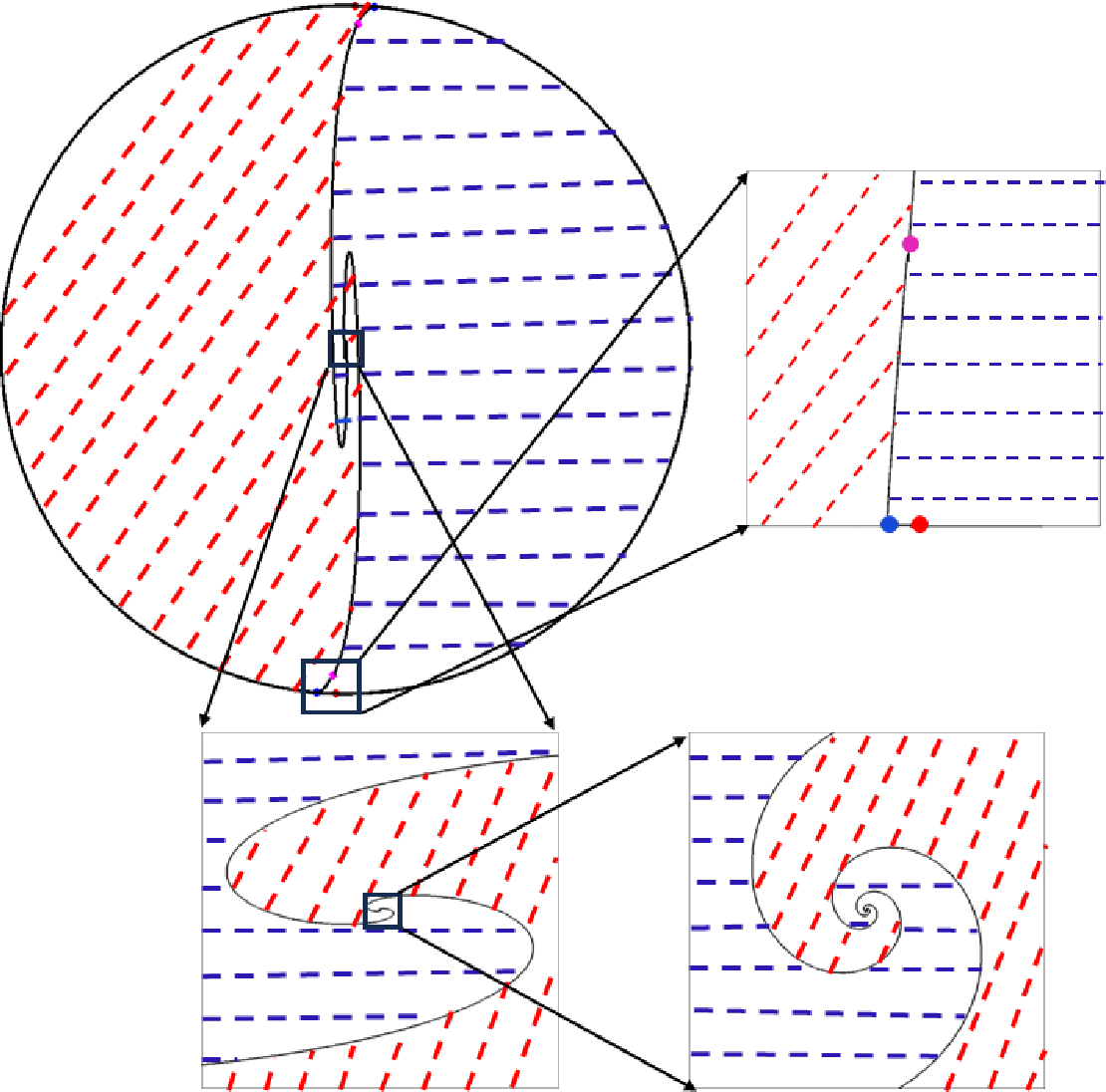}\label{fig_arrest_sketch_b}}
 	\caption{Projected views of the unit hemisphere illustrate the boundaries\,(in black) demarcating the basins of attraction, corresponding to the different invariant sets, for (a) $De =0.214 \,\in\,(De_{c2},De_{c3})$ and (b) $De =0.23  >De_{c3}$; the views correspond to an oblate spheroid with $\kappa =0.05$, for $\epsilon =-2/3$. The closed curve within the basin boundaries in (a) is the stable limit cycle. The magnified view in (a) shows a pair of basin boundaries, terminating at one of the interior saddle points\,(magenta) lying very close to the flow-gradient plane; the latter plane contains a pair of nodes, one stable\,(red) and the other unstable\,(blue). The bottom pair of views in (b), at successively higher levels of magnification, highlight the infinitely interlaced nature of the basins of attraction in the neighborhood of the vorticity-axis fixed point.} \label{fig_arrest_sketch} 
\end{figure}

It is important to determine the implications of the three bifurcations, described above, for the long-time rotational motion of a spheroid. The nature of thin oblate spheroid rotation for $De < De_{c1}(\kappa)$, for $\epsilon$ in the polymeric range, has already been clarified via figure \ref{fig_kappa_epsilon} - the oblate spheroid spirals towards a stable limit cycle starting from almost any initial orientation, with exceptions being the vorticity-aligned orientation, and orientations in the flow-gradient plane. Although rotation arrest occurs for $De_{c1} < De < De_{c2}$, the fixed points that emerge are all unstable, and hence, the oblate spheroid still converges onto the stable limit cycle for long times; as for the prolate spheroid in figure \ref{fig_multi_fail_b}, the drift towards the limit cycle now has a monotonic character to begin with. For $De_{c2} < De < De_{c3}$, a pair of stable nodes in the flow-gradient coexist with a pair of stable limit cycles centered at the vorticity axis fixed points. Finally, for $De_{c3} < De\,(\ll 1)$, the only stable fixed points are the pair of stable nodes in the flow-gradient plane, and a spheroid starting from almost any initial orientation must converge to one of these nodes - the blue curve in figure \ref{fig_multi_fail_d}, in fact, shows the spheroid drifting monotonically towards a stable node in the flow-gradient plane since the chosen $De\,(=0.45)$ is greater than $De_{c3}$. For both $De_{c2} < De < De_{c3}$ and $De > De_{c3}$, the stable manifolds of the pair of interior saddle points serve as the basin boundaries on each unit hemisphere. In the former interval, these invariant curves originate at the two unstable nodes, and separate the basins of attraction of the stable limit cycle, from those pertaining to the pair of stable nodes, on each unit hemisphere, as shown in figure \ref{fig_arrest_sketch_a}; magnified views show the stable manifolds converging onto one of the interior saddle points. In the latter interval, the stable manifolds are a pair of trajectories that spiral out from the vorticity axis, demarcating the basins of attraction corresponding to the individual stable nodes on each unit hemisphere, as shown in figure \ref{fig_arrest_sketch_b}. The pair of magnified views below the main figure show that the spiralling character of the basin boundaries for $De > De_{c3}$ leads to basins that are infinitely interlaced sufficiently close to the vorticity axis.

\section{\bf Conclusions}\label{conclu}

We have given a comprehensive description of the viscoelastic orientation dynamics of a spheroid with aspect ratio $\kappa$, in an ambient simple shear flow, for small $De$. The viscoelastic fluid is modeled as a second-order fluid, in which case the the small-$De$ orientation dynamics is both a function of $\kappa$ and the second-order fluid parameter $\epsilon$. For $\epsilon$ in the range typical of polymer solutions, and for sufficiently small $De$, a prolate spheroid of any aspect ratio drifts towards the vorticity axis for long times. Oblate spheroids, with $\kappa$ greater than an $\epsilon$-dependent threshold, drift towards the flow-gradient plane for sufficiently small $De$. Those with smaller $\kappa$ drift towards a stable limit cycle instead; the threshold aspect ratio varies from $0.2$\,(for $\epsilon = -0.7$) to $0.454$\,(for $\epsilon = -0.5$) over the polymeric range. The above predictions, specifically that of a stable limit cycle\,(kayaking mode) for disk-shaped particles, are in agreement with the early experiments of Mason and co-workers discussed in \S\ref{introsec:expt}. 

Our results deviate from those of \citet{arjun2023}, taken in the limit $c\,De \rightarrow 0$. In this limit, \citet{arjun2023} found a stable limit cycle in the vicinity of the vorticity axis, and a slender fiber with any initial orientation was predicted to approach this limit cycle for long times. This finding is also in disagreement with the conclusions of \citet{leal1975}, and \citet{arjun2023} attributed the  disagreement to the neglect of higher-order singularities that are $O(1/\kappa^2)$ smaller for non-aligned orientations, but that become important during the long period of near-flow alignment, when the leading order line distribution of Stokeslets becomes asymptotically small. The said limit cycle was predicted to result from a balance of oppositely signed torque contributions arising from the leading order Stokeslet distribution, and the aforementioned higher order singularities\,(both contributions are $O(De)$, for $De \rightarrow 0$, and the size of the limit cycle is expected to be $De$-independent in this limit). However, our results are exact, and the expression for the Stokesian velocity field, in terms of vector spheroidal harmonics, must evidently account for both the leading order Stokeslet distribution, and the contribution of higher order singularities, during the aligned phase; indeed,  for $\kappa \gg 1$, $\bm{u}_{1s} \sim O(\ln \kappa)^{-1}$, corresponding to the line distribution of Stokeslets, and $\bm{u}_{2s}-\bm{u}_{5s} \sim O(\kappa^{-2})$, corresponding to higher order singularities not included within the usual slender body theory framework. Yet, our results do not indicate the presence of any intermediate limit cycle for $\epsilon$ in the polymeric range. It is shown in \S\ref{appB} that an independent calculation of the viscoelastic torque, using slender body theory and a simpler Fourier space approach, gives results that are identical to those of \citet{leal1975}, but for a factor of $4$; these results also match up to the large-$\kappa$ limit of the exact expressions given in \S \ref{Problemform:visco} and \S \ref{appA}. Further, the stability of the prolate spinning mode found here is also consistent with the computations of \citet{davino2014}, who found prolate spheroids with aspect ratios ranging from $2$ to $16$ to drift towards the spinning mode, for sufficiently small $De$, starting from any initial orientation; note that the highest aspect ratio examined\,($16$) is likely large enough for an approach based on slender body theory to be accurate. The said authors used a Giesekus fluid with the additional parameter $\alpha$ that multiplies the term, quadratic in the polymeric stress, in the constitutive equation; $\alpha = 0\,(\epsilon = -1/2)$ corresponds to an Oldroyd-B fluid. It may be shown that $\displaystyle N_2/N_1 = -\alpha/2$, and for small $De$, the Giesekus fluid therefore maps to a second-order fluid with $\epsilon = 1/(\alpha - 2)$; $\alpha = 0.2$ used by \citet{davino2014} corresponds to $\epsilon \approx -0.56$ in the polymeric range.

For $De$ exceeding a small but finite threshold of $O(1/\kappa)$, slender prolate spheroids stop rotating about the vorticity axis. Exceptional cases of orientations within the flow-gradient plane exhibit a complete rotation arrest, although these arrested states are unstable to off-plane perturbations - perturbed spheroids exhibit an initially monotonic, and eventually spiralling, drift towards the vorticity axis, again consistent with the observations of \citet{bartram1975}. Flat oblate spheroids restricted to the flow-gradient plane, with $\kappa$ smaller than an $\epsilon$-dependent threshold\,(that, as mentioned above, varies from $0.2$ to $0.454$ in the polymeric range), exhibit an analogous rotation arrest above a threshold $De$ of $O(\kappa)$; off-plane perturbations now lead to a drift towards a stable limit cycle\,(small amplitude kayaking mode) 
around the vorticity axis. 
In contrast to prolate spheroids, however, the primary rotation-arrest for oblate spheroids is followed by secondary and tertiary bifurcations at higher $De$, also of $O(\kappa)$, and therefore within the purview of the theory. These latter bifurcations allow for the emergence of stable rotation-arrested states in the flow-gradient plane. Interestingly, this sequence of bifurcations resembles an analogous sequence of bifurcations found by \citet{arjun2023}, at finite $De$, for a prolate spheroid. As evident from the discussion of the literature in section \ref{intro}, work on oblate spheroids has been much more limited in comparison to that on prolate spheroids, and we hope that our findings stimulate future experimental and computational efforts.

It is important to mention that the $O(1/\kappa)$ scaling of the threshold $De$ above, for slender spheroids, arises due to the viscoelastic torque being $O(1)$ in the large-$\kappa$ limit, this in turn being due to to a logarithmically dominant contribution from the matching region compensating for the $O(\ln \kappa)^{-1}$ scaling of the disturbance velocity field. The details are provided in Appendix \S\ref{appB}, where we also show consistency with the results of \citet{leal1975} by evaluating the $\kappa$-dependent coefficient therein. However, this leading order contribution is proportional to $(1+2\epsilon)$, and therefore, vanishes for an Oldroyd-B fluid\,(or, in experimental terms, for Boger fluids). From (\ref{eq4p2}), the resulting $De_c$ is then found to be $O(\ln \kappa/\kappa)$, which is logarithmically larger, so as to compensate for the logarithmic smallness of the torque for $\epsilon = -1/2$; the latter was evident in the logarithmically small estimate for the orbital drift, $\Delta C \sim De/[\sqrt{\xi_0-1}\ln (\xi_0-1)]$, provided in section \ref{Results:epsilon_kappa}. On the whole, the above aspects again emphasize the importance of a non-zero $N_2$ to the small-$De$ orientation dynamics; a zero $N_2$ renders the elastic effects logarithmically smaller in the aspect ratio. It is worth adding that the above is also consistent with the $O(\ln \kappa)^{-1}$ torque scaling found by \citet{arjun2023}, for the small-$De$ limit of an Oldroyd-B fluid. Nevertheless, as already indicated, our detailed predictions both below and beyond rotation arrest differ from the said effort, and the reasons for this discrepancy are not clear at the moment.

One of the key conclusions that emerges from the analysis here is that stationary orientations in the flow-vorticity plane, intermediate between the flow and vorticity directions, that have been observed in both experiments \citep{iso1996a,iso1996b} and computations \citep{davino2014} of prolate spheroids in a simple shear flow, appear outside the purview of a small-$De$ analysis. The only such orientations found here correspond to the interior saddle points shown in figure \ref{fig_rotation_eps_m2by3}. Such orientations arise for a sufficiently flat oblate spheroid, and only after a secondary bifurcation. Importantly, however, a saddle point represents an unstable equilibrium, and is therefore susceptible to small perturbations. Any perturbation will lead to the (oblate)\,spheroid eventually exhibiting a kayaking mode about the vorticity axis, or adopting a stationary orientation in the flow-gradient plane. Thus, an explanation of the aforesaid flow-vorticity plane equilibria would seem to be necessitate an examination of finite-$De$ orientation dynamics in the so-called ultra-dilute limit. This corresponds to the polymer concentration being small enough that the flow field stretching and aligning the polymeric molecules still has a Newtonian character; although, unlike the ordered fluid paradigm examined here, the relaxation time of the polymers can be comparable to, or larger than, the flow time scale\,\citep{RemmelgasLeal_1998}. This approach has been used in multiple scenarios before, including the efforts of \citet{vivek2015} for a sedimenting fiber, and those of \citet{harlen1993} and \citet{arjun2023}, discussed above and in the introduction, which examine orientation dynamics of a slender fiber in simple shear flow of an Oldroyd-B fluid. The polymeric stress at a given point now begins to depend on the history of stretch encountered by a polymer, as it is convected along fluid pathlines. It has recently been shown by \citet{banerjee2021} that the configuration of Stokesian pathlines around a spheroid, rotating along a Jeffery orbit other than the spinning mode, can have a very intricate structure. The structure is that of a chaotic repeller, with the residence time of a fluid element in a given neighbhorhood of the spheroid exhibiting a fractal dependence on initial conditions. The implications for the nature of the polymeric stress field, and thence, spheroid rotation, for large $De$ in the ultra-dilute limit, is an interesting avenue to explore in future. 

As mentioned in the introduction, the linear-flow orientation dynamics analyzed here will also serve as an input towards both rheology and shear-induced migration calculations. The inertial case has already been analyzed along these lines - the calculation of the leading order inertial rheology, of a suspension of spheroids, was done in \cite{vivek2016} and \cite{Marath_Rapids_2017}, while the migration calculation has recently appeared in \citet{prateek2023}. Extensions of these results to the inertio-elastic regime will be reported in the near future.
 \section*{Acknowledgment}
This work was partially supported by the SERB grant CRG/2020/004137.
 
\section*{Declaration of Interests}
The authors report no conflict of interest.
\appendix
\section{The viscoelastic and inertial shape functions for a spheroid in simple shear flow}\label{appA}

\subsection{\bf The viscoelastic aspect-ratio functions} \label{shape_fun_visco}

Each of the aspect-ratio-dependent functions in (\ref{eq41}) and (\ref{eq42}) can be written as the following sum of co-rotational and quadratic contributions:
\begin{equation}\label{eqA1}
     F_{i}(\xi_{0},\epsilon)= \epsilon F_{i,c}(\xi_{0})+ (1+\epsilon) F_{i,q}(\xi_{0},\epsilon),
 \end{equation}
  \begin{equation}\label{eqA2}
     G_{i}(\xi_{0},\epsilon)=\epsilon G_{i,c}(\xi_{0})+ (1+\epsilon)G_{i,q}(\xi_{0}),
 \end{equation}
 where the suffix $i$ takes values from $1$ to $6$, and from $1$ to $4$, for the $F_i$'s and $G_i$'s, respectively. For a prolate spheroid, the co-rotational contributions are given by:
{\allowdisplaybreaks
    \begin{align}\nonumber
        &F_{1,c}(\xi_{0})= \bigg(\xi_{0}^2 \big(270 \xi_{0}^6-591 \xi_{0}^4+397 \xi_{0}^2-74\big)-3 \xi_{0} \big(\xi_{0}^2-1\big)^2 \big(90 \xi_{0}^6+13 \xi_{0}^4-62 \xi_{0}^2\\\nonumber
        &+15\big) (\coth ^{-1} \xi_{0})^3+\big(-810 \xi_{0}^9+1683 \xi_{0}^7-950 \xi_{0}^5+21 \xi_{0}^3+52 \xi_{0}\big) \coth ^{-1}\xi_{0}+\big(810 \xi_{0}^{10}\\\nonumber
        &-1593 \xi_{0}^8+559 \xi_{0}^6+479 \xi_{0}^4-277 \xi_{0}^2+22\big) (\coth ^{-1} \xi_{0})^2\bigg)/\bigg(16 \big(1-2 \xi_{0}^2\big)^2 \big(9 \xi_{0} \big(\xi_{0}^2-1\big)^3\\\nonumber
        &\big(3 \xi_{0}^2-1\big) (\coth ^{-1} \xi_{0})^3-3 \xi_{0}^2 \big(9 \xi_{0}^4-21 \xi_{0}^2+10\big)+\xi_{0} \big(81 \xi_{0}^6-216 \xi_{0}^4+159 \xi_{0}^2-28\big) \coth ^{-1}\xi_{0}\\\label{eqA3}
        &-3 \big(27 \xi_{0}^8-81 \xi_{0}^6+79 \xi_{0}^4-27 \xi_{0}^2+2\big) (\coth ^{-1} \xi_{0})^2\big)\bigg),
    \end{align}}
\begin{align}\label{eqA4}
        F_{2,c}(\xi_{0})&=0,
    \end{align}
{\allowdisplaybreaks
    \begin{align}\nonumber
        &F_{3,c}(\xi_{0})=  \bigg(\xi_{0}^2 \big(90 \xi_{0}^6-225 \xi_{0}^4+155 \xi_{0}^2-22\big)-3 \xi_{0} \big(\xi_{0}^2-1\big)^2 \big(30 \xi_{0}^6-5 \xi_{0}^4-26 \xi_{0}^2+9\big) (\coth ^{-1} \xi_{0})^3\\\nonumber
        &+\big(-270 \xi_{0}^9+645 \xi_{0}^7-402 \xi_{0}^5+19 \xi_{0}^3+12 \xi_{0}\big) \coth ^{-1}\xi_{0}+\big(270 \xi_{0}^{10}-615 \xi_{0}^8+289 \xi_{0}^6+161 \xi_{0}^4\\\nonumber
        &-115 \xi_{0}^2+10\big) (\coth ^{-1} \xi_{0})^2\bigg) \bigg/\bigg(16 \big(1-2 \xi_{0}^2\big)^2 \big(9 \xi_{0} \big(\xi_{0}^2-1\big)^3\big(3 \xi_{0}^2-1\big) (\coth ^{-1} \xi_{0})^3-3 \xi_{0}^2 \big(9 \xi_{0}^4\\ \nonumber
        &-21 \xi_{0}^2+10\big)+\xi_{0} \big(81 \xi_{0}^6-216 \xi_{0}^4+159 \xi_{0}^2-28\big) \coth ^{-1}\xi_{0}-3 \big(27 \xi_{0}^8-81 \xi_{0}^6+79 \xi_{0}^4-27 \xi_{0}^2\\\label{eqA5}
        &+2\big) (\coth ^{-1} \xi_{0})^2\big)\bigg),
    \end{align}}
{\allowdisplaybreaks
     \begin{align}\label{eqA6}
         &F_{4,c}(\xi_{0})=F_{3,c}(\xi_{0}),\\\label{eqA7}
          & F_{5,c}(\xi_{0})= F_{6,c}(\xi_{0})=-\frac{F_{3,c}(\xi_{0})}{2},\\\label{eqA8}
          & G_{1,c}(\xi_{0})=0,\\\label{eqA9}
     &G_{2,c}(\xi_{0})=0,
 \end{align}}
 {\allowdisplaybreaks
     \begin{align}\nonumber
      &  G_{3,c}(\xi_{0})=\bigg( -90 \xi_{0}^8+225 \xi_{0}^6-155 \xi_{0}^4+22 \xi_{0}^2+3 \big(\xi_{0}^2-1\big)^2 \big(30 \xi_{0}^6-5 \xi_{0}^4-26 \xi_{0}^2\\\nonumber
        &+9\big) \xi_{0} (\coth ^{-1} \xi_{0})^3+\big(270 \xi_{0}^8-645 \xi_{0}^6+402 \xi_{0}^4-19 \xi_{0}^2-12\big) \xi_{0} \coth ^{-1}\xi_{0}-\big(270 \xi_{0}^{10}-615 \xi_{0}^8\\\nonumber
        &+289 \xi_{0}^6+161 \xi_{0}^4-115 \xi_{0}^2+10\big) (\coth ^{-1} \xi_{0})^2\bigg) \bigg/\bigg(4 \big(1-2 \xi_{0}^2\big)^2 \big(9 \xi_{0} \big(\xi_{0}^2-1\big)^3 \big(3 \xi_{0}^2-1\big)(\coth ^{-1} \xi_{0})^3\\\nonumber
        & -3 \xi_{0}^2 \big(9 \xi_{0}^4-21 \xi_{0}^2+10\big)+\xi_{0} \big(81 \xi_{0}^6-216 \xi_{0}^4+159 \xi_{0}^2-28\big) \coth ^{-1}\xi_{0}-3 \big(27 \xi_{0}^8-81 \xi_{0}^6+79 \xi_{0}^4\\\label{eqA10}
        &-27 \xi_{0}^2+2\big) (\coth ^{-1} \xi_{0})^2\big)\bigg),
     \end{align}}
 {\allowdisplaybreaks
 \begin{align}\label{eqA11}
        G_{4,c}(\xi_{0})=-G_{3,c}(\xi_{0}).
     \end{align}}
 The quadratic contributions are given by:
{\allowdisplaybreaks
    \begin{align}\nonumber
       &F_{1,q}(\xi_{0})=4\bigg(-54 \xi_{0}^8+5775 \xi_{0}^6-10873 \xi_{0}^4+5158 \xi_{0}^2+3 \big(\xi_{0}^2-1\big)^2 \big(18 \xi_{0}^6+373 \xi_{0}^4-842 \xi_{0}^2\\\nonumber
        &+171\big) \xi_{0} (\coth^{-1}\xi_{0})^3+\big(162 \xi_{0}^8-10539 \xi_{0}^6+21578 \xi_{0}^4-12561 \xi_{0}^2+1356\big) \xi_{0} \coth^{-1}\xi_{0}\\\nonumber
        &+\big(-162 \xi_{0}^{10}+3753 \xi_{0}^8-5995 \xi_{0}^6+1477 \xi_{0}^4+913 \xi_{0}^2+14\big) (\coth^{-1}\xi_{0})^2\bigg) \bigg/ \bigg(64 \big(1-2 \xi_{0}^2\big)^2 \\\nonumber
        &\big(-3 \xi_{0}^2+3 \big(\xi_{0}^2-1\big) \xi_{0} \coth^{-1}\xi_{0}+2\big) \big(-3 \xi_{0}^3+3 \big(\xi_{0}^2-1\big)^2 \coth^{-1}\xi_{0}+5 \xi_{0}\big)\\\label{eqA12}
        & \big(\big(3 \xi_{0}^2-1\big) \coth^{-1}\xi_{0}-3 \xi_{0}\big)\bigg),
    \end{align}}
\begin{align}\label{eqA13}
         F_{2,q}(\xi_{0})=0,
 \end{align}
{\allowdisplaybreaks
    \begin{align}\nonumber
       & F_{3,q}(\xi_{0})=4\bigg(-90 \xi_{0}^8+2097 \xi_{0}^6-4487 \xi_{0}^4+2474 \xi_{0}^2+3 \big(\xi_{0}^2-1\big)^2 \big(30 \xi_{0}^6+91 \xi_{0}^4-326 \xi_{0}^2\\\nonumber
        &+165\big) \xi_{0} (\coth^{-1}\xi_{0})^3+\big(270 \xi_{0}^8-4101 \xi_{0}^6+9030 \xi_{0}^4-6431 \xi_{0}^2+1236\big) \xi_{0} \coth^{-1}\xi_{0}\\\nonumber
        &+\big(-270 \xi_{0}^{10}+1911 \xi_{0}^8-3109 \xi_{0}^6+1483 \xi_{0}^4-17 \xi_{0}^2+2\big) (\coth^{-1}\xi_{0})^2\bigg) \bigg/ \bigg(64 \big(1-2 \xi_{0}^2\big)^2 \\\nonumber
        &\big(-3 \xi_{0}^2+3 \big(\xi_{0}^2-1\big) \xi_{0} \coth^{-1}\xi_{0}+2\big) \big(-3 \xi_{0}^3+3 \big(\xi_{0}^2-1\big)^2 \coth^{-1}\xi_{0}+5 \xi_{0}\big)\\\label{eqA14}
        & \big(\big(3 \xi_{0}^2-1\big) \coth^{-1}\xi_{0}-3 \xi_{0}\big)\bigg),
    \end{align}}
{\allowdisplaybreaks
    \begin{align}\label{eqA15}
    &F_{4,q}(\xi_{0})=F_{3,q}(\xi_{0}),\\\label{eqA16}
    &F_{5,q}(\xi_{0})= F_{6,q}(\xi_{0})=-\frac{F_{3,q}(\xi_{0})}{2},\\\label{eqA17}
     &G_{1,q}(\xi_{0})=0,\\\label{eqA18}
    & G_{2,q}(\xi_{0})= 0,
\end{align}}
{\allowdisplaybreaks
    \begin{align}\nonumber
       & G_{3,q}(\xi_{0})=4\bigg( \xi_{0}^2 \big(90 \xi_{0}^6-2097 \xi_{0}^4+4487 \xi_{0}^2-2474\big)-3 \xi_{0} \big(\xi_{0}^2-1\big)^2 \big(30 \xi_{0}^6+91 \xi_{0}^4\\\nonumber
        &-326 \xi_{0}^2+165\big) (\coth^{-1}\xi_{0})^3+\big(-270 \xi_{0}^9+4101 \xi_{0}^7-9030 \xi_{0}^5+6431 \xi_{0}^3-1236 \xi_{0}\big) \coth^{-1}\xi_{0}\\\nonumber
        &+\big(270 \xi_{0}^{10}-1911 \xi_{0}^8+3109 \xi_{0}^6-1483 \xi_{0}^4+17 \xi_{0}^2-2\big)(\coth^{-1}\xi_{0})^2\bigg) \bigg/ \bigg(16 \big(1-2 \xi_{0}^2\big)^2 \\\nonumber
        &\big(-3 \xi_{0}^2+3 \big(\xi_{0}^2-1\big) \xi_{0} \coth^{-1}\xi_{0}+2\big) \big(-3 \xi_{0}^3+3 \big(\xi_{0}^2-1\big)^2 \coth^{-1}\xi_{0}+5 \xi_{0}\big) \\\label{eqA19}
        &\big(\big(3 \xi_{0}^2-1\big) \coth^{-1}\xi_{0}-3 \xi_{0}\big) \bigg),
     \end{align}}
\begin{align}\label{eqA20}
          G_{4,q}(\xi_{0})=-G_{3,q}(\xi_{0}).
  \end{align}
 The corresponding contributions for an oblate spheroid may be obtained from the above expressions by using the transformation $\xi_{0} \leftrightarrow \mathrm{i} \bar{\xi_{0}}$, $d \leftrightarrow -\mathrm{i} d$. 

\begin{figure}
		\centering
		\subfigure[]{\includegraphics[scale=0.325]{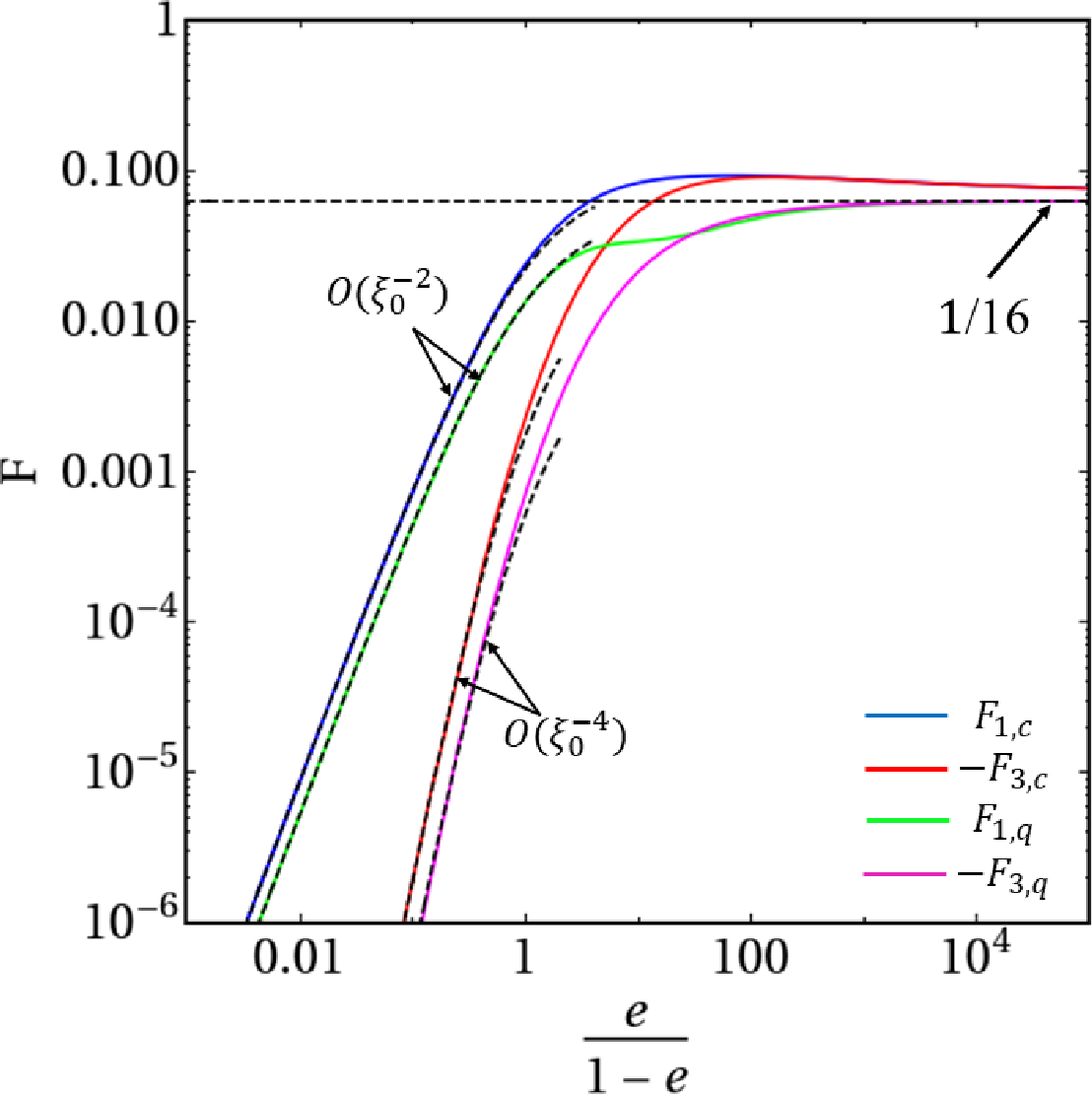}\label{fig_shape_fun_prolate_a}}
		\subfigure[]{\includegraphics[scale=0.325]{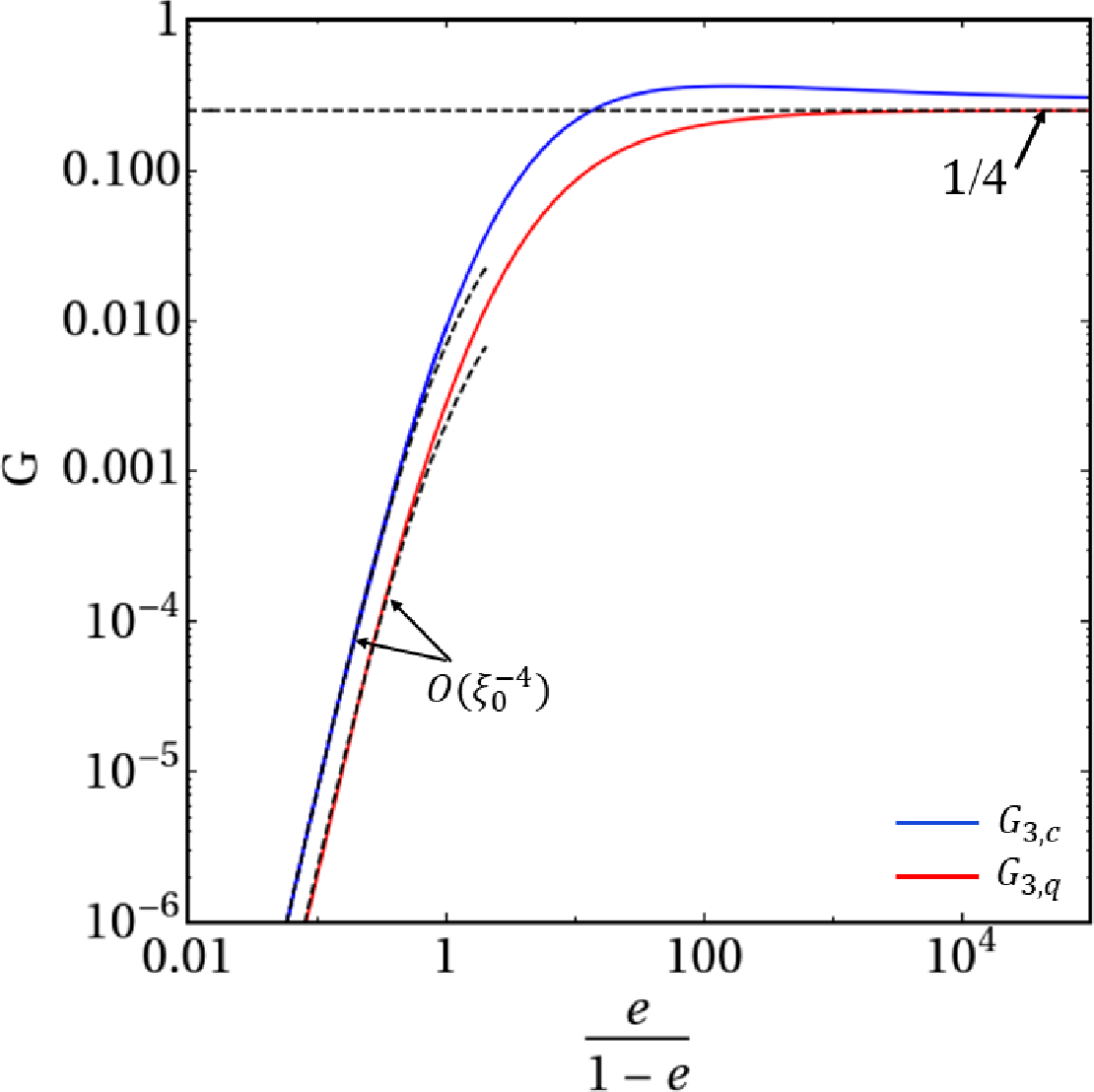}\label{fig_shape_fun_prolate_b}}
		\caption{The viscoelastic aspect-ratio functions plotted against $e/(1-e)$, where $e = 1/\xi_0$ represents the eccentricity of the prolate spheroid. (a) Functions $F_{i,c}$ and $F_{i,q}$ which contribute to $(\dot{\theta}_{j})_{De}$, and (b) functions $G_{i,c}$ and $G_{i,q}$, which contribute to $(\dot{\phi}_{j})_{De}$. The black dashed lines correspond to the limiting cases of slender fiber and the near sphere asymptotes.}\label{fig_shape_fun_prolate}
\end{figure}

\begin{figure}
		\centering
		\subfigure[]{\includegraphics[scale=0.325]{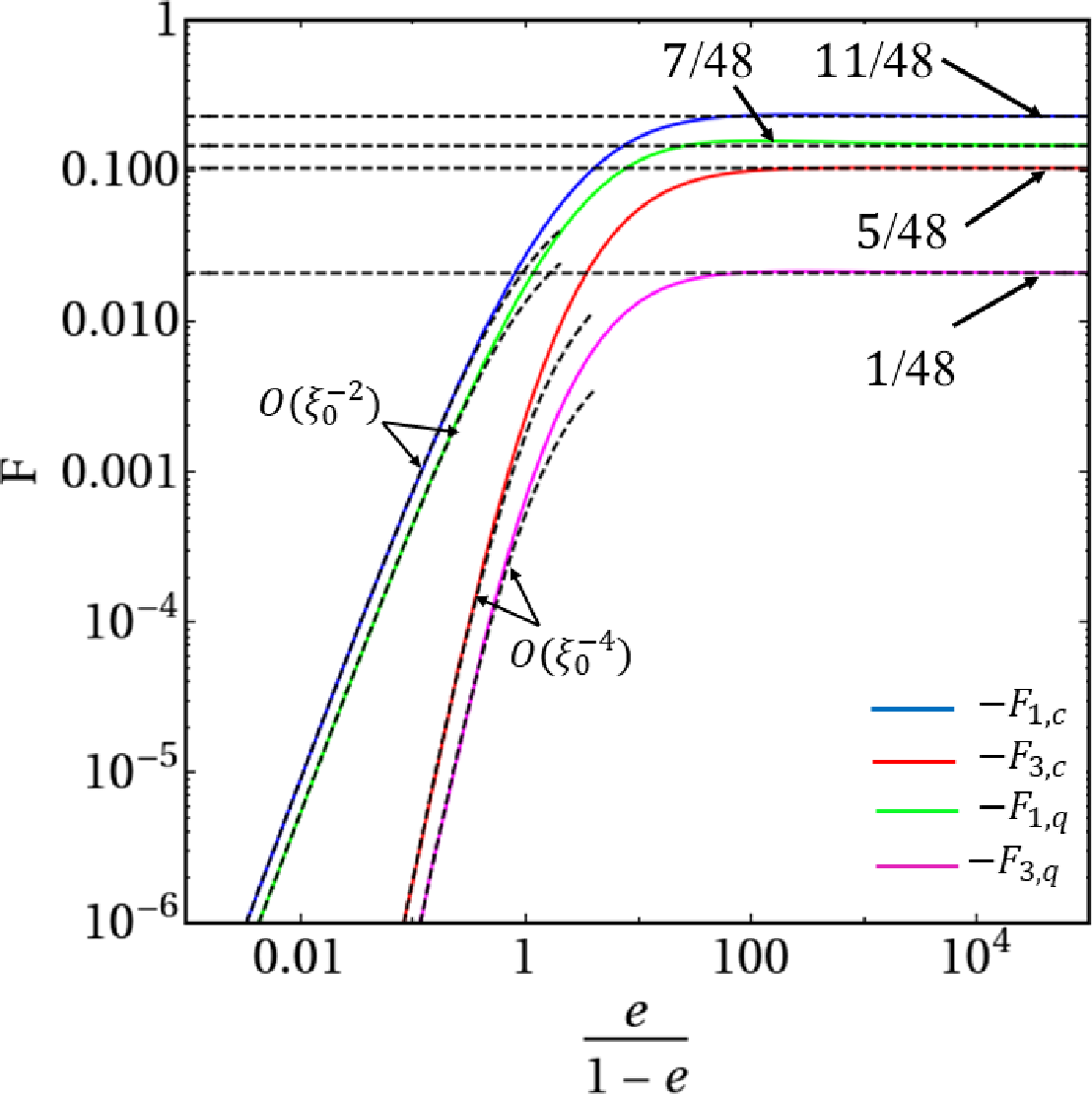}\label{fig_shape_fun_oblate_a}}
		\subfigure[]{\includegraphics[scale=0.325]{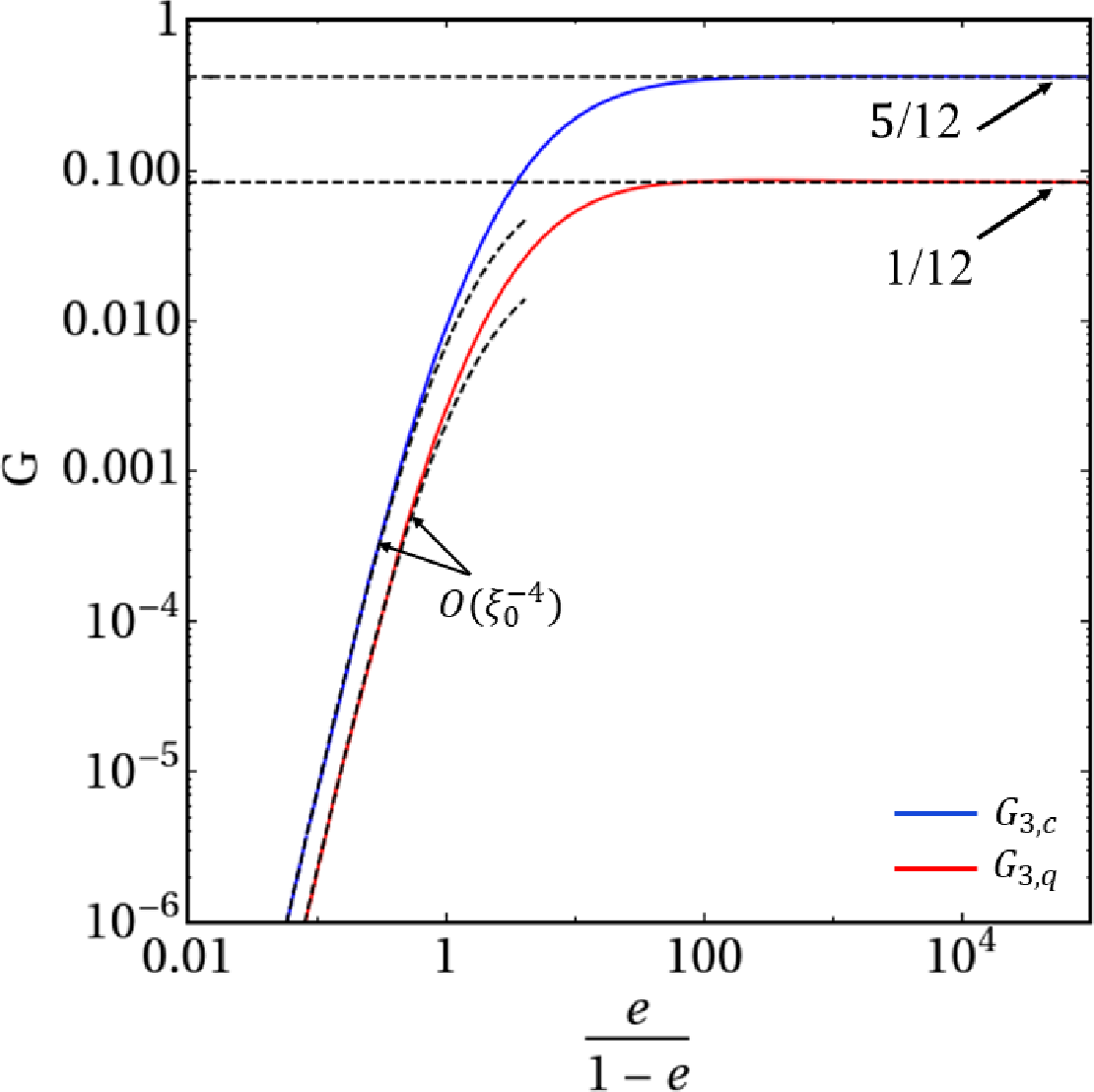}\label{fig_shape_fun_oblate_b}}
		\caption{The viscoelastic shape-dependent functions are plotted against $e/(1-e)$, where $e = 1/\xi_0$ represents the eccentricity of the oblate spheroid. (a) Functions $F_{i,c}$ and $F_{i,q}$ which contribute to $(\dot{\theta}_{j})_{De}$, and (b) functions $G_{i,c}$ and $G_{i,q}$, which contributes to $(\dot{\phi}_{j})_{De}$. The black dashed lines indicate the limiting cases corresponding to a flat disk and the near sphere asymptotes}\label{fig_shape_fun_oblate}
\end{figure}

 One may find simplified forms of the above exact 
 expressions in various asymptotic limits. In the near-sphere limit\,($\xi_{0} \rightarrow \infty$), one finds $F_{1,c} \rightarrow \pm 5/56 \xi_{0}^{2}$, $F_{3,c} \rightarrow -395/14112 \xi_{0}^{4}$, $G_{3,c} \rightarrow 395 /3528 \xi_{0}^{4}$,  $F_{1,q} \rightarrow \pm 3/56 \xi_{0}^{2}$, $F_{3,q} \rightarrow -59 /7056 \xi_{0}^{4}$  and $G_{3,q} \rightarrow 59 /1764 \xi_{0}^{4}$. All contributions go to zero, as expected, since a sphere cannot drift by symmetry; the plus and minus signs correspond to approaching the sphere from the prolate and oblate sides, respectively. In the slender-fiber limit ($\xi_{0} \rightarrow 1 $), one finds $F_{1,c} \rightarrow 1/16$, $F_{3,c} \rightarrow -1/16$, $G_{3,c} \rightarrow 1/4$,  $F_{1,q} \rightarrow 1/16$, $F_{3,q} \rightarrow -1/16$  and $G_{3,q} \rightarrow 1/4$. In the flat-disk limit (setting $\xi_{0} \rightarrow 1 $ after the prolate-oblate transformation), one finds $F_{1,c} \rightarrow -11/48$, $F_{3,c} \rightarrow -5/48$, $G_{3,c} \rightarrow 5/12$,  $F_{1,q} \rightarrow -7/48$, $F_{3,q} \rightarrow -1/48$  and $G_{3,q} \rightarrow 1/12$. Figure \ref{fig_shape_fun_prolate} illustrates the aspect-ratio-dependent functions for a prolate spheroid as a function of $e/(1-e)$ where the eccentricity $e=1/\xi_{0}$. Figure \ref{fig_shape_fun_prolate_a} plots the corotational and quadratic contributions of the functions that enter the expression for $(\dot{\theta}_{j})_{De}$, while figure \ref{fig_shape_fun_prolate_b} plots those that enter $(\dot{\phi}_{j})_{De}$, both for a prolate spheroid. While all aspect-ratio functions approach $O(1)$ values for $e/(1-e) \rightarrow \infty$, as evident from the slender-body limiting forms given above, the convergence is slow and has a logarithmic character. It may be shown that the quadratic stress contributions in the slender fiber limit scale as $O(1)+O(\ln \kappa/\kappa^2)+\cdots$, and the slow convergence therefore arises from the corotational contributions which scale as $O(1)+O(1/\ln \kappa)$ in this limit. Figures \ref{fig_shape_fun_oblate_a} and \ref{fig_shape_fun_oblate_b} plot the oblate analogs of the aspect-ratio functions shown in figures \ref{fig_shape_fun_prolate_a} and \ref{fig_shape_fun_prolate_b}. In this case all aspect-ratio-dependent show a rapid transition from the near-sphere to the flat-disk asymptote. 


\subsection{\bf The inertial aspect-ratio functions}\label{shape_funs_inertia}

Although the inertial orientation dynamics was originally analyzed in \cite{vivek2016} with the coefficients of the trigonometric terms in the inertial analogs of (\ref{eq41}) and (\ref{eq42}) being given in closed form, the essential point about there being only four independent aspect ratio functions underlying these coefficients, as in (\ref{eq40}), has only been mentioned here. Hence, for the sake of completeness, we give the closed-form expressions for the inertial analogs of the $\beta_i$'s\,($i$ varying from $1$ to $4$) in (\ref{eq40}), for a prolate spheroid, below:
{\allowdisplaybreaks
    \begin{align}\nonumber
        &\beta_{1}=\bigg(-9 \xi_{0} \big(\xi_{0}^2-1\big)^3 \big(60 \xi_{0}^8-100 \xi_{0}^6+121 \xi_{0}^4-98 \xi_{0}^2+25\big) (\coth ^{-1}\xi_{0})^3+\xi_{0}^2 \big(540 \xi_{0}^{10}+180 \xi_{0}^8\\\nonumber
        &-3903 \xi_{0}^6+5092 \xi_{0}^4-2279 \xi_{0}^2+342\big)+3 \big(\xi_{0}^2-1\big)^2 \big(540 \xi_{0}^{10}-540 \xi_{0}^8+145 \xi_{0}^6-400 \xi_{0}^4+245 \xi_{0}^2\\\nonumber
        &-22\big) (\coth ^{-1}\xi_{0})^2+\big(-1620 \xi_{0}^{13}+2160 \xi_{0}^{11}+4017 \xi_{0}^9-8531 \xi_{0}^7+4783 \xi_{0}^5-853 \xi_{0}^3\\\nonumber
        &+44 \xi_{0}\big) \coth ^{-1}\xi_{0}\bigg) \bigg/ \bigg(30 \xi_{0}^2 \big(2 \xi_{0}^2-1\big)^3 \big(-3 \xi_{0}^2+3 \big(\xi_{0}^2-1\big) \xi_{0} \coth ^{-1}\xi_{0}+2\big) \\\label{eqA21}
        &\big(-3 \xi_{0}^3+3 \big(\xi_{0}^2-1\big)^2 \coth ^{-1}(\xi_{0})+5 \xi_{0}\big) \big(\big(3 \xi_{0}^2-1\big) \coth ^{-1}\xi_{0}-3 \xi_{0}\big)\bigg),
    \end{align}}
{\allowdisplaybreaks
    \begin{align}\nonumber   
       &\beta_{2}=\bigg(3 \xi_{0} \big(30 \xi_{0}^4-27 \xi_{0}^2+5\big) \big(\xi_{0}^2-1\big)^4 (\coth ^{-1}\xi_{0})^3+\xi_{0}^2 \big(-90 \xi_{0}^8+111 \xi_{0}^6+98 \xi_{0}^4-133 \xi_{0}^2+18\big)\\\nonumber 
       &-\big(270 \xi_{0}^8-453 \xi_{0}^6+294 \xi_{0}^4-85 \xi_{0}^2+6\big) \big(\xi_{0}^2-1\big)^2 (\coth ^{-1}\xi_{0})^2+\xi_{0} \big(270 \xi_{0}^{10}-663 \xi_{0}^8+493 \xi_{0}^6\\\nonumber 
       &-117 \xi_{0}^4+29 \xi_{0}^2-12\big) \coth ^{-1}\xi_{0}\bigg) \bigg/ \bigg(10 \big(\xi_{0}-2 \xi_{0}^3\big)^2 \big(-3 \xi_{0}^2+3 \big(\xi_{0}^2-1\big) \xi_{0} \coth ^{-1}\xi_{0}+2\big)\\\label{eqA22}
       & \big(-3 \xi_{0}^3+3 \big(\xi_{0}^2-1\big)^2 \coth ^{-1}\xi_{0}+5 \xi_{0}\big) \big(\big(3 \xi_{0}^2-1\big) \coth ^{-1}\xi_{0}-3 \xi_{0}\big)\bigg),
    \end{align}}
{\allowdisplaybreaks
    \begin{align}\nonumber
   &\beta_{3}= \bigg(2 \big(\xi_{0}^2-1\big) \big(-21 \xi_{0}^5+20 \xi_{0}^3+9 \big(\xi_{0}^2-1\big)^3 \xi_{0} (\coth ^{-1}\xi_{0})^2+3 \big(4 \xi_{0}^6-\xi_{0}^4-4 \xi_{0}^2+1\big) \coth ^{-1}\xi_{0}\\\label{eqA23}
       &-3 \xi_{0}\big)\bigg) \bigg/ \bigg(5 \big(\xi_{0}-2 \xi_{0}^3\big)^2 \big(-3 \xi_{0}^2+3 \big(\xi_{0}^2-1\big) \xi_{0} \coth ^{-1}\xi_{0}+2\big) \big(-3 \xi_{0}^3+3 \big(\xi_{0}^2-1\big)^2 \coth ^{-1}\xi_{0}+5 \xi_{0}\big)\bigg),
    \end{align}}
{\allowdisplaybreaks
    \begin{align}\nonumber
     &\beta_{4}=\bigg(2 \big(\xi_{0}^2-1\big) \big(9 \xi_{0} \big(\xi_{0}^2-1\big)^2 \big(5 \xi_{0}^6-7 \xi_{0}^4+7 \xi_{0}^2-3\big) (\coth ^{-1}\xi_{0})^2+\xi_{0} \big(45 \xi_{0}^8+57 \xi_{0}^6-181 \xi_{0}^4\\\nonumber
     &+82 \xi_{0}^2-7\big)+3 \big(-30 \xi_{0}^{10}+32 \xi_{0}^8+4 \xi_{0}^6+9 \xi_{0}^4-20 \xi_{0}^2+5\big) \coth ^{-1}\xi_{0}\big)\bigg) \bigg/ \bigg(15 \xi_{0}^2 \big(2 \xi_{0}^2-1\big)^3 \\\label{eqA24}
     &\big(-3 \xi_{0}^2+3 \big(\xi_{0}^2-1\big) \xi_{0} \coth ^{-1}\xi_{0}+2\big) \big(-3 \xi_{0}^3+3 \big(\xi_{0}^2-1\big)^2 \coth ^{-1}\xi_{0}+5 \xi_{0}\big)\bigg)
    \end{align}}
The corresponding expressions for an oblate spheroid can be obtained by first reverting to the dimensional form of the angular velocity, and then applying the transformation $\xi_{0} \leftrightarrow \mathrm{i} \bar{\xi}_{0}$, $d \leftrightarrow -\mathrm{i} d$.
\subsection{\bf The shape functions in the orbital drift interpretation}\label{ItoJ_expr}
Herein, we give the expressions for the $I_i$'s and $J_i$'s that appear in the orbital drift $\Delta $C\,(equation (\ref{eq3p8}) in the main manuscript): 
\begin{eqnarray}\label{eqA25}
    I_{1}&=&2 \pi,\\\label{eqA26}
     I_{2}&=&\frac{2 \pi  (\kappa -1)}{\kappa +1},\\\label{eqA27}
    I_{3}&=&2 \pi  \left( \frac{2}{\sqrt{\left(C^2+1\right) \left(C^2 \kappa ^2+1\right)}}-1\right), \\\label{eqA28}
    I_{4}&=&2 \pi  \left(\frac{\kappa -1}{\kappa +1}\right)^2,\\ \nonumber
    I_{5}+I_{6}&=&-\bigg(4 \pi  \big(2 \kappa ^2 \big(3 \sqrt{\big(C^2+1\big) \big(C^2 \kappa ^2+1\big)}-8 C^2-6\big)+(4 \kappa +1) \sqrt{\big(C^2+1\big) \big(C^2 \kappa ^2+1\big)}\\ \nonumber
    && +4 \big(4 C^2+1\big) \kappa ^3 \sqrt{\big(C^2+1\big) \big(C^2 \kappa ^2+1\big)}+\kappa ^4 \big(\sqrt{\big(C^2+1\big) \big(C^2 \kappa ^2+1\big)}-16 \big(C^4+C^2\big)\\ \label{eqA29}
    && -2\big)-2\big) \bigg) \bigg/\bigg(\big(\kappa ^2-1\big)^2 \sqrt{\big(C^2+1\big) \big(C^2 \kappa ^2+1\big)}\bigg),
\end{eqnarray}
\begin{eqnarray} \label{eqA30}
      J_{1}&=&\frac{\pi  (\kappa -1)}{\kappa +1}, \\ \label{eqA31}
      J_{2}&=&-\frac{\pi  \left(-4 \sqrt{\left(C^2+1\right) \left(C^2 \kappa ^2+1\right)}+C^2 (\kappa +1)^2+4\right)}{C^2 \left(\kappa ^2-1\right)}, \\\label{eqA32}
      J_{3}&=&\pi  \left(\frac{\kappa -1}{\kappa +1}\right)^2,\\ \nonumber
     J_{4}&=&-\bigg(\pi  \big(8 C^4 \kappa ^3+C^2 \big((\kappa +1)^4-8 \kappa ^2 \sqrt{\big(C^2+1 \big) \big(C^2 \kappa ^2+1\big)}\big)-4 \big(\kappa ^2+1\big) \big(\sqrt{\big(C^2+1\big) \big(C^2 \kappa ^2+1\big)}\\ \label{eqA33}
    &&-1 \big)\big)\bigg) \bigg/ \bigg(C^2 \left(\kappa ^2-1\right)^2 \bigg).
\end{eqnarray}
Note that since $F_5(\xi_{0})=F_{6}(\xi_{0})$, only the combination $I_{5}+I_{6}$, given above, enters the expression for $\Delta C$.

\subsection{\bf The regime boundaries in the $\kappa-\epsilon$ plane}\label{appC}

One of the boundaries in figure \ref{fig_kappa_epsilon} is the sphere-line\,($\kappa = 1$) that separates the tumbling and spinning regimes\,(except in the neighborhood of the point $(-3/8,0)$) on the prolate and oblate sides of the $\kappa-\epsilon$ plane. All other bounding curves separate either of Regimes $1$ or $2$, from Regime $3$ or $4$. Thus, these boundaries coincide with the emergence\,(disappearance) of a stable or unstable limit cycle from\,(into) either the vorticity axis\,($C=0$) or the flow-gradient plane\,($C = \infty$). Therefore, to obtain the functional forms for these bounding curves, we first obtain the leading order approximation for $\Delta C$ in the limits $C \rightarrow 0$ and $C \rightarrow \infty$. Equating the leading order term to zero then yields the boundary, in terms of $\epsilon$ expressed as a function of $\xi_0$, that corresponds to the emergence of a zero crossing in the vicinity of the spinning\,($C=0$; $\epsilon_0(\xi_0)$) or tumbling\,($C=\infty$; $\epsilon_\infty(\xi_0)$) mode. 

Considering the limiting form of $\Delta C$ for $C \rightarrow 0$ gives:
{\allowdisplaybreaks
    \begin{align}\nonumber
        \epsilon_0(\xi_0)&=\bigg(6 \xi_{0}^5-322 \xi_{0}^3+2 \big(\xi_{0}^2-1\big) \coth ^{-1}\xi_{0} \big(-6 \xi_{0}^4+128 \xi_{0}^2+3 \big(\xi_{0}^4+10 \xi_{0}^2-21\big) \xi_{0} \coth ^{-1}\xi_{0} \\ \nonumber
        &-1\big)+318 \xi_{0}\bigg) \bigg/ \bigg(\xi_{0} \big(9 \xi_{0}^4+303 \xi_{0}^2-314\big)+3 (\xi_{0}^{2}-1)\coth ^{-1}\xi_{0} \big(-6 \xi_{0}^4-86 \xi_{0}^2+3 \big(\xi_{0}^4-6 \xi_{0}^2\\\label{eqA34}
        &+13\big) \xi_{0} \coth ^{-1}\xi_{0}+2\big) \bigg),
    \end{align}
}on the prolate half of the plane; the extension of this boundary into the oblate half is obtained using the transformation $\xi_{0} \leftrightarrow \mathrm{i} \bar{\xi_{0}}$, $d \leftrightarrow -\mathrm{i}d$. The prolate and oblate portions of (\ref{eqA34}), taken together, corresponds to the curve in figure \ref{fig_kappa_epsilon}, one branch of which begins at $\kappa =6.963$ in the limit $\epsilon \rightarrow - \infty$, and asymptotes to $\epsilon = -1$ for $\kappa \rightarrow \infty$; the other branch originates at $\kappa =6.963$ for $\epsilon \rightarrow \infty$, and asymptotically approaches $\epsilon=-1/3$ for $\kappa\rightarrow 0$, passing through the point $(\kappa,\epsilon) \equiv (1,-3/8)$ in the process.

Similarly, considering the leading order approximation for $\Delta C$ for $C \rightarrow \infty$ gives:
{\allowdisplaybreaks
    \begin{align}\nonumber
        &\epsilon_\infty(\xi_0)=\bigg(8 \xi_{0} \big(\xi_{0}^2-1\big)^{3/2} \big(-9 \xi_{0}^6+483 \xi_{0}^4-477 \xi_{0}^2+\big(27 \xi_{0}^6-888 \xi_{0}^4+1025 \xi_{0}^2-162\big) \xi_{0} \coth ^{-1}\xi_{0}\\\nonumber
        &+3 \big(3 \xi_{0}^8+26 \xi_{0}^6-102 \xi_{0}^4+94 \xi_{0}^2-21\big) \xi_{0} (\coth ^{-1}\xi_{0})^3-\big(27 \xi_{0}^8-327 \xi_{0}^6+242 \xi_{0}^4+57 \xi_{0}^2+1\big) \\\nonumber
        &(\coth ^{-1}\xi_{0})^2\big)-\big(2 \xi_{0}^2-1\big) \big(-54 \xi_{0}^8+129 \xi_{0}^6-647 \xi_{0}^4+566 \xi_{0}^2+3 \big(\xi_{0}^2-1\big)^2 \big(18 \xi_{0}^6-25 \xi_{0}^4-34 \xi_{0}^2\\\nonumber
        &+81\big) \xi_{0} (\coth ^{-1}\xi_{0})^3+\big(162 \xi_{0}^8-441 \xi_{0}^6+1378 \xi_{0}^4-1683 \xi_{0}^2+588\big) \xi_{0} \coth ^{-1}\xi_{0}-\big(162 \xi_{0}^{10}\\\nonumber
        &-495 \xi_{0}^8+833 \xi_{0}^6-743 \xi_{0}^4+241 \xi_{0}^2+2\big) (\coth ^{-1}\xi_{0})^2\big)\bigg) \bigg/\bigg(4 \xi_{0} \big(\xi_{0}^2-1\big)^{3/2} \big(-3 \xi_{0}^2 \big(9 \xi_{0}^4+303 \xi_{0}^2\\\nonumber
        &-314\big)+\xi_{0} \big(81 \xi_{0}^6+1620 \xi_{0}^4-2037 \xi_{0}^2+332\big) \coth ^{-1}\xi_{0}+9 \xi_{0} \big(3 \xi_{0}^8-22 \xi_{0}^6+64 \xi_{0}^4-58 \xi_{0}^2+13\big) \\\nonumber
        &(\coth ^{-1}\xi_{0})^3+\big(-81 \xi_{0}^8-513 \xi_{0}^6+519 \xi_{0}^4+69 \xi_{0}^2+6\big) (\coth ^{-1}\xi_{0})^2\big)+2 \xi_{0} \big(2 \xi_{0}^2-1\big) \big(-27 \xi_{0}^7\\\nonumber
        &+54 \xi_{0}^5-315 \xi_{0}^3+9 \big(\xi_{0}^2-1\big)^2 \big(3 \xi_{0}^6-3 \xi_{0}^4-5 \xi_{0}^2+13\big) (\coth ^{-1}\xi_{0})^3-3 \big(27 \xi_{0}^8-72 \xi_{0}^6+126 \xi_{0}^4\\\label{eqA35}
        &-124 \xi_{0}^2+43\big) \xi_{0} (\coth ^{-1}\xi_{0})^2+\big(81 \xi_{0}^8-189 \xi_{0}^6+657 \xi_{0}^4-837 \xi_{0}^2+292\big) \coth ^{-1}\xi_{0}+284 \xi_{0}\big)\bigg),
    \end{align}
}again on the prolate side, with the oblate-side continuation obtained as described above. This corresponds to the bounding curve in figure \ref{fig_kappa_epsilon} that has the form $\epsilon = (2 \pi)\kappa^{-1}/(-1168 + 3 \pi (-8 + 39 \pi)) $ for $\kappa \rightarrow 0$, and approaches $\epsilon = -1/2$ in the limit $\kappa \rightarrow \infty$, again cutting the sphere-line at $\epsilon = -3/8$. It also intersects (\ref{eqA34}) at a second point $(-0.288,2.485)$.

\section{The small-$De$ orientation dynamics of a slender fiber in simple shear flow}\label{appB}

An independent approach to validate the arbitrary-aspect-ratio calculation, detailed in the main manuscript, is to analytically obtain the viscoelasticity-induced drift of a prolate spheroid, for large $\kappa$, using slender body theory\,\citep{cox1970,batchelor1970}. A calculation along these lines for the inertial case was carried out by \citet{ganesh2005}, who showed that the drift in this large-aspect-ratio limit was $O(Re/\ln \kappa)$. The simplicity of  the slender-body analytical framework relies on the outer-region contribution to fiber rotation being dominant. The outer and inner regions in this framework correspond, respectively, to scales of $O(L)$ and and $O(b)$ around the fiber. The dominance of the outer region implies that the fiber, for purposes of the drift calculation, can be treated as a line distribution of forces, which lends itself to a Fourier space approach. Now, scaling arguments in \citet{ganesh2005} showed that the leading order contribution to the volume integral in the reciprocal theorem formulation does come from the outer region, with the inner-region contribution being algebraically small in comparison; the logarithmic small disturbance field, when used in the linearized inertial terms in the outer region, leads to the $(\ln \kappa)^{-1}$ factor in the estimate above for the inertial angular velocity.

The reciprocal theorem framework of \S\ref{Problemform:recipro} yields the following expression for the angular velocity, $\bm{\Omega}^{(1)}$, of a torque-free fiber in viscoelastic simple shear flow:
\begin{equation}\label{eqB1}
    (\bm{\Omega}^{(2)} \wedge \bm{p}) \cdot (\bm{\Omega}^{(1)} \wedge \bm{p}-\bm{\Gamma} \cdot \bm{p})= \frac{3 De \ln \kappa}{8 \pi} \int_{V} \bm{\sigma}_{NN}^{\prime(1)} \bm{:} \nabla \bm{u}^{(2)} dV,
\end{equation}
where $\bm{p}$ denotes the fiber orientation, and $\bm{\Omega}^{(2)}$ is the angular velocity of the fiber in the test problem that now rotates about an axis perpendicular to $\bm p$. As already stated in the main manuscript, $\bm{\Gamma}$ is the transpose of the velocity gradient tensor for simple shear, with $\bm{\sigma}_{NN}^{\prime (1)}$ being the disturbance second-order fluid stress. 

Unlike the inertial case, the feasibility of a Fourier space approach is not obvious since the viscoelastic integrand in (\ref{eqB1}) involves, in effect, the cube of the velocity gradient\,($(\nabla \bm{u}^{(1)})^2. \nabla \bm{u}^{(2)}$) which would seem to favor the inner region - the velocity field is $O(\ln \rho)$ for $b \ll \rho \ll L$, $\rho$ being the distance from the fiber in the plane transverse to $\bm{p}$; as a result, the velocity gradient diverges as $O(1/\rho)$ in the same interval, appearing to favor contributions from smaller $\rho$. To better characterize the contributions from the different asymptotic regions, we start by noting that all disturbance fields are $O(\ln \kappa)^{-1}$ in the outer region. The non-Newtonian stress includes both terms linear and quadratic in $\bm{u}^{(1)}$, which therefore lead to $O(\ln \kappa)^{-2}$ and $O(\ln \kappa)^{-3}$ contributions to the integral in (\ref{eqB1}); the corresponding contributions to $\bm{\Omega}^{(1)}$ are $O(\ln \kappa)^{-1}$ and $O(\ln \kappa)^{-2}$. In the matching region, corresponding to $b \ll \rho \ll L$, $\bm{u}^{(1)}$ and $\bm{u}^{(2)}$ are $O(\ln \rho/\ln \kappa)$, as noted above. As a result, the linear and nonlinear terms in $\bm{\sigma}_{NN}^{\prime(1)}$ are $O(1/\rho \ln \kappa)$ and $O(1/\rho^2 \ln \kappa)$. Since $\nabla \bm{u}^{(2)} \sim O(1/\rho\ln \kappa)$, this leads to contributions to the integrand in (\ref{eqB1}) that diverge as $O(1/\rho^2)$ and $O(1/\rho^3)$, respectively. When combined with $dV \sim \rho d\rho ds$, $s$ being the axial coordinate, the linear terms lead to a logarithmically divergent integral\,($\propto (\ln \kappa)^{-2}\textstyle\int d\rho/\rho$), and the nonlinear terms to an algebraically divergent one\,($\propto (\ln \kappa)^{-3}\textstyle\int d\rho/\rho^2$). The latter integral turns out to be identically zero - from among the five Stokesian components in (\ref{eq:distvel_actual}), only $\bm{u}_{1s} \sim O(\ln \kappa)^{-1})$ is relevant in the slender-fiber limit, with the other four components being $O(\kappa^{-2})$. However, the nonlinear combination of $\bm{u}_{1s}$ with itself cannot induce a rotation on account of symmetry. Thus, only the logarithmically divergent integral from the linear terms survives, with the nonlinear (quadratic)\,terms only contributing to fiber rotation at $O(\ln \kappa/\kappa^2)$. But, the said divergence is only an apparent one, being truncated when $\rho \sim O(b)$\,(or $O(\kappa^{-1})$, in units of $L$) corresponding to the inner region. The resulting $O(\ln \kappa)$ estimate for the integral leads to an $O(1)$ contribution to the angular velocity, from the linearized terms in $\bm{\sigma}_{NN}^{\prime (1)}$, and corresponding to the matching region; smaller $O(\ln \kappa)^{-1}$ contributions arise from the inner and outer regions.

The dominance of the matching-region contribution, to leading logarithmic order, implies that the rotating fiber can still be treated as a line distribution of forces. The negligible volume associated with this singular line-forcing implies that volume integral in (\ref{eqB1}) can be extended to involve all of physical space, which allows for a Fourier space approach; as will be seen below, the resulting Fourier-space integral is a divergent one, although this divergence may be sensibly truncated to yield a finite answer. The Fourier-space approach involves first applying the convolution theorem to the integral in (\ref{eqB1}), so one obtains:
\begin{equation}\label{eqB2}
  (\bm{\Omega}^{(2)} \wedge \bm{p}) \cdot (\bm{\Omega}^{(1)} \wedge \bm{p}-\bm{\Gamma} \cdot \bm{p})= \frac{3 De \ln \kappa}{8 \pi}\int \hat{\bm{\sigma}}_{NN}^{\prime(1)}(-\bm{k}) \bm{:} (2\pi \mathrm{i}) \bm{k} \hat{\bm{u}}^{(2)}(\bm{k}) d\bm{k},
\end{equation}
which is the starting point of our calculations; here, $\hat{f}(\bm k) = \textstyle\int \mathrm{e}^{-2\pi \mathrm{i} \bm{k} \cdot \bm{r}}f(\bm r) dV$ defines the Fourier transform. In accordance with the scaling arguments above, to leading logarithmic order, one only need consider the linear terms in $\bm{\sigma}_{NN}^{\prime(1)}$, and non-Newtonian component of the Fourier transformed disturbance stress field, $\hat{\bm{\sigma}}_{NN}^{\prime(1)}(\bm{k})$ in (\ref{eqB2}), is therefore given by:
\begin{align}
\hat{\bm{\sigma}}_{NN}^{\prime(1)}(\bm{k}) =& 2 \epsilon   \Bigg( \frac{\partial \hat{\bm{e}}^{\prime(1)}(\bm{k})}{\partial t} +(\bm{\Gamma}^{\dagger} \cdot \bm{k}) \cdot \nabla_{k} \hat{\bm{e}}^{\prime(1)}(\bm{k}) +\bm{W} \cdot \hat{\bm{e}}^{\prime(1)}(\bm{k})+(\bm{W} \cdot \hat{\bm{e}}^{\prime(1)}(\bm{k}))^{\dagger} \nonumber\\
&+\hat{\bm{w}}^{\prime(1)} \cdot \bm{E}+(\hat{\bm{w}}^{\prime(1)}(\bm{k}) \cdot \bm{E})^{\dagger}  \Bigg)+4 (1+\epsilon) \left(\hat{\bm{e}}^{\prime(1)}(\bm{k}) \cdot \bm{E}+\bm{E} \cdot \hat{\bm{e}}^{\prime(1)}(\bm{k})\right), \label{eqB3}
\end{align}
where $\hat{\bm{e}}^{\prime(1)}(\bm{k}) =\pi \mathrm{i}\left(\bm{k} \hat{\bm{u}}^{\prime(1)}(\bm{k}) +  \hat{\bm{u}}^{\prime(1)}(\bm{k}) \bm{k} \right)$ and $\hat{\bm{w}}^{\prime(1)} = \pi \mathrm{i}\left(\bm{k} \hat{\bm{u}}^{\prime(1)}(\bm{k}) -  \hat{\bm{u}}^{\prime(1)}(\bm{k}) \bm{k} \right)$.
The analysis that follows is essentially a simpler Fourier-space alternative to the original \citet{leal1975} calculation, while also serving to correct the said author's results by a factor of $4$. 

The disturbance velocity field in the actual problem, to be used in (\ref{eqB3}), is given by \citep{ganesh2005}:
\begin{equation}\label{eqB4}
    \bm{u}^{\prime(1)}(\bm{k})=\int^{1}_{-1} \bm{G}_{0}(\bm{r}-s\bm{p}) \cdot \bm{f}(s;\bm{p}) ds,
\end{equation}
in the Stokes limit, with $\bm{f}(s;\bm{p})=\frac{-2\pi }{\ln \kappa} (\bm{ E: p p})\bm{p} s$ being the axial force density; note that this is the large-$\kappa$ limiting form of $\bm{u}_{1s}$. The test velocity field due to the Stokesian (transverse)\,rotation of the fiber, to be used in (\ref{eqB2}), is given by 
\begin{equation}\label{eqB5}
    \bm{u}^{(2)}(\bm{k})=\int^{1}_{-1} \bm{G}_{0}(\bm{r}-s\bm{p}) \cdot \tilde{\bm{f}}(s;\bm{p}) ds,
\end{equation}
with $\tilde{\bm{f}}(s;\bm{p}) =\frac{4 \pi}{\ln \kappa} (\bm{\Omega}^{(2)} \wedge \bm{p})s $ being the transverse force density induced by the rotating fiber. In (\ref{eqB4}) and (\ref{eqB5}), $\bm{G}_{0}(\bm{r}) = \frac{1}{8 \pi}\left(\frac{\bm{I}}{r}+\frac{\bm{rr}}{r^3} \right)$ is the Oseen-Burgers tensor. The Fourier transformed actual and test velocity fields are then given by
\begin{equation}\label{eqB6}
    \hat{\bm{u}}^{\prime(1)}(\bm{k})= \frac{4 \pi \mathrm{i}}{\ln \kappa} j_{1}(2 \pi \bm{k \cdot p})  (\bm{ E: p p})\,\,\hat{\bm{G}}_{0}(\bm{k}) \bm{\cdot p},
\end{equation}
    \begin{equation}\label{eqB7}
        \hat{\bm{u}}^{(2)}(\bm{k})= \frac{-8 \pi \mathrm{i}}{\ln \kappa} \, j_{1}(2 \pi \bm{k \cdot p}) \, 
        (\bm{\Omega}^{(2)} \wedge \bm{p}) \cdot \hat{\bm{G}}_{0}(\bm{k}),
    \end{equation}
where $j_{1}(z)$ is the spherical Bessel function of the first kind, of order $1$, with $\hat{\bm{G}}_{0}(\bm{k})=\frac{1}{(2\pi)^{2}}\left( \frac{\bm{I}}{k^{2}}-\frac{\bm{kk}}{k^4}\right)$. Substituting (\ref{eqB7}) in (\ref{eqB2}), we get 
\begin{equation}\label{eqB8}
\begin{aligned}
&(\bm{\Omega}^{(2)} \wedge \bm{p}) \cdot (\bm{\Omega}^{(1)} \wedge \bm{p}-\bm{\Gamma} \cdot \bm{p})\\
&= \frac{3 De \ln \kappa}{8 \pi}  \int \hat{\bm{\sigma}}_{NN}^{\prime(1)}(-\bm{k}) \bm{:} (2\pi \mathrm{i}) \bm{k} \left( \frac{-8 \pi \mathrm{i}}{\ln \kappa} \, j_{1}(2 \pi \bm{k \cdot p}) \, 
        (\bm{\Omega}^{(2)} \times \bm{p}) \cdot \hat{\bm{G}}_{0}(\bm{k}) \right) d\bm{k}.
\end{aligned}
\end{equation}
Since $\bm{\Omega}^{(2)}$ in (\ref{eqB8}) is only constrained to be transverse to $\bm{p}$, but is otherwise arbitrary, one obtains:
\begin{equation}\label{eqB9}
\begin{aligned}
\dot{\bm{p}}=\bm{\Omega}^{(1)} \wedge \bm{p}=(\bm{\Gamma} \cdot \bm{p}) \cdot (\bm{I} - \bm{p}\bm{p}) + 6 \pi  De   \int \hat{\bm{\sigma}}_{NN}^{\prime(1)}(-\bm{k}) \bm{:}  \bm{k}\hat{\bm{G}}_{0}(\bm{k}) j_{1}(2 \pi \bm{k \cdot p})  d\bm{k},
\end{aligned}
\end{equation}
where the first term corresponds to the Jeffery rotation of a slender fiber. It is convenient to separately consider the corotational and quadratic stress contributions defined in (\ref{eqB3}).
We begin with the integral for the quadratic contribution, given by:
\begin{equation}\label{eqB10}
    \begin{aligned}
    &\int \hat{\bm{\sigma}}_{NNQ}^{\prime(1)}(-\bm{k}) \bm{:}  \bm{k}\hat{\bm{G}}_{0}(\bm{k}) j_{1}(2 \pi \bm{k \cdot p}) d\bm{k}\\
        &=-4\pi \mathrm{i}(1+\epsilon) \int  \left( \left(\bm{k} \hat{\bm{u}}^{\prime(1)}(-\bm{k}) +  \hat{\bm{u}}^{\prime(1)}(-\bm{k}) \bm{k} \right) \cdot \bm{E} + \bm{E} \cdot \left(\bm{k} \hat{\bm{u}}^{\prime(1)}(-\bm{k}) +  \hat{\bm{u}}^{\prime(1)}(-\bm{k}) \bm{k} \right)\right) \\
        &\bm{:} \bm{k} \, 
        \hat{\bm{G}}_{0}(\bm{k})  \,  j_{1}(2 \pi \bm{k \cdot p}) d\bm{k}.\\
    \end{aligned}
\end{equation}
Substituting (\ref{eqB6}) in (\ref{eqB10}) leads to:
\begin{equation}\label{eqB11}
    \begin{aligned} 
    \int \hat{\bm{\sigma}}_{NNQ}^{\prime(1)}(-\bm{k}) \bm{:}  \bm{k}  \,\hat{\bm{G}}_{0}(\bm{k}) \, j_{1}(2 \pi \bm{k \cdot p}) d\bm{k}
=\frac{-(1+\epsilon) (\bm{ E: p p})}{ \pi^2 \ln \kappa} \left( \bm{Q}_1+\bm{Q}_2+\bm{Q}_3+\bm{Q}_4 \right),
    \end{aligned}
\end{equation}
where, with $\bm{p} = \bm{1}_{3}$, one has:
 \begin{equation}\label{eqB12}
     \begin{rcases}
         Q_{1i}=\int   E_{jp} k^{2} j_{1}^{2}(2 \pi k_{3})  \left( \frac{\delta_{3j}}{k^{2}}-\frac{k_{3}k_{j}}{k^4}\right)   \left( \frac{\delta_{ip}}{k^{2}}-\frac{k_{i}k_{p}}{k^4}\right)   d\bm{k},\\
        {Q}_{2i}=\int   E_{jp} k_{j} k_{k} j_{1}^{2}(2 \pi k_{3})  \left( \frac{\delta_{3k}}{k^{2}}-\frac{k_{3}k_{k}}{k^4}\right)   \left( \frac{\delta_{ip}}{k^{2}}-\frac{k_{i}k_{p}}{k^4}\right)   d\bm{k},\\
         {Q}_{3i}=\int   E_{kj} k_{j} k_{k} j_{1}^{2}(2 \pi k_{3})  \left( \frac{\delta_{3p}}{k^{2}}-\frac{k_{3}k_{p}}{k^4}\right)   \left( \frac{\delta_{ip}}{k^{2}}-\frac{k_{i}k_{p}}{k^4}\right)   d\bm{k},\\
        {Q}_{4i}=\int   E_{jp} k_{p} k_{k} j_{1}^{2}(2 \pi k_{3})  \left( \frac{\delta_{3j}}{k^{2}}-\frac{k_{3}k_{j}}{k^4}\right)   \left( \frac{\delta_{ip}}{k^{2}}-\frac{k_{i}k_{p}}{k^4}\right)   d\bm{k}.\\
     \end{rcases}
 \end{equation}
The above integrals are evaluated in spherical polar coordinates with $\bm p = {\bm 1}_3$ as the polar axis. Analytical evaluation of the angular integrals leads to a residual integral over $k$, of the form $\textstyle\int_0^\infty f(k) dk$, where $f(k)$ is finite at $k = 0$, while being $O(1/k)$ for $k \rightarrow \infty$, pointing to a logarithmic divergence in the latter limit. This large-$k$ divergence mirrors the small-$\rho$ divergence mentioned in the context of the scaling estimates above. Specifically, truncating the latter divergence at a $\rho$ of $O(\kappa^{-1})$ led one to the leading order matching-region contribution. The analog of this truncation in Fourier space would be to terminate the $k$-integration at $k_{max} \sim O(\kappa)$. The resulting $\ln \kappa$ factor cancels out the $\ln \kappa$ in (\ref{eqB11}), leading to the following $O(1)$ quadratic contribution in the slender fiber limit:
\begin{equation}\label{eqB13}
    \begin{aligned}
      &\int \hat{\bm{\sigma}}_{NNQ}^{\prime(1)}(-\bm{k}) \bm{:}  \bm{k}\hat{\bm{G}}_{0}(\bm{k}) j_{1}(2 \pi \bm{k \cdot p}) d\bm{k}  =- \frac{(1+\epsilon)}{6 \pi }({\bm E}:\bm{p}\bm{p})(\bm{E} \cdot\bm{p}).
    \end{aligned}
\end{equation}
It is worth noting that, after evaluation of the integral over the azimuthal angle alone, the divergence of the 2D Fourier integral over $k$ and $\theta$\,(the polar angle) arises in the limit $k \rightarrow \infty, \theta\rightarrow \frac{\pi}{2}$. This asymptotically distant portion of the equatorial plane, in Fourier space, corresponds to the asymptotically thin cylindrical annulus, $\rho \rightarrow 0$, in physical space\,(the inner region in the slender body theory framework).

Next, considering the corotational stress contribution in (\ref{eqB9}), and using (\ref{eqB3}), one obtains:
\begin{equation}\label{eqB14}
    \begin{aligned}
        &\int \hat{\bm{\sigma}}_{NNC}^{\prime(1)}(-\bm{k}) \bm{:}  \bm{k}  \,\hat{\bm{G}}_{0}(\bm{k}) \, j_{1}(2 \pi \bm{k \cdot p}) d \bm{k}
       =2 \epsilon \int \Bigg( \underbrace{ \frac{\partial \hat{\bm{e}}^{\prime(1)}(-\bm{k})}{\partial t} }_\text{term 1}-\underbrace{(\bm{\Gamma}^{\dagger} \cdot \bm{k})  \cdot \nabla_{\bm{k}}\hat{\bm{e}}^{\prime(1)}(-\bm{k})}_\text{term 2} \\
&\underbrace{+\bm{W} \cdot \hat{\bm{e}}^{\prime(1)}(-\bm{k})+(\bm{W} \cdot \hat{\bm{e}}^{\prime(1)}(-\bm{k}))^{\dagger}+\hat{\bm{w}}^{\prime(1)}(-\bm{k}) \cdot \bm{E}+(\hat{\bm{w}}^{\prime(1)}(-\bm{k}) \cdot \bm{E})^{\dagger}}_\text{term 3}  \Bigg) \bm{:}  \bm{k}  \,   \hat{\bm{G}}_{0}(\bm{k}) \, j_{1}(2 \pi \bm{k \cdot p}) d\bm{k},\\
    \end{aligned}
\end{equation}
where, to leading order in $De$, the time rate of change can be evaluated using the Jeffery relation\,(the first term in (\ref{eqB9}) denoted $\dot{\bm{p}}_{jeff}$ herein). We address each term in (\ref{eqB14}) individually. Considering term $1$:
\begin{equation}\label{eqB15}
    \begin{aligned}
       & 2 \epsilon \int   \frac{\partial \hat{\bm{e}}^{\prime(1)}(-\bm{k})}{\partial t}  \bm{:}  \bm{k}  \, 
        \hat{\bm{G}}_{0}(\bm{k}) \, j_{1}(2 \pi \bm{k \cdot p})d\bm{k}\\
       & =2 \epsilon (-\pi i)\int   \left( \bm{k} \frac{\partial \hat{\bm{u}}^{\prime(1)}(-\bm{k})}{\partial t}  +  \frac{\partial \hat{\bm{u}}^{\prime(1)} (-\bm{k})}{\partial t} \bm{k}\right)  \bm{:}  \bm{k}
        \hat{\bm{G}}_{0}(\bm{k}) \, j_{1}(2 \pi \bm{k \cdot p}) d\bm{k}
        =\frac{-\epsilon  }{2 \pi^{2} \ln \kappa} (\bm{T}_1+\bm{T}_2), \\
    \end{aligned}
\end{equation}
where
\begin{equation}\label{eqB16}
    \begin{rcases}
    &T_{1i}=\int  E_{mn} k^{2} \left( \frac{\delta_{lp}}{k^{2}}-\frac{k_{l}k_{p}}{k^4}\right)  \frac{\partial}{\partial t} \left[ p_{m}p_{n} p_{l} j_{1}(2 \pi k_{r} p_{r}) \right]   j_{1}(2 \pi k_q  p_q) \left( \frac{\delta_{ip}}{k^{2}}-\frac{k_{i}k_{p}}{k^4}\right) d\bm{k},    \\ 
    &T_{2i}=\int  E_{mn} k_{p}  k_{k}\left( \frac{\delta_{lk}}{k^{2}}-\frac{k_{l}k_{k}}{k^4}\right)  \frac{\partial}{\partial t} \left[  p_{m}p_{n} p_{l} j_{1}(2 \pi k_{r} p_{r})   \right]  j_{1}(2 \pi k_q  p_q) \left( \frac{\delta_{ip}}{k^{2}}-\frac{k_{i}k_{p}}{k^4}\right) d\bm{k}.\\ 
    \end{rcases}
\end{equation}
Performing the integrations, with the final logarithmically divergent $k$-integral being truncated in the manner described above, one obtains:
\begin{equation}\label{eqB17}
    \begin{aligned}
        &2 \epsilon \int   \frac{\partial \hat{\bm{e}}^{\prime(1)}(-\bm{k})}{\partial t}  \bm{:}  \bm{k}  \, 
        \hat{\bm{G}}_{0}(\bm{k}) \, j_{1}(2 \pi \bm{k \cdot p}) d\bm{k} =\frac{\epsilon  }{24 \pi } ({\bm E}:\bm{p}\bm{p}) \dot{\bm{p}}_{jeff}.
        \end{aligned}
\end{equation}
Next, term $2$ is written as:
\begin{equation}\label{eqB18}
    \begin{aligned}
-2 \epsilon \int  (\bm{\Gamma}^{\dagger} \cdot \bm{k})  \cdot \nabla_{\bm{k}}\hat{\bm{e}}^{\prime(1)}(-\bm{k})    \bm{:}  \bm{k}  \, \hat{\bm{G}}_{0}(\bm{k}) \, j_{1}(2 \pi \bm{k \cdot p}) d\bm{k}=\frac{ \epsilon }{2 \pi^{2} \ln \kappa}(\bm{ E} :\bm{ p p})( \bm{A}_1+ \bm{A}_2),
    \end{aligned}
\end{equation}
 where
 \begin{equation}\label{eqB19}
     \begin{rcases}
     &A_{1i}= \int  \Gamma^{\dagger}_{jm}  k_{m}   k_{k}  j_{1}(2 \pi k_{3}) \,  \left( \frac{\delta_{ip}}{k^{2}}-\frac{k_{i}k_{p}}{k^4}\right) \frac{\partial}{\partial k_{j}} \bigg( k_{k} \, j_{1}(2 \pi k_{3}) \,\left( \frac{\delta_{3p}}{k^{2}}-\frac{k_{3}k_{p}}{k^4}\right)   \bigg)   d\bm{k},\\
     &A_{2i}= \int  \Gamma^{\dagger}_{jm}  k_{m}   k_{k}  j_{1}(2 \pi k_{3}) \,  \left( \frac{\delta_{ip}}{k^{2}}-\frac{k_{i}k_{p}}{k^4}\right) \frac{\partial}{\partial k_{j}} \bigg( k_{p} \, j_{1}(2 \pi k_{3}) \,\left( \frac{\delta_{3k}}{k^{2}}-\frac{k_{3}k_{k}}{k^4}\right)   \bigg)   d\bm{k}.\\
     \end{rcases}
 \end{equation}
Evaluating the integrals leads to:
 \begin{equation}\label{eqB20}
    \begin{aligned}
-2 \epsilon \int  (\bm{\Gamma}^{\dagger} \cdot \bm{k})  \cdot \nabla_{\bm{k}}\hat{\bm{e}}^{\prime(1)}(-\bm{k})    \bm{:}  \bm{k}  \, \hat{\bm{G}}_{0}(\bm{k}) \, j_{1}(2 \pi \bm{k \cdot p}) d\bm{k} =\frac{-\epsilon  }{8 \pi }({\bm E}:\bm{p}\bm{p}) (\bm{\Gamma}\cdot \bm{p}).
    \end{aligned}
\end{equation}
Finally, term $3$ is written in the form:
\begin{equation}\label{eqB21}
    \begin{aligned}
        2 \epsilon &\int \Bigg(\bm{W} \cdot \hat{\bm{e}}^{\prime(1)}(-\bm{k})+(\bm{W} \cdot \hat{\bm{e}}^{\prime(1)}(-\bm{k}))^{\dagger}+\hat{\bm{w}}^{\prime(1)}(-\bm{k}) \cdot \bm{E}+(\hat{\bm{w}}^{\prime(1)}(-\bm{k}) \cdot \bm{E})^{\dagger}  \Bigg) \\
        &\bm{:}  \bm{k} \, 
        \hat{\bm{G}}_{0}(\bm{k}) \, j_{1}(2 \pi \bm{k \cdot p})  d\bm{k}  =\frac{-\epsilon (\bm{ E} :\bm{ p p}) }{2 \pi^{2} \ln \kappa} (\bm{S}_1+\bm{S}_2+\bm{S}_3+\bm{S}_4+\bm{S}_5+\bm{S}_6+\bm{S}_7+\bm{S}_8),
    \end{aligned}
\end{equation}
where
\begin{equation}\label{eqB22}
    \begin{rcases}
        &S_{1i}=\int W_{kj} k_{j} k_{k} j_{1}^{2}(2 \pi k_{3}) \,\left(  \frac{\delta_{3p}}{k^{2}}-\frac{k_{3}k_{p}}{k^4}\right)    
        \left(  \frac{\delta_{ip}}{k^{2}}-\frac{k_{i}k_{p}}{k^4}\right)  d\bm{k},\\
    &S_{2i}=\int  W_{kj} k_{p} k_{k} j_{1}^{2}(2 \pi k_{3}) \,\left(  \frac{\delta_{3j}}{k^{2}}-\frac{k_{3}k_{j}}{k^4}\right)    
        \left(  \frac{\delta_{ip}}{k^{2}}-\frac{k_{i}k_{p}}{k^4}\right)  d\bm{k},\\
        &S_{3i}=-\int W_{jp} k^{2}  j_{1}^{2}(2 \pi k_{3}) \,\left(  \frac{\delta_{3j}}{k^{2}}-\frac{k_{3}k_{j}}{k^4}\right)    
        \left(  \frac{\delta_{ip}}{k^{2}}-\frac{k_{i}k_{p}}{k^4}\right)  d\bm{k},\\
    &S_{4i}=-\int  W_{jp} k_{j} k_{k} j_{1}^{2}(2 \pi k_{3}) \,\left(  \frac{\delta_{3k}}{k^{2}}-\frac{k_{3}k_{k}}{k^4}\right)    
        \left(  \frac{\delta_{ip}}{k^{2}}-\frac{k_{i}k_{p}}{k^4}\right)  d\bm{k},\\
        &S_{5i}=\int E_{jp}  k^{2} j_{1}^{2}(2 \pi k_{3}) \,\left(  \frac{\delta_{3j}}{k^{2}}-\frac{k_{3}k_{j}}{k^4}\right)    
        \left(  \frac{\delta_{ip}}{k^{2}}-\frac{k_{i}k_{p}}{k^4}\right)  d\bm{k},\\
    &S_{6i}=\int  E_{jp} k_{j} k_{k} j_{1}^{2}(2 \pi k_{3}) \,\left(  \frac{\delta_{3k}}{k^{2}}-\frac{k_{3}k_{k}}{k^4}\right)    
        \left(  \frac{\delta_{ip}}{k^{2}}-\frac{k_{i}k_{p}}{k^4}\right)  d\bm{k},\\
        &S_{7i}=-\int  E_{jk} k_{j} k_{k} j_{1}^{2}(2 \pi k_{3}) \,\left(  \frac{\delta_{3p}}{k^{2}}-\frac{k_{3}k_{p}}{k^4}\right)    
        \left(  \frac{\delta_{ip}}{k^{2}}-\frac{k_{i}k_{p}}{k^4}\right)  d\bm{k},\\
        &S_{8i}=\int E_{kj} k_{p}  k_{k} j_{1}^{2}(2 \pi k_{3}) \,\left(  \frac{\delta_{3j}}{k^{2}}-\frac{k_{3}k_{j}}{k^4}\right)    
        \left(  \frac{\delta_{ip}}{k^{2}}-\frac{k_{i}k_{p}}{k^4}\right)  d\bm{k}.\\
    \end{rcases}    
\end{equation}
Again, evaluating the integrals leads to:
\begin{equation}\label{eqB23}
    \begin{aligned}
        2 \epsilon  \int & \Bigg( \bm{W} \cdot \hat{\bm{e}}^{\prime(1)}(-\bm{k})+(\bm{W} \cdot \hat{\bm{e}}^{\prime(1)}(-\bm{k}))^{\dagger}+\hat{\bm{w}}^{\prime(1)}(-\bm{k}) \cdot \bm{E}+(\hat{\bm{w}}^{\prime(1)}(-\bm{k}) \cdot \bm{E})^{\dagger}  \Bigg) \\
       & \bm{:}  \bm{k}  \, 
        \hat{\bm{G}}_{0}(\bm{k}) \, j_{1}(2 \pi \bm{k \cdot p}) d\bm{k}  =\frac{-\epsilon  }{12 \pi} ({\bm E}:\bm{p}\bm{p}) 
 (\bm{W} \cdot \bm{p}+\bm{E} \cdot \bm{p}).
        \end{aligned}
        \end{equation}
Substituting equations (\ref{eqB13}),  (\ref{eqB17}), (\ref{eqB20}) and (\ref{eqB23}) into (\ref{eqB9}), one obtains
\begin{equation}\label{eqB24}
\begin{aligned}
\dot{\bm{p}}&=(\bm{\Gamma} \cdot \bm{p}) \cdot (\bm{I} - \bm{p}\bm{p}) + De ({\bm E}:\bm{p}\bm{p})  \left(\epsilon\left( \frac{\dot{\bm{p}}_{jeff}  }{4}   -\frac{3  \bm{\Gamma} \cdot \bm{p} }{4 }   -\frac{ \bm{W}\cdot \bm{p}+\bm{E}\cdot \bm{p}}{2}  \right)- (1+\epsilon)  \,\bm{E}\cdot \bm{p} \right), \\
&= (\bm{\Gamma} \cdot \bm{p}) \cdot (\bm{I} - \bm{p}\bm{p}) + De ({\bm E}:\bm{p}\bm{p}) \left( \epsilon \left( -\bm{E}\cdot \bm{p}-\frac{({\bm E}:\bm{p}\bm{p})\bm{p}}{4} \right)- (1+\epsilon)  \,\bm{E}\cdot \bm{p}\right),
\end{aligned}
\end{equation}
on using the expression for $\dot{\bm p}_{jeff}$. The components of $\dot{\bm p}$ may be written in terms of $\dot{\theta}_{j}$ and $\dot{\phi}_{j}$, in which case (\ref{eqB24}) takes the form:
 \begin{eqnarray}\label{eqB25}
     &&\dot{\theta}_{j} = \sin \theta_{j} \cos \theta_{j} \sin \phi_{j} \cos \phi_{j}+De (1+2 \epsilon)\left( \cos \theta_{j} \cos^{2} \phi_{j}\sin^{3} \theta_{j}  \sin^{2} \phi_{j}  \right),\\ \label{eqB26}
    &&\dot{\phi}_{j} = -\sin^{2} \phi_{j} +De\frac{(1+2 \epsilon)}{8} \left(  \sin^{2} \theta_{j}  \sin 4 \phi_{j}\right),
 \end{eqnarray}
which are in agreement with the exact expressions, given by (\ref{eq41}-\ref{eq42}) in \S\ref{Problemform:visco} of the main manuscript, when the large-$\kappa$ limiting forms of the various $F_i$'s and $G_i$'s are used therein.

Note that the truncation of the logarithmic divergent $k$-integrals at a $k_{max}$ of $O(\kappa)$, mentioned above, implies that the leading order correction to (\ref{eqB24}) is $O(\ln \kappa)^{-1}$. The logarithmically small estimate of the neglected terms is consistent with the generic slow approach, towards the slender body limiting forms, already seen earlier - for instance, refer to figure \ref{fig_shape_fun_prolate} in \S\ref{appA}, where the prolate aspect-ratio functions, when plotted as a function of $e/(1-e)$, exhibit a very slow approach to their respective slender-fiber plateaus; and figures \ref{fig6a}-\ref{fig6f}, in the main manuscript, which plot the orbital drift as a function of $C$ and which show a perceptible separation between slender fiber asymptote\,(based on (\ref{eqB24}) above), and the exact drift, even for $\kappa = 10^4$.

The equations for slender-fiber rotation, to $O(De)$, and to leading logarithmic order, were originally determined by \cite{leal1975}\,(see equations ($58a$) and ($58b$) therein). Translated to the present notation, they are of the form: 
\begin{eqnarray}\label{eqB27}
    &&(\dot{\theta}_{j})_{De}=-2 K M_1  (2 \epsilon +1) \sin ^3 \theta_{j} \cos \theta_{j} \sin ^2 \phi_j \cos ^2 \phi_j,\\ \label{eqB28}
    &&(\dot{\phi}_{j})_{De}=K M_1  (2 \epsilon+1) \sin ^2\theta_{j} \sin \phi_j  \cos \phi_j (\sin^2\phi_j-\cos^2\phi_j), 
\end{eqnarray}
where $K=-\frac{3De}{16 \ln \kappa}$  and $M_1=\int_{-1+\kappa^{-1}}^{1+\kappa^{-1}}dx_1\int_{R(x_1)}^{c_1}\frac{x_{1}^{2}}{\rho} d\rho$ in the limit $\kappa \gg 1$, with $R(x_{1})$ describing the longitudinal profile of the fiber. As is well known, slender body theory allows for an arbitrary longitudinal profile, provided only that the characteristic scale along the longitudinal direction\,(that now plays the role of $L$) is much larger than the fiber radius. For a slender spheroid, $R = \kappa^{-1} \sqrt{1-x_{1}^{2}}$, and $M_1=\frac{2 \ln \kappa}{3}$, so the prefactor in (\ref{eqB27}-\ref{eqB28}) simplifies to $K M_{1} =-De/8$, leading to the reduced forms: 
\begin{eqnarray}\label{eqB29}
     &&(\dot{\theta}_{j})_{De}=\frac{1}{4} (2 \epsilon +1) \sin ^3 \theta_{j} \cos \theta_{j} \sin ^2 \phi_j \cos ^2 \phi_j, \\ \label{eqB30}
    &&(\dot{\phi}_{j})_{De}= \frac{1}{32} (2 \epsilon+1) \sin ^2\theta_{j} \sin 4 \phi_j.
\end{eqnarray}
These differ from (\ref{eq41}-\ref{eq42}), and the $O(De)$ contributions in (\ref{eqB25}-\ref{eqB26}) by a factor of $4$.
	\bibliographystyle{jfm}
	\bibliography{jfm.bib}
	
\end{document}